\documentclass[12pt,preprint]{aastex}
\usepackage{natbib}

\shorttitle{Massive YSOs in LMC: Identification and Classification}
\shortauthors{Seale et al.}

\begin{document}

\title{The Evolution of Massive YSOs in the LMC: \\Part I. Identification and Spectral Classification}

\author{Jonathan P. Seale, Leslie W. Looney, You-Hua Chu, Robert A. Gruendl}
\affil{Astronomy Department, University of Illinois \\ Department of Astronomy, MC-221 \\ 1002 W. Green Street \\ Urbana, IL 61801}

\author{Bernhard Brandl}
\affil{Leiden Observatory, Leiden University\\ Niels Bohrweg 2 \\ 2300 RA Leiden, The Netherlands}

\author{C.-H. Rosie Chen}
\affil{Astronomy Department, University of Vigrinia \\ Department of Astronomy \\P.O. Box 400325 \\ Charlottesville, VA 22904-4325}

\author{Wolfgang Brandner}
\affil{Max-Planck-Institut f\"{u}r Astronomie,  Planet \& Star Formation \\ K\"{o}nigstuhl 17 \\ 69117 Heidelberg, Germany}

\author{Geoffrey A. Blake}
\affil{Astronomy Department, Caliornia Institute of Technology \\ Caltech, MC 105-24 \\ 1200 East California Blvd \\ Pasadena, CA 91125 }

\begin{abstract}

We present and categorize \textit{Spitzer} IRS spectra of 294 objects in the Large Magellanic Cloud (LMC) to create the largest and most complete catalog of massive young stellar object (YSO) spectra in the LMC. Target sources were identified from infrared photometry and multi-wavelength images indicative of young, massive stars highly enshrouded in their natal gas and dust clouds. Several objects have been spectroscopically identified as non-YSOs and have features similar to more evolved stars such as red supergiants, asymptotic giant branch (AGB), and post-AGB stars. Our sample primarily consists of 277 objects we identify as having spectral features indicative of embedded YSOs. The remaining sources are comprised of 7 C-rich evolved sources, 8 sources dominated by broad silicate emission, and 1 source with multiple broad emission features. Those with YSO-like spectra show a range of spectral features including polycyclic aromatic hydrocarbon emission, deep silicate absorption, fine-structure lines, and ice absorption features. Based upon the relative strengths of these features, we have classified the YSO candidates into several distinct categories using the widely-used statistical procedure known as principal component analysis. We propose that these categories represent a spectrum of evolutionary stages during massive YSO formation. Using our catalog we put statistical constraints on the relative evolutionary timescale of processes involved in massive star formation. We conclude that massive pre-main sequence stars spend a majority (possibly as high as $90\%$) of their massive, embedded lives emitting in the UV. Half of the sources in our study have features typical of compact HII regions, suggesting that massive YSOs can create a detectable compact HII region half-way through the formation time present in our sample. This study also provides a check on commonly used source-selection procedures including the use of photometry to identify YSOs. We determine a high success rate ($>95\%$) of identifying objects with YSO-like spectra can be achieved through careful use of infrared color magnitude diagrams, spectral energy distributions, and image inspections.

\end{abstract}


\section{Introduction}
During all stages of their lives, stars play an important role in the evolution of their host galaxy.  Massive stars, in particular, have a strong influence on galactic properties \citep[e.g.][]{kenn05}, as they are the principal source of heavy elements and UV radiation.  A single star can influence the structure and chemistry of its local interstellar medium \citep[e.g.][]{vand98,kwon06,seal08}, and the combined effects of many evolving stars can shape galactic superstructures spanning 1000's of parsecs \citep[e.g.][]{mccr87,teno88}. More specifically, during their formation, massive stars inject energy into the ISM via  powerful jets and outflows \citep[e.g.][]{shep96}. As they evolve toward and onto the main-sequence, massive stars become sources of copious amounts of UV radiation that energizes the surrounding medium creating compact HII regions \citep[see reviews by][]{gara99,beut07}. OB associations, groups of massive stars, can generate enough UV radiation to ionize large regions of the interstellar medium (ISM). Finally, massive stars end their relatively short lives with supernovae that stir up the ISM by their strong explosions. The combined effects of supernovae and fast stellar winds can form large superbubbles or giant interstellar shells that propagate star formation to distances of hundreds or thousands of parsecs \citep[e.g.][]{mccr87,chu05,chen09,book09}. In addition, it is now commonly speculated that most stars -- including our own Sun \citep[e.g.][]{loon06} -- form in clusters, and within these clusters, massive stars dominate the overall luminosity, greatly affecting any nearby star or planet formation. Clearly, massive stars play a considerable role in galactic ISM structure and overall stellar content.

To understand the formation of massive stars, it is necessary to identify massive young stellar objects (YSOs) throughout a galaxy. YSOs are surrounded by dense envelopes of gas and dust that serve as the young stars' mass reservoirs for accretion \citep[see reviews by][]{zinn07,mcke07}. Photospheric optical and UV radiation from the embedded central star absorbed by hundreds of magnitudes of extinction from the circumstellar gas and dust envelope is re-radiated in the infrared (IR).  Massive YSOs spend a large fraction of their relatively short lives in these embedded stages and are therefore best identified by their point-source IR radiation \citep{chur02}.  The observed IR emission, reprocessed stellar radiation, reveals information about the state of the circumstellar material \citep{vand98}. 

While the formation of low-mass stars can be studied in great detail and at high spatial resolution in nearby Galactic star forming regions \citep[e.g.][]{hart05,luhm99,tobi07,seal08}, massive stars are much rarer, making observational studies more difficult. High-mass star formation studies within the Milky Way are plagued by numerous problems -- a low number of observable sources, dust obscuration along the Galactic plane, and distance ambiguities leading to high luminosity uncertainties.

The formation of individual massive stars in other galaxies could not be resolved until recently. The advent of the \textit{Spitzer Space Telescope} \citep[\textit{SST};][]{wern04} has enabled large surveys to identify YSOs in the Galaxy.  Its superb angular resolution makes it possible to identify YSOs even in nearby galaxies, such as the Large and Small Magellanic Clouds (LMC \& SMC).  The LMC is an excellent laboratory to study high-mass star formation as it is located at a small and known distance of 50 kpc, where 1$''$ corresponds to 0.25 pc \citep{feas99}; furthermore, its low foreground extinction and nearly face-on orientation provides a clear view throughout the entire galaxy. Thus, surveys of high-mass YSOs in the LMC have been conducted by \citet{whit08} and \citet{grue09}.

We have used the \textit{SST}'s Infrared Spectrometer \citep[IRS;][]{houc04} to obtain mid-IR spectra of nearly 300 massive YSO candidates in the LMC to assess and confirm their nature.  These observations represent the most extensive and complete collection of massive YSO spectra to date. These infrared spectra are rich with spectral features from a wide variety of species, some of which are unobservable at other wavelengths. Among the most interesting features are ice-phase molecular absorptions, deep silicate absorptions, polycyclic aromatic hydrocarbon (PAH) emission, and ionic fine-structure emission lines.

Such a large spectral survey allows meaningful comparisons with theoretically or empirically derived formation and evolutionary paths of massive stars, as reviewed by \citet{zinn07}, \citet{beut07}, \citet{krum07}, and \citet{gara99}. Star formation typically begins with the collapse of a high-density core within a large clump of a molecular cloud. A protostar forms and acquires its mass by accreting from its natal cloud via a circumstellar disk. The forming star heats the dense massive core, turning it into a hot dense massive core. Eventually, hydrogen burning will begin. It is commonly accepted that massive stars, unlike their low-mass counterparts, begin hydrogen burning while still accreting material. At this stage, the circumstellar material including the accretion disk may become photoionized, a possible origin of the so-called hypercompact HII region \citep{keto07}. Initially, the accretion flow is able to quench the ionizing flux, but once accretion slows or becomes geometrically unable to contain the UV flux, the ionizing radiation can escape into the surrounding regions creating ultracompact/compact HII regions and eventually, larger classical HII regions. Over time, stellar winds and radiation will disperse the circumstellar material, revealing a newly-formed high-mass star.

In this paper we will use the term ``YSO'' to describe the young central object-disk-envelope system that will eventually become a main-sequence star. The term ``protostar'' is used more specifically to describe an object that will eventually become a main-sequence star but has yet to begin hydrogen burning. The sample of YSOs studied here may all be hydrogen-burning YSOs that are still highly embedded and may be accreting. Additionally, a large majority of them probably have massive central objects ($\ga$10$M_{\sun}$; see \S 3.5). Our sample of sources have mid-IR emission from warm dust, and thus are highly embedded but already contain a central heating source. True high-mass protostars, i.e. pre-hydrogen burning accreting sources with masses above 10$M_{\sun}$, likely only exist very briefly. Since the YSOs in our \textit{SST} IRS spectral catalog are likely high-mass, most (possibly all) have already begun hydrogen burning.

The spectra of our objects show great variations that may reflect an evolutionary sequence during a high-mass YSO's formation. This paper discusses the spectral features present in the IRS spectra and ties them to a possible evolutionary scheme. In \S 2, we introduce the catalog of sources and describe the observations used in this spectroscopic study. The results of a spectral categorization are presented in \S 3 along with our classification method. Finally, in \S 4 we discuss our proposed connection between spectral classification and evolutionary stage.

\section{Observations and Reduction}

\subsection{Targets}

Given the benefits of studying the LMC, it is of no surprise that this galaxy has been the focus of intense star formation research in recent years. Of particular interest is the \textit{Spitzer} survey of the LMC \citep[SAGE;][]{meix06}, which has enabled two catalogs of high-mass YSOs in the LMC, produced independently by \citet{whit08} and \citet[][hereafter GC09]{grue09}.  Their different selection criteria and approaches have resulted in significant discrepancies between these two catalogs (see GC09 for details).  We adopt the results of GC09, a precursor of the IRS observations reported in this paper.  A brief description of GC09's methodology follows.  GC09 have extracted point-source photometry from all available observations of the LMC made with \textit{SST}'s InfraRed Array Camera \citep[IRAC;][]{fazi04} in the 3.6, 4.5, 5.8, and 8.0 $\mu$m bands and Multiband Imaging Photometer for \textit{Spitzer} \citep[MIPS;][]{riek04} in the 24 and 70 $\mu$m bands. The [4.5]$-$[8.0] versus [8.0] color-magnitude diagram of all LMC sources was used to identify objects with IR excesses.  The nature of each object with IR excess was then assessed by simultaneously examining its images and spectral energy distribution (SED) from optical to mid-IR wavelengths.  These IR objects were classified into six broad categories: Asymptotic Giant Branch (AGB) and post-AGB stars, planetary nebulae, background galaxies, diffuse sources, normal stars, and YSOs. Objects that could not be classified unambiguously were usually assigned a primary (with higher probability) classification and a secondary (possible but with lower probability) classification. The YSOs are distributed throughout the galaxy and are seen to both cluster near some of the numerous large HII complexes in the LMC and to form in more isolated environments (see Figure 14 of GC09).

For the \textit{SST} IRS observations, we have selected 294 targets from the GC09 catalog. Of these, 269 were identified as `Definite' or `Probable' YSOs including all of the catalog's `Definite' or `Probable' YSOs with [8.0]$<$8.0. In addition, we selected 10 objects with primary classification as AGB/post-AGB stars, and 2 that were probably diffuse sources.  The latter 12 targets were included in our IRS observing program because they had properties that are unusual for their primary classifications, making classification as YSOs an attractive alternative. For example, seven of the observed AGB/post-AGB sources belong to a subgroup dubbed `ERO' (Extremely Red Object) by GC09.  These EROs are redder than sources identified as typical AGB and post-AGB stars, but their SEDs are also unlike the YSOs in that they peak between 8 and 24 $\mu$m.  Our IRS spectra have shown that these EROs are Extreme Carbon Stars, and are reported in detail in a discovery paper by \citet{grue08}. Finally, 13 of the targets were previously uncategorized because they lacked IRAC photometry in the 4.5 $\mu$m band, but they are bright sources in the MIPS 24 and 70 $\mu$m bands and are in gas/dust-rich environments, suggesting that these sources are possibly YSOs.

\subsection{Observations}

The IRS on board \textit{SST} is composed of four modules, two of which provide low spectral resolution (R $\sim60-127$) between 5.2 and 37.9 $\mu$m, and two provide high spectral resolution (R $\sim$ 600) between 9.9 and 36.8 $\mu$m. To cover the entire $\sim$5 to 37 $\mu$m wavelength range, we used the IRS modules SL (short wavelength, low-resolution), SH (short wavelength, high-resolution), and LH (long wavelength, high-resolution). The SL module has two possible configurations: SL1 for the wavelength range 7.6--14.6 $\mu$m and SL2 for 5.2--8.4 $\mu$m.  The high-resolution modules, SH and LH, each consist of 10 spectral orders and cover the wavelength ranges 9.9--19.3 and 18.9--36.9 $\mu$m, respectively. A more detailed description of the IRS is presented by \citet{houc04}.

In order to efficiently observe as many LMC YSOs as possible, we used the GC09 YSO catalog to identify groups of targets with similar 8 and 
24 $\mu$m brightnesses within 2$^\circ$ of one another. Each group of targets could then be observed with the same setup during a single AOR that treated the group as a fixed `cluster.' To ensure that a background estimate was available to remove the zodiacal light contribution from the SH and LH modules, we added one off-position to each cluster. This off-position was chosen by searching for the point with lowest surface brightness in the \textit{IRAS} 12 and 25 $\mu$m maps within the 2$^\circ$ constraint imposed by the targets.  The exposure times for the SL1, SL2, SH and LH modules were chosen such that the observations would produce a contiguous spectrum from 5-37 $\mu$m with a signal-to-noise of $\sim$30.

Seven sources were observed with only the SH and LH modules by our program. Five of these (053142.4$-$683454.3, 053943.3$-$693854.6, 053943.8$-$693834.0, 053944.3$-$693847.5, and 053945.9$-$693839.2) did not include the SL modules because the detectors would have saturated for the shortest possible exposures. The two remaining sources, 053941.9$-$694612.0 and 053959.3$-$694526.3, had previous observations with the SL1 and SL2 modules available in the \textit{SST} archive and could be observed as part of the same cluster. For these latter two sources, we have downloaded and combined the archival data with our SH and LH observations.

\subsection{Data Reduction}

The observations were processed by the \textit{Spitzer} Science Center's pipeline reduction software. We began each spectral extraction process with the basic calibrated data (BCD) product. The campaign-specific rogue pixel masks were applied to the BCD data and removed using the interactive IDL \textsf{IRSCLEAN} package (version 1.9).  We further identify and remove transient rogue pixels by examining variations between frames, also using the \textsf{IRSCLEAN} package.  Multiple exposures at each slit position were then median averaged together.

The subtraction of a background was carried out differently for the high- and low-resolution modules because of their different slit sizes: 3.7$\arcsec$ $\times$ 57$\arcsec$  for SL1, 3.6$\arcsec$ $\times$ 57$\arcsec$ for SL2, 4.7$\arcsec$ $\times$ 11.3$\arcsec$ for SH, and 11.1$\arcsec$ $\times$ 22.3$\arcsec$ for LH. For SL1 and SL2, the slit is long enough that a local background along the slit can be determined and subtracted from the source. This is achieved by differencing the two nod positions. For the SH and LH modules, the slit is too short for a source-independent background; thus the background subtraction is performed by subtracting the cleaned BCD data from a nearby background position. 

High-resolution spectra were extracted from the full SH and LH apertures using the \textsf{SMART} software package \citep{higd04}.  For a full aperture extraction, any flux within the bounds of each order at a given wavelength is extracted and summed.  The low-resolution SL spectra were extracted using the tapered column point source extraction package within \textsf{SMART}.  Because of the large number of spectra, an automated procedure was used to detect a source in a cleaned, background-subtracted spectral image and extract its spectrum with an extraction window whose width varies to accommodate the wavelength-dependent point spread function (i.e., wider at longer wavelengths). Fifty-eight sources could not be detected by the automated procedure either because they appeared extended or resided in a crowded environment, resulting in multiple sources within the slit. These sources were extracted manually using the interactive portion of \textsf{SMART}.  Here the position and FWHM of the source on the spectral image were visually determined and the spectrum extracted using a tapered column.

To combine (or ``stitch'' together) spectra from different modules into a single contiguous spectrum, several corrections need to be made. First, the fringes in the high-resolution 1-D spectra, originating from the detectors of the SH and LH modules, are removed using the IDL package \textsf{IRSFRINGE} (version 1.1). Next, flux discontinuities between modules are corrected by applying scaling factors to spectra from different modules. Flux discontinuities may arise from differences in slit size and orientation, background-subtraction, and spectral extraction. To account for the latter two effects, we assume that the local background-subtraction in the SL1 and SL2 spectra are more reliable, and further scale SL1 to SL2 (scaling factor is $\sim$1; average of 0.98 with standard deviation of 0.14). The LH spectrum is then combined with the SH spectrum after applying a scale factor determined from the ratio of fluxes in the wavelength region where the two overlap ($\sim$18.9--19.3 $\mu$m). Finally, the SH + LH combined spectrum is scaled to the SL1 + SL2 combined spectrum based on the ratio of fluxes in the region where the two spectra overlap ($\sim$9.9--14.6 $\mu$m).  Typical values of the net scaling factors range between 0.5 and 2.0.

\subsection{Cautionary Remarks on Stitching Spectra}

The scaling factors can be used to produce a contiguous spectrum, but may also introduce inadvertent effects.  Several considerations need to be born in mind when interpreting these spectra.

If the full aperture were uniformly illuminated in both the SH and LH modules, the scale factor between them would simply be the ratio of their 
aperture extraction areas. If the observed source is point-like, then no scaling factor would be required. The fact that a scaling factor is required suggests an extended and dusty nature of the objects or a significant contribution from background emission.  Furthermore, the scaling factor is determined from the dust continuum, but the lines may behave differently. For example, the fine-structure lines may originate from a small compact core, while the continuum is extended; therefore, the lines may not have required scaling.

The necessity of module stitching is partly a result of each module's different aperture size and orientation which result in different spatial regions being probed.  The stitching process is further complicated by possible differences in the spatial extent of the regions emitting at different wavelengths.  For example, the emission at longer wavelengths may originate from cooler, more extended gas and dust surrounding a smaller, warmer region which dominates the emission at shorter wavelengths.  Alternatively, since the spectrum at shorter wavelengths contains many spectral emission features associated with dust, the spatial extent of the emitting region at short wavelengths may be equal or even larger than that of the region emitting at longer wavelengths.  To get a sense of how strongly the stitching is affected by spatial differences in emitting areas, we compared $F^{phot}_{24}$/$F^{spec}_{24}$ (the ratio of MIPS photometric 24 $\mu$m fluxes to IRS 24 $\mu$m spectroscopic fluxes) to the scaling factor applied to the 24 $\mu$m wavelength region. We find that any source with $F^{phot}_{24}$/$F^{spec}_{24}\ga$1.5 has a scaling factor less than 1. In other words, the long wavelength portion of the spectrum was scaled down to match the short wavelength portion causing the 24 $\mu$m spectroscopic flux value to underestimate the photometric value. This is consistent with the picture described above where emission contributing to the short wavelength portion of the spectrum originates from a smaller region than the long wavelength portion.

\section{Results \& Classification}

The spectra are first separated by-eye into two primary categories that reflect their level of `embedded-ness.' The first category's members all have continua that rise into the red end of the spectrum, while the second category's spectra are much flatter or even begin to fall at long wavelengths. The second category also contains sources with slightly rising spectra with a strong emission feature at 10 $\mu$m. The first category are likely highly embedded objects such as YSOs while the second are likely either less-embedded, more-evolved young stars or highly evolved stars such as AGBs, post-AGBs or red supergiants (RSGs). The embedded sources are additionally organized via principal component analysis (PCA, see \S 3.2) into 6 groups according to their dominant spectral features. Note that the details of the use of PCA follows in \S 3.2 and 3.3, but first, we will discuss the spectra and their features more generally. The less-embedded object spectra are further subdivided by-eye into three groups based upon their very different dominant features. It is more efficient to classify these spectra by-eye since the spectra are very different among the three groups and do not suffer from the same by-eye biases that the embedded source classifications do. In this manner, the surveyed sources are separated into 9 different spectral classifications, and a sample of the spectra for each category is presented in Figures 1a--1c and 2a. The spectral catalog can be accessed in its entirety (Figures 1d--1ee, Figures 2b--2c) in the electronic version of this paper. A summary of the source classification is given in Table 1.

Our categories closely resemble a portion of a classification system originally developed by \citet{krae02} to classify spectra from the \textit{Infrared Space Observatory} \citep[ISO;][]{kess96} Short Wavelength Spectrometer (SWS), hereafter referred to as the KSPW system. In the KSPW system, objects are first classified by their continuum shape into six broad groups with subclasses assigned based upon specific spectral features. Under the original KSPW system, \citet{krae02} classified over 900 objects of various astronomical origin into a total of 59 classes. All the sources we have identified as embedded would be broadly classified into the KSPW Group 5, sources with SEDs dominated by dust emission with a peak wavelength longer than 45 $\mu$m. Since \textit{Spitzer} IRS spectra cover a different wavelength range than ISO, and because our catalog focuses on a narrow portion of the large spectrum of categories established in the KSPW system, we have developed a modified version of the categories originally developed by \citet{krae02}. In Table 1, we list both our own classification of each source along with the corresponding classification in the KSPW system. Classification by the KSPW scheme was performed by-eye, and every source was classified three times with no knowledge of prior categorization to ensure an unbiased classification. There is generally a one-to-one correlation between the two classifications, but because classifications in our system are performed automatically via PCA for the embedded sources while the KSPW system is done by-eye, there are some discrepancies.

\subsection{Embedded YSO Spectral Features}

The mid-IR is rich in spectral features including bands due to molecular ices, PAHs, silicates, and atoms, thus spectroscopic observations provide an abundance of information about the physical and chemical properties of a forming star and its environment. The spectra of the coldest and therefore youngest YSOs peak around 100 $\mu$m \cite[e.g.][]{vand04}, corresponding to a blackbody temperature of several tens of Kelvins. In environments at high densities and low temperatures such as these, absorption from ices (particularly CO$_2$) and silicates are detected in the spectra. Older YSOs begin to show signs of UV radiation from a central ionizing source including UV-pumped fluorescing PAHs and fine-structure lines from compact HII regions.

The six groups of spectra in the first category, whose continua rise with wavelength throughout the IRS spectral range, are described below and examples are shown in Figures 1a--1c with major spectral features marked. (1) The S group, consisting of 12 sources, are dominated by silicate absorption features including the prominent 10 $\mu$m absorption feature and sometimes including the 18 $\mu$m absorption feature. (2) The SE group, consisting of 5 sources, have both strong silicate absorption features (similar to S group) and strong fine-structure emission lines from ionized gas. (3) The P group, consisting of 100 sources, show prominent PAH emission features. The sources in this group may also show some absorption from silicates, particularly at 10 $\mu$m, but PAH emission features at 6.2, 7.7, 8.6, and 11.3 $\mu$m make it difficult to identify the silicate absorption unambiguously. For example, strong PAH emission features on either side of a weak 10 $\mu$m silicate feature may mimic a strong silicate absorption feature. (4) The PE group, consisting of 142 sources, show both strong PAH emission features and fine-structure lines. Similar to the previous group, silicate absorption features are also difficult to identify in the presence of strong PAH emission features. (5) The E group consists of only 2 sources whose dominant features are fine-structure lines originating from ionized gas. This group appears similar to the PE group and may be an artifact of the thresholds used in the PCA. (6) The F group, consisting of 11 sources, do not meet the criteria for the above five groups, but are spectrally similar to embedded YSOs. They may show weak indications of PAHs, silicates, or fine-structure lines but are generally nearly featureless.

The largest of the six groups is the PE group, which is characterized by both strong PAH and fine-structure line emission. PAH emission features at 6.2, 7.7, 8.6, 11.3, 12.7, 14.2, and 16.2 $\mu$m, first identified in the mid 1970's in the spectra of compact HII regions and planetary nebulae (e.g. Soifer et al 1976), originate from the C-C and C-H stretching and bending modes of PAHs excited by UV radiation. Fine-structure emission lines such as [SIV] 10.5 $\mu$m, [NeII] 12.8 $\mu$m, [NeIII] 15.5 $\mu$m, [SIII] 18.7 $\mu$m and 33.5 $\mu$m, and [SiII] 34.8 $\mu$m originate from ionized gas and can be used to assess the hardness of the UV radiation field. The presence of both PAH and fine-structure line emission clearly suggests the presence of an ionizing source with UV radiation.

The deep silicate feature at 10 $\mu$m and the shallower 18 $\mu$m feature in the S and SE groups are generally smooth, indicating silicate dust in its amorphous form; however, several sources show a more structured 10 $\mu$m feature (Figure 3), possibly a result of dust processing (van Dishoeck 2004). There appears to be a range of strengths to the structure in our sample, possibly representing evolutionary changes from environmental heating. However, there are also known emission lines (e.g. H2 S(3) at 9.64 $\mu$m) which may be contributing to the non-smooth silicate profile. This phenomenon will be further investigated in the future. A majority of the sources dominated by silicate absorption also have a prominent solid CO$_2$ absorption at 15.2 $\mu$m. The infrared vibrational bands of CO$_2$ are known to be very sensitive to the environment, and are therefore excellent tracers of thermal and chemical conditions within the embedding molecular core. Similar to the 10 $\mu$m silicate band, the CO$_2$ absorption feature can be either double-peaked in a smoother single-peaked form (Figure 3). The relative strength of the double/single peaked components is indicative of the environment the ice was formed in \citep{vand04} and will be explored in future papers. 

\subsection{Principal Component Analysis}

For the embedded sources, we organize the spectra into groups with common spectral features by using PCA to quantify the strength of the various spectral features described above. PCA is a technique that allows for the recognition of patterns in spectra and can be used to express a set of spectra in ways that highlight both their similarities and their differences \citep{murt87}. PCA has previously been used in the spectral classification of stars \citep[e.g.][]{murt87}, QSOs \citep[e.g.][]{fran92}, and galaxies \citep[e.g.][]{conn95}. Given the large number of spectra in our sample and the great number of possible spectral features in every spectrum, it can be difficult to visually recognize patterns in the absence, presence, or strength of spectral features. PCA removes by-eye classification guesswork and can quantify the differences between spectra, making it an impartial and unbiased method to assign classifications to spectra. We use PCA to identify and quantify the strength of several spectral features including silicate absorption, PAH emission, and fine-structure lines. We then classify the spectra into several groups based upon the strengths of these features. What follows is the PCA recipe used in this study. For a more complete description of the process and proper implementation of PCA, see \citet{murt87}.

Suppose we have a catalog of $N$ spectra ($i = 1, 2$ ... $N$), each covering the same wavelength range sampled at $M$ ($j = 1, 2$ ... $M$) identical wavelengths. Each spectrum can be represented by a vector $S_{i}$ in an $M$-dimensional space containing the flux at $M$ wavelengths where $i$ corresponds to the specific source. We can then construct an $N \times M$ matrix, $A$, in which each row contains the $M$ flux measurements for each of the $N$ sources. In this manner, each row represents a single point in the $M$-dimensional $A$-space. If $M$ is a small number, i.e. if $A$ were a low-dimensional space, it may be possible to recognize groups and classifications of spectra by their location in $A$-space. But when covering a large wavelength range, with even moderate resolution many spectral features, the number of dimensions becomes too great to visualize without the aid of a statistical method. PCA allows us to reconstruct the $N$-dimensional space into a projection of only a handful of principal coordinates/components. PCA is an orthogonal linear transformation of data (in this case, spectra) to a new coordinate system such that the greatest variance in the data can be attributed to the first coordinate, the second most variance to the second coordinate, and so on. Each new coordinate is called a component, with those contributing the most variance being the principal components. The new coordinates can be expressed graphically by treating each coordinate as a spectrum, in the same way an object's spectrum can be represented as a vector.

To implement PCA, we must normalize all the spectra to isolate variances due to spectral features and minimize variances from bolometric luminosity differences. The first step is to scale all the spectra to the same total integrated luminosity over the wavelength range, 5.5 to 37 $\mu$m.  Next, we use a spline interpolation to place all spectra in the same wavelength grid. A final normalization is performed to create a dataset whose mean is zero. To do so, the mean of each dimension is subtracted from each flux measurement in that dimension. If $f_{ij}$ is the flux measurement for the $i$\textsuperscript{th} source in the $j$\textsuperscript{th} dimension, the full normalized spectra matrix A is constructed as follows:

$$A_{ij}=f_{ij}-\frac{1}{N}\sum_{i'=1}^{N}{f_{i'j}}$$

We can then compute the $M \times M$ covariance matrix, $C$, of the matrix containing all the spectral data:

$$C_{jk}=\frac{1}{N}\sum_{i=1}^{N}{A_{ji}A_{ki}}$$

\noindent where both $j$ and $k$ range between 1 and $M$, inclusive. 

We can then calculate $M$ eigenvectors and the $M$ corresponding eigenvalues of this covariance matrix. The eigenvectors $e_{m}$ and eigenvalues $\lambda_{m}$ obey the identity equation

$$Ce_{m}=\lambda_{m}e_{m}$$

The eigenvector with the largest absolute value of the paired eigenvalue represents the dimension containing the largest variance, the first principal component; the eigenvector with the second largest eigenvalue is the dimension with the second largest variance, the second principal component, and so on. The fractional variance from each principal component is $\frac{\lambda_{i}}{\sum_{i}^{}{\lambda_{i}}}$. A matrix of all $M$ eigenvectors represents a complete set of orthogonal axes which completely describe the entire dataset. If the summed fractional variance of the first few principal components is $\sim$1, then the spectra can be well represented by projecting the spectra onto these first few principal components. We find that often $90\%$ or more of the variance can be described by the first few principal components.

There are several weaknesses to using PCA including its assumption of linearity. This problem is amplified by the nature of our source catalog. Since many of our embedded sources are likely small clusters of YSOs, features from several different sources are mixed together into a single spectrum. In fact, within the HII complex N44, \citet{chen09} found $65\%$ of the candidate YSOs to be resolved multiples. Therefore features that may show a strong correlation within a single source may not have as strong a correlation when originating from a clustered object. In this study, we are using PCA to extract and quantify spectral features. Since we are only using PCA to broadly categorize spectra by their major spectral features, we accept the assumption of linearity as valid to first order.

\subsection{Automated Classification of Embedded YSOs}

In order to use PCA to quantify the strength of different spectral features, four wavelength regions are chosen that contain the strongest components of a particular species type -- one region highlights the strong 10 $\mu$m absorption feature, two highlight PAH emissions between 5 and 10 $\mu$m and between 10 and 13 $\mu$m, and a final region between 10 and 20 $\mu$m encompasses four ion fine-structure lines. PCA is performed on each range to quantify the contribution from various features. Each of these four wavelength regions goes through the procedure described above to find the principal components in that region. The five sources without complete spectral coverage were excluded from this analysis.

The wavelength segment between 7 and 15 $\mu$m is used to determine the strength of each spectrum's 10$\mu$m silicate absorption. The second principal component of this wavelength range (Silicate PC of Figure 4), primarily responsible for silicate absorption at 10 $\mu$m, accounts for $21\%$ of the variance in this wavelength range, and is used to assess the strength of a spectrum's silicate absorption. To get a measure of the amount of silicate absorption in a single spectrum, we do a projection of the vector form of the spectrum into the principal component containing the silicate feature, in this case the second principal component of the 7 to 15 $\mu$m segment. Figure 5 displays what we call the Silicate Projection of each source, dot-products of the embedded sources' spectra with the Silicate PC. A value of zero means there is no contribution from this eigenvector in the spectrum (the spectrum's vector is perpendicular to the component's vector), while positive and negative values indicate positive and negative contributions, respectively; more positive dot-products mean larger silicate absorptions. Because PC2 is primarily the result of silicate absorption, we define all sources with a Silicate PC projection greater than 0.2 to have strong silicate absorption.

There are two tests implemented to establish the presence of PAH emission. PCA is applied to two wavelength regions, each containing strong bands of PAH emission. The first, between 5 and 10 $\mu$m, contains PAH features at 6.2, 7.7, and 8.6 $\mu$m, and the second, between 10 and 13 $\mu$m, has a feature at 11.3 $\mu$m. Between 5 and 10 $\mu$m, the second principal component, which contains the negative of three PAH bands, is responsible for $26\%$ of the spectral variance. Similarly, component 2 of the 10 to 13 $\mu$m range contains the negative of a PAH feature which is responsible for $33\%$ of the variation. These components are shown in Figure 4. We again project the spectra of the sources into these components to determine the contribution from each eigenvector. Since the principal components with PAHs contain negatives of the features, objects with strong PAH features should have small or negative projections onto both these principal components. We set a cut-off of 0.33 for both ranges, and define any source with a second component projection of less than 0.33 in either wavelength range to have strong PAH emission. Figure 5 shows the dot products (PAH Projections) of all the embedded source spectra onto both the PAH principal components. Note that while most objects that demonstrate strong PAH emission from one test also pass the second PAH emission test, a minority of sources pass only one test. The relative strengths of different PAH features will depend on the specific composition of the emitting PAH \citep{tiel04} and will be explored in future studies.

A final PCA test is implemented to test for spectral contributions from fine-structure lines. Between 10 and 20 $\mu$m, there are four lines attributed to [S IV], [Ne II], [Ne III], and [S III]. The third principal component in this wavelength range, which contains the negative of the fine-structure lines and is plotted in Figure 4, accounts for only $5\%$ of the spectral variance. We define sources showing strong fine-structure emission lines as having a projection into the third component of less than 0.2. Figure 5 plots the Fine-Structure Line Projections, the projections onto the third eigenvector, and indicates the boundary between sources we define as with and without fine-structure lines.

To test the ability of PCA to correctly identify related spectral features and produce their proper shape, we have compared the principal components to sources with known spectral origins. The silicate principal component has a direct counterpart in the spectrum of the embedded massive YSO W33 A. The silicate principal component along with W33 A's 7 $\mu$m-15 $\mu$m spectrum are shown in Figure 4. The spectral data of W33 A (Gerakines et al. 1999) were obtained from the ISO archive via the NASA Astrophysics Data System's\footnote[1]{http://adswww.harvard.edu/} online data function. The ISO data were reduced using SMART, multiple exposures were averaged, and the spectrum was smoothed to the resolution of the \textit{Spitzer} IRS. The infrared spectra of W33 A -- a YSO embedded within a cold, dense envelope -- is known to be dominated by deep ice and silicate absorption features. Note that the broad feature at 10 $\mu$m in the principal component matches in location and breadth the deep silicate absorption seen in the spectrum of W33 A. We further compared the PAH principal components to the ISO spectrum of HD 44179. Spectra were obtained and processed for HD 44179 in the same manner as for W33 A. HD 44179, associated with the Red Rectangle nebula, is the prototype object showing PAH emission features. Now identified as a post-AGB star, HD 44179 illuminates the surrounding nebula and fluoresces PAH grains in the IR. The structured peaks seen in the PAH principal components match well features at 6.2, 7.7, 8.6, and 11.3 $\mu$m identified as PAH emission in the Red Rectangle (Figure 4).

Applying the four PCA cut-offs allowed us to quantitatively assess the contribution different spectral features have to a spectrum. The results of the analysis are given in Table 1. There are 54 sources which pass the 10 $\mu$m silicate absorption test, 243 which pass at least one PAH test, and 152 which pass the fine-structure line test. Using the results of the PCA, we divide the embedded objects into the 6 groups described earlier: 1) sources displaying only strong silicate absorption, 2) sources with only strong silicate absorption and fine-structure lines, 3) those with dominant PAH features, 4) objects with both strong PAH features and fine-structure lines, 5) sources with only fine-structure emission lines, and 6) largely featureless spectra. Note that although 54 sources pass the 10 $\mu$m silicate absorption test, only 12 and 5 objects are in the S and SE groups, respectively. This is due to the earlier-stated reason that the blending of PAH emission with silicate absorption makes the two features difficult to distinguish. Any source identified as having PAH emission by definition can only belong to the P or PE groups. The five embedded sources which do not have full spectral coverage cannot be properly classified into any particular group because PCA cannot be performed. Therefore, despite showing fine-structure lines, they have been left unclassified and are represented in Table 1 as embedded objects with unknown classification.

\subsection{Classification of Non-embedded Sources}

By-eye classification is implemented for the less-embedded sources whose spectra are largely flatter or fall at longer wavelengths (Figure 2). The first non-embedded group (group O) all have a broad emission feature at $9-11$ $\mu$m superimposed on a cool dust continuum. The $\sim$10 $\mu$m feature corresponds to emission from silicates typical of oxygen-rich objects including O-rich AGBs and the disks of Herbig Ae/Be stars. The slope of the spectrum at the long wavelength end of the spectrum should be an indicator for how much gas and dust is around the source; since the long wavelength part of the mid-IR is dominated by thermal radiation from dust, spectra which are flat or rise into the far-IR must be surrounded by neighboring gas and dust. Objects without much circumstellar dust would be expected to have spectra which fall into the far-IR. Indeed, of the eight O group sources, two of the three with spectra that fall at long wavelengths were suggested by GC09 to be either possible AGBs or stars with an IR excess, stars which should no longer be associated with much circumstellar material. We probed the dusty environment of these sources by inspecting their IRAC images which are sensitive to both PAH emission and thermal dust emission and assessed their locations with respect to molecular clouds with CO observations from the NANTEN CO(J=1--0) survey \citep{fuku02}. Source 053642.42--700746.7, located on the outer edge of a molecular cloud, is the only source with a falling spectrum located within a dusty complex and a molecular cloud. This is in great contrast to embedded sources which are found to reside in molecular clouds $\sim$80$\%$ of the time and virtually always sit in a larger cloud of dust. The five O group sources with flatter spectra may be highly-evolved pre-main sequence stars such as a Herbig Ae/Be stars -- all are associated with either dusty filaments or larger dusty structures, and four of the five are located within molecular clouds, suggesting they still reside with the clouds they formed from.

The seven objects in the second non-embedded group are the previously mentioned EROs. Their spectra are a mostly featureless blackbody continuum consistent with warm dust emission. Six of the spectra show a broad absorption feature at 11 $\mu$m coinciding with the SiC feature seen in Galactic extreme carbon stars. All the EROs additionally have a C$_2$H$_2$  absorption feature at 13.7 $\mu$m, confirming the C-rich nature of these objects \citep{grue08}.

The final group (group B) contains only one source, 054054.25--693318.7. It has a step-like appearance produced by broad emission features at 8, 11, 21, and 25 $\mu$m superimposed on a continuum that peaks at $\sim$30 $\mu$m. Previously-identified sources with similar spectra have been classified as F or G supergiants or possible proto-planetary nebulae \citep{krae02}. Its identification as a non-YSO is strengthened by the fact it is not located within a molecular cloud.

\subsection{Comparison to Previous Classifications}

The catalog from which our sources were chosen (GC09) was developed via multi-wavelength photometry and image inspection. We compared our classifications with those originally proposed by GC09. Table 2 summarizes the results of this comparison. Of the 294 sources in this study, 269 had been identified as Definite or Probable YSOs by GC09. Of the 269 candidate GC09 YSOs, 262 of them have here been confirmed spectroscopically to have spectra indicative of embedded YSOs. Six remaining sources are members of the O group which may be Herbig Ae/Be stars, AGBs, or post-AGBs, and the seventh is a member of the B group. Most LMC AGBs are excluded from the GC09 candidate YSO catalog based upon mid-IR colors bluer than YSOs. GC09 identified other candidate AGBs by their SEDs which tend to peak in the IRAC bands. We find that the spectra of members of the O Group are generally flat, explaining why they may have not been excluded from the GC09 candidate YSO catalog. Clearly, with a success rate of $\ga95\%$, the CMD, SED, and image inspection criteria used by GC09 to distinguish sources with YSO-like spectra from non-YSOs constitute a robust procedure.

GC09 gave sources 052219.71--654319.0, 052335.54--675235.6, and 054033.97--692509.9 the primary classification of AGB star; additionally, 045245.62--691149.5 received a secondary classification of AGB. Source 052335.54--675235.6 has strong PAH features and is likely an embedded YSO. This object has been previously identified in the literature as IRAS 05237--6755, is listed in the online database SIMBAD\footnote[2]{http://simbad.u-strasbg.fr/simbad/} as a star, and has been studied previously as a candidate AGB. We agree with previous studies by \citet{reid90} and \citet{loup97} which concluded via color criterion that this source is not an AGB. Source 045245.62--691149.5, a member of the F group, has a mostly featureless spectrum with extremely weak silicate absorption and even weaker PAH emission. GC09 classified it as a `Probable' YSO but could not rule out the possibility of it being an AGB. Previously identified as an AGB by \citet{wood92}, we find that it has spectral properties more consistent with a massive embedded YSO. This is further supported by its location within both a molecular cloud and a complex of dusty material seen prominently in the 5.8 and 8.0 $\mu$m IRAC images. GC09 gave 054033.97--692509.9 a primary classification of AGB, and we can confirm that it has the spectroscopic properties consistent with an evolved star. Finally, the spectrum of 052219.71--654319.0 is likely a result of a miss-pointing by \textit{SST}; its spectrum is completely feature-less and is unlike any other previously-identified spectra.

Five of our observed objects are GC09 candidate YSOs with a secondary classification of normal star. All five of these sources have now been spectroscopically confirmed to have spectra indicative of embedded YSOs. One GC09 candidate YSO, 045713.47--661900.1, has a secondary classification of background galaxy. Here, we have classified it in the PE Group. The 13 sources we observed left unclassified by GC09 due to the lack of data in at least one IRAC band are classified as follows: 3 are classified in the P Group, 9 in the PE group, and 1 in the O group. All seven EROs were photometrically identified by GC09 and have here been confirmed spectroscopically to represent a rare class of evolved star.

Two sources have GC09 primary classifications of `diffuse' source. Diffuse sources have SEDs and images consistent with local enhancements of dust emission (see \S 5.4 of GC09). Possible characteristics of diffuse objects include extended or non-point-source IRAC images and the absence of a near-IR or MIPS counterpart. Since diffuse sources have IRAC photometry that is dominated by PAH emission, it is unsurprising that the four diffuse sources were here classified in the PE group. It is probable that both these sources truly are only local enhancements of the ISM and not YSOs. Sources 045205.39--665513.8 and 052204.85--675800.3 have no optical or near-IR counterparts, and both are located on prominent dust filaments seen in IRAC images and are extended in the direction of the filament. Two other sources have GC09 secondary classifications of diffuse sources. The two sources, 053554.94--693903.0 and 054137.57--711902.0 are spatially extended in the mid-IR, although both have tentative near-IR counterparts -- a weak K-band counterpart exists for the former, and a K-band point-source with a coincident compact H$\alpha$ emission region is slightly off-set from the the latter. We concur with GC09 that 045205.39--665513.8 and 052204.85--675800.3 are likely diffuse sources, while 053554.94--693903.0 and 054137.57--711902.0 may be YSOs. It is important to note that these likely diffuse objects are spectroscopically indistinguishable from YSOs based solely upon thier IRS spectra. Clearly a careful inspection of SEDs, multi-wavelength images, and spectra is required to separate sources which may be YSOs from those that are not.

The GC09 candidate YSOs are broadly separated into two luminosity categories -- those with [8.0]$>$8.0 and those with [8.0]$<$8.0. The 8.0 $\mu$m magnitude is likely a good first-order proxy for a YSO's mass so that the brightest sources ([8.0]$<$8.0) are high-mass while the dimmer sources are less massive. This is supported by several recent studies. GC09 found that the luminosity function of the [8.0] band of their Definite and Probable YSOs with [8.0]$<$8.0 is consistent with recent power-law determinations of the high-mass part of the initial mass function (see Fig. 17 of GC09). Chen et al. (2008) found in a case study of massive YSOs in the LMC's N44 that SEDs of YSOs with [8.0]$<$8.0 can be well-fit by radiative transfer models where the central YSO has a mass $\ga$9 $M_{\sun}$ \citep[Table 7 of][]{chen09}. But their results also demonstrate that good fits to the SEDs can be achieved by using a variety of central source masses. Thus the direct application of YSO masses based upon luminosities is likely most valid when considering a large sample of sources. We observed a total of 38 objects with [8.0]$>$8.0. Of those, 36 were classified as definite YSOs by GC09 and 2 as probable YSOs. The dimmest sources do not show any large biases toward a particular spectral feature. The 38 [8.0]$>$8.0 sources are classified by PCA as follows: 1 source in the S group(3$\%$), 16 in the P group(42$\%$), and 21 in the PE group (55$\%$). This distribution is consistent within the limits of small number statistics with the entire PCA-based classification catalog which is composed of 4$\%$ S Group, 2$\%$ SE Group, 37$\%$ P Group, 52$\%$ PE Group, 1$\%$ E Group, and 4$\%$ F Group sources.

\section{Spectral Evolution of an Embedded Massive YSO}

Since massive YSOs can take thousands of years to form, it is observationally impossible to watch a single star develop. Instead, studies are done statistically on a large sample of sources. Our catalog of sources represents the most reliable census of massive YSOs in the LMC, and given the large number of sources and the unbiased nature of their identification, we have captured YSOs at various stages of evolution. This allows us to develop a scheme for the evolution of massive YSO spectra and to tie the spectra to the astrophysics involved in star formation. Here, we primarily focus on the evolution of the most massive YSOs, and so are only concerned with those embedded sources with [8.0]$<$8.0. We refer to these sources as massive based upon the rough mass lower-limit described in \S 3.5. This mass cut-off is certainly not exact, and it should be noted that when we refer to massive YSOs, we are referring not to a mass-based cut-off, but a luminosity-based one.

\subsection{Complications of the Spectral-Evolutionary Connection}

The assessment of a YSO's evolutionary state based upon its IRS spectra is primarily complicated by the possibility of spectral features originating from regions external to the YSO's immediate star-disk-envelope system. Given the large size of the IRS slits (as wide as 11.1$\arcsec$ -- 2.6 pc at the LMC distance), any radiation from an object's environment that falls within the slit will contribute to its spectrum. As evidenced by the apparent spatially-extended nature of these objects, we suspect this type of spectral contamination from the environment may play an important role. 

In particular, there may be contamination from fine-structure lines that originate external to the YSOs. To better assess the origin of the fine-structure lines, we have compared the strength of a fine-structure line in the spectrum of the IRS's full-slit to the strength of the same line when a spectrum is extracted at the position on the slit farthest from the source. The SH module has a slit size of 5$\times$2 pixels; in the first nod position, the source is centered on the first 2$\times$2 pixels, while in the second nod position the source is centered on the last 2$\times$2 pixels. We have therefore extracted a spectrum for every object with fine-structure lines along a 1-pixel wide strip on the side of the slit opposite the source. At the distance of the LMC, the `off-source' position is $\sim$2 pc from the object. Nominally, the 1$\times$2 pixels at these slightly `off-source' positions would not detect emission from a point source. Consequently, if the target source is coincident with a compact fine-structure line-emitting object, then there is a concentration of fine-structure emission on the source, and the equivalent width per pixel of a fine-structure line from the full slit spectrum will be larger than the equivalent width per pixel at the `off-source' slit position. However, if the ratio of the fine-structure line's equivalent width per pixel of the full slit and the `off-source' position is equal to or less than one, the full slit spectrum used in our classification is likely contaminated by strong fine-structure line emission external to the source. The [SIII] line at 18.7 $\mu$m is ideal for this investigation since its location on a relatively flat region of the spectrum allows for an easy determination of the equivalent width. We have calculated this ratio for every fine-structure line source (SE, PE, and E groups) and have provided this [SIII] concentration ratio in Table 1. It may be that objects with [SIII] concentration ratios of nearly or less than unity are being mis-identified as YSOs capable of ionizing their surroundings and producing fine-structure lines. Given the spatial resolution of the IRS, determining the exact nature of each source may require the less contaminated spectra that will be possible with the \textit{James Webb Space Telescope} \citep{gard06}.

To test the severity of environmental contamination from the presence of external dust within the slit, we compared the flux values at 24 $\mu$m from background subtracted MIPS photometry and IRS spectroscopy. Comparing the 24 $\mu$m fluxes in this manner can be a measure of the radiative contamination from thermally-emitting dust. We find that the sources with the smallest photometric to spectroscopic ratios are the dimmest sources. This implies an expected finding, that environmental contamination from dust is the most severe for the dimmest sources.

A second way the environment can affect an object's spectra is through the external pumping of UV radiation onto the outer regions of the object, a situation that is unrecognizable using the [SIII] concentration ratio. Such a scenario is certainly the case for the previously discussed diffuse sources 045205.39--665513.8 and 052204.85--675800.3. We demonstrated that these sources identified as local enhancements of dust -- which are without a central heating source -- can have spectra that mimic YSOs with photodissociation regions and compact HII regions. The PAHs within the local dust knot are most likely being UV-pumped from neighboring O and B stars, while the fine-structure lines likely originate from the larger HII complexes in which they reside. Similar external radiation effects almost certainly contaminate the spectra of many objects in our sample.

Given that a majority of our sources are located within or on the edges of large HII complexes, spectral contamination from ionized gas is expected to be prevalent. This problem is difficult to identify due to the limited spatial resolution of the observations and is harder to correct since the radiation is a phenomenon local to the YSO and cannot be corrected by background subtraction. For every source, we have investigated its environment through careful examination of H$\alpha$, IRAC, MIPS, and CO maps. We obtained the H$\alpha$ images from the Magellanic Cloud Emission Line Survey (MCELS; Smith et al. 1999) and downloaded available data from the NANTEN CO(J=1--0) survey of the LMC \citep{fuku02}. The results of this examination are given in Table 1. Each source is checked for four different environmental situations: whether or not it is located within a molecular cloud delineated by CO data, possible coincidence with a small ($<$1pc radius) H$\alpha$ source, if it is located within a larger diffuse H$\alpha$ emission region, and whether or not it is located within large-scale dust complexes including filaments, rings, or more irregular shapes. We find that $\sim$80$\%$ are located within CO clouds, approximately 1/3 of the sources are coincident with a compact HII region, nearly every embedded source is located within a dusty structure, and $\sim$3/4 are within regions of diffuse H$\alpha$ emission. A complete assessment of an object's true nature must include considerations of its spectrum along with possible environmental sources of spectral contamination.

It is important to note that many of the 277 embedded object spectra likely are the combined spectra of several YSOs in a tight cluster that are spatially unresolvable by \textit{Spitzer} IRS. Should there be one source that dominates the IR-spectrum, then the following analysis is fairly strong, but if there are multiple equally-contributing YSOs, the analysis is less robust. We have inspected near-IR images of every object to look for signs of multiplicity. We obtained J and K$_{s}$ band observations from the Two Micron All Sky Survey \citep[2MASS;][]{skru06} and with the higher-resolution Infrared Side Port Imager \citep[ISPI;][]{vander04} camera on the CTIO Blanco 4m telescope. Table 1 reports our results. Sources are considered candidates for multiplicity if they clearly show several point-sources or are extended in the J and K$_{S}$ bands where IRAC images show a single source. A total of 162 embedded sources (58$\%$) are identified as being possible multiple sources, in good agreement with the 65$\%$ finding of \citet{chen09} within the N44 complex. However, a majority of the possible multiple sources appear to have a single source that dominates the luminosity, a finding that is consistent with the known steep power law between mass and luminosity that allows massive stars to dominate the luminosity of a cluster. We conclude that treating the YSOs as single massive YSO systems is valid to first order.

\subsection{Evolutionary Groups}

The objects in the S group represent the earliest stages of a massive YSO's life after attaining enough mass to be included in our sample. The extremely deep absorption from silicates indicates that the sources are highly embedded in dense envelopes of material, with the deepest absorptions originating from the most-embedded sources. An alternate view is that the strength of the silicate absorption depends on the sight line from which the YSO is being viewed. Because of the inherent high radiation pressure that can repel accreting material, massive star formation is poorly described by a spherically symmetric envelope model. Instead, the most recent models have included axis-symmetric \citep{york02,whit03a,whit03b} or even asymmetric, ``clumpy'' envelope geometries \citep{krum09,inde06}. In the case of axis-symmetric models, the strength of the silicate feature will depend on the viewing angle with respect to a YSO's bipolar outflow. Edge-on (parallel to the plane of the disk) sight lines show prominent silicate absorption profiles, while highly inclined sight lines down the ``throat'' of a bipolar outflow show little or no silicate absorption \citep{robi06}. Similarly, \citet{inde06} showed that the 10 $\mu$m silicate feature can vary greatly in strength between different lines of sight through a clumpy envelope. In fact, the voids between the clumps allow near IR radiation to escape the envelope, resulting in little or no silicate absorption along a sight line through such a void. Given that recent modeling predicts that a significant percentage of sight lines produce silicate absorption features, the small number of objects with silicate absorption relative to the number without it suggests that silicate absorption strengths are likely not exclusively line of sight dependent. Although it is possible that S and SE group objects may be identified as P, PE, or E group objects if viewed from a different location, it should be noted that the most embedded YSOs, those enshrouded in the most material, are more likely to have sight lines through either dense clumps (in the asymmetric models) or the densest regions of the envelope not cleared by outflows (in the axis-symmetric models). 

Nine of the eleven massive objects in the S stage additionally have a prominent 15.2 $\mu$m CO$_2$ ice feature, likely originating from the cold outer region of the YSO's envelope. These absorption featues resemble those seen in galactic sources W33 A \citep{gibb98} and Elias 29 \citep{boog00}, both YSOs deeply embedded in the clouds from which they formed. While silicate absorptions can arise from diffuse cloud spectra, features identified with ices are seen exclusively in dense clouds; with ice sublimation temperatures of $\sim$100 K or less, the absorption must occur in the outer, colder regions of the stellar envelope \citep{vand04}. There is a large variation in the strength of the ice band, and it is therefore an excellent diagnostic of environmental changes. In particular, the large number of S group sources with ice bands and the scarcity of sources in other groups with ice features demonstrate a gradual heating of the protostellar envelope. While $70\%$ of the S and SE sources have a prominent CO$_2$ absorption, only $\lesssim10\%$ of the later-stage objects display the same feature. This finding supports the idea of gradual envelope heating due to an increase in the ratio of stellar radiation to circumstellar mass resulting from the dissipation of the envelope from accretion, outflow, and stellar wind. There also appears to be differences between sources in the line shape of the ice feature, a variation that may reflect evolutionary changes within the S group itself.

Members of the SE group likely represent a more evolved form of the highly embedded sources in the S group. Fine-structure lines likely originate from compact or ultracompact HII regions. Small HII regions can form around single massive YSOs after they begin producing a copious amount of UV radiation. SE objects are still highly embedded in circumstellar material but also show indications of a compact HII region. This suggests that the creation of a compact or ultracompact HII region can begin very early in the life of a YSO, before it has cleared much circumstellar material. The strength of SE group fine-structure lines varies by several orders of magnitude between sources. If the emission is originating from a small HII region surrounded by a larger gas and dust shell, then the line emission may be attenuated by the intervening material by an amount determined by the optical depth of the shell. We have compared six different fine-structure line strengths for the five SE sources to the strength of the silicate feature, a measure of the column density of the surrounding material. Silicate feature optical depths are determined by fitting the spectra to model silicate spectra using the IDL PAHFIT code \citep{smit07}. For SE objects, we find a very strong inverse correlation between the fine-structure line strengths and the silicate optical depth, suggesting that the fine-structure emission is being screened and attenuated by the surrounding dust shell. However, given the small number of sources in the SE group (5), the implicit assumption of equal intrinsic line strengths (intrinsic line strengths will vary between sources with differing ionization fluxes, densities, region diameters, and abundances), and the fact that the correlation disappears after including sources of all groups, the correlation may not represent a true phenomenon. If the fine-structure-silicate correlation within the SE group is real, then the lack of a correlation among the other groups implies that either the optical depths of the dust shells are not large in these groups, or that the fine-structure emission in the other groups does not originate from within a dust shell at all.

Because SE objects show both the characteristics of very young (S group) and more evolved (E and PE group) objects, SE sources could represent the transition from a highly embedded to a less-embedded YSO, objects that are beginning to clear away their parent molecular cloud. Since members of the S group show no signs of UV radiation, they are either true massive protostars (accreting, pre-hydrogen burning massive YSOs) or are accreting at a rate and with a geometry that is able to contain the UV radiation. In either scenario, the massive YSO is unable to substantially disturb its natal cloud in the S stage. Although there are only 17 total sources in either the S or SE groups, 52 sources ($\sim$20$\%$) pass the PCA test for silicate absorption. A majority of these objects are members of the P and PE spectroscopic groups, both of which have features indicative of UV-emitting central objects. As discussed earlier, several PAH emission bands are superimposed on the 10 $\mu$m silicate absorption feature, making the strength of the silicate feature difficult to quantify in P and PE spectra. Even in the scenario that PCA is overestimating the number of silicate absorption sources by improperly separating the blended silicate and PAH features, sources with strong silicate absorption represent a minority of sources. Assuming a typical formation time of $10^5$ years (Zinnecker \& Yorke 2007), we estimate that it takes a massive YSO on the order of a few thousand to a few ten thousand years to clear its surroundings to the point where the silicate absorption feature is no longer apparent in its spectrum.

Three of the five SE objects are coincident with small compact/ultracompact HII regions seen in the H$\alpha$ images of MCELS. The presence of a compact H$\alpha$ source strengthens our assertion that the fine-structure lines present in the spectra originate from a small region ionized by the source itself. The two objects without corresponding compact HII regions, 045206.97--665516.4 and 053536.88--691214.1, are both located within areas of high H$\alpha$ emission from a larger, more extended structure. It is therefore possible that the fine-structure lines do not originate from the source itself and were not properly background subtracted. In fact, the two sources without a visible coincident compact H$\alpha$ source also have the lowest [SIII] concentration ratios of the SE group. Object 045206.97--665516.4 has a ratio less than 1, indicating that the strength of the [SIII] line is stronger at distances farther from the source and that the fine-structure line emission is likely at least in part either an environmental contamination or the result of an extended fine-structure line-emitting structure. Although the sample size is small, there seems to be a correlation between the presence of a compact H$\alpha$ source and the existence of fine-structure lines in the spectra -- $60\%$ of the SE sources have accompanying compact HII region, compared with only $25\%$ of the S group objects. Of the ten massive S objects, only three, 045111.39--692646.7, 045550.62--663434.6, and 052646.61--684847.2 show any sign of compact H$\alpha$ emission on the source, and the emission is relatively faint.

The P, PE, and E groups likely represent the later stages of a YSO's embedded lifetime. Of the 239 massive embedded objects, 217 of them have either PAH emission, fine-structure lines, or both. Since both PAH and fine-structure lines are indicative of a UV-emitting central object, massive YSOs spend a substantial portion of their lives emitting in the UV. Objects with UV-produced features represent $\sim90\%$ of our catalog, suggesting that massive embedded YSOs emit in the UV during a substantial proportion of their lives, possibly as high as $90\%$. However, note that given the possibility of environmental spectral contamination, the length of this stage may in fact be shorter.

The strengths of fine-structure lines vary by up to 6 orders of magnitude between sources, and PAH emission power can vary by as much as 5 orders of magnitude. If fine-structure line and PAH band photons travel along a line of sight through dense circumstellar material, the intrinsic line strengths may be attenuated by a factor set by the optical depth at each line's wavelength. Given a high enough optical depth, objects with inherently strong PAH or fine-strcutre line emission may have the features completely attenuated by surrounding material, causing them to be improperly categorized in our spectral classification scheme. As previously discussed, we find no correlation within the P, PE, or E groups between any of the fine-structure lines and the optical depth. Similarly, there appears to be no recognizable correlation between the 7.7 $\mu$m and 11.2 $\mu$m PAH bands' strengths and the strength of the silicate absorption feature. We fit these PAH features using Lozentzian profiles on a first order polinomial baseline. This suggests that the optical depth of dust surrounding the fine-structre line and PAH-emitting regions are generally not large. We conclude and that viewing angle does not have a large effect on the strength of PAH bands or fine-structure lines, and should not effect classification.

Among the 144 PE and E group objects, the [SIII] line concentration ratios range between 0.9 and 408 with a median value of 1.4. A median value greater than 1 indicates that for a large majority of our spectra, the fine-structure line is stronger at slit positions closer to the source, suggesting that most the PE and E objects have fine-structure line peaks on the source and are less likely to be solely environmental in nature. However any object with a concentration ratio near or less than 1 is more likely to be affected by the local environment. There are a total of four sources with [SIII] concentration ratios less than 1.0; three of these are located within large complexes of strong H$\alpha$ emission and lack a compact H$\alpha$ object. This implies that the fine-structure lines of these sources are more likely environmental in nature. Of all the sources located within large HII complexes, the median [SIII] concentration ratio is 1.36. Conversely, sources not located within larger HII complexes tend to have larger [SIII] concentration ratios: the median [SIII] concentration ratio of sources not located within large HII complexes is 3.37. The strong correlation between a source's location within an HII complex and its [SIII] concentration ratio suggests that this ratio is a good measure of the amount of environmental fine-structure line contamination. Note that ratios greater than 1.0 are strongly favored, even with sources within HII complexes. This implies that although environmental fine-structure line contamination likely plays a heavy role, most sources have ratios consistent with being coincident with a compact fine-structure line-emitting object. PE and E group objects with compact source-like [SIII] concentration ratios represent approximately one half the sample of massive YSO candidates in this study. A compact or ultracompact HII region is likely to have formed around half of the YSOs in our catalog, suggesting that massive YSOs can create a region whose fine-structure lines are \textit{Spitzer} IRS-detectable approximately half-way through the massive stages of its formation.

The spectra of P and PE group objects resemble those typical of photodissociation regions (PDRs); in our sample of massive YSOs, the PAH emission pumped by the YSO's own UV radiation probably originates from such a region around the forming star. High spatial resolution ISO observations of $\rho$ Oph W within the $\rho$ Oph molecular cloud indicate that the PAH bands and fine-structure lines are common to hot dust illuminated by O or B-type stars \citep{boul00}. The central object heats its surroundings to high temperatures, dissociating molecules, and exciting PAH vibrational levels leading to fluorescent PAH emission. The objects in the P stage, when embedded YSOs display only strong PAH emission, may be younger than those of the PE or E stages. Although both PAHs and fine-structure emission lines are indicators of UV radiation, PAHs can absorb and reprocess longer wavelength radiation than is necessary to create an HII region. In fact, PAH spectra are produced most efficiently when the incident UV radiation is intense but relatively soft -- even below the Lyman limit. P group objects may then represent YSOs who are hot enough to produce soft UV radiation, but have not yet become hot enough to produce many higher energy photons that can ionize its surroundings. The presence of PAH emission and lack of fine-structure lines can also be explained if a hot, hard UV-producing star's radiation is being quenched by a high accretion rate, preventing a detectable compact HII region from forming. In either scenario, objects in the P stage therefore are the transition objects that bridge objects that cannot form detectable compact HII regions and those that can. Externally-produced PAH emission is likely a large contaminant to both the P and PE spectra, but even given this environmental effect on the spectra, the absence of fine-structure lines in the P group indicates a source unable to produce a detectable compact HII region. 

Seventeen of the 100 total P objects passed the PCA test for silicate absorption, indicating they may have a silicate feature at 10 $\mu$m. These may be the youngest of the P group, and could represent a subgroup of sources that have just begun clearing their surroundings and are still embedded in a thick molecular envelope. Alternatively, these objects could in fact be examples of objects whose line of sight passes through a dense region of the envelope, creating a strong silicate feature. The fact that there is no recognizable correlation between either PAH or fine-structure line strength and optical depth also suggests that the PAH features and fine structure lines are not formed in a region inside a shell of circumstellar material as would be expected in a spherically-symmetric model. This is consistent with previously described models which allow radiation to scatter and escape via outflows or voids between clumps. In such a scenario, the 10 $\mu$m absorption feature originates from a dense cloud or clump that is both self-screening and attenuating radiation originating from behind the cloud/clump, while PAH and fine-structure line emission can originate from regions external to this line of sight (e.g. outflow cavity walls, other clumps, extended HII regions) that are being illuminated/ionized by the central source.

PE group objects represent the most evolved form of YSO in our proposed evolutionary scheme. PAH emission and fine-structure lines indicate that a high level of UV radiation is originating from the central source and has created a compact HII region surrounding the object. While the fine-structure lines originate from this HII region, the PAH emission likely originates from the PDR surrounding the HII region. PAHs are depleted within the boundary of the HII region, destroyed by the intense hard-UV radiation. P group objects may be an extreme PE source -- where a full-sized compact HII region cannot form due to high accretion rates but a PDR can form outside the very small central ionized region. E group objects likely represent a type of PE object, ones with strong fine-structure lines and relatively weak PAH emission. Although the 2 objects in the E group do not pass either PCA test for PAH emission, by-eye inspections of both spectra show clear signs of PAH emission. Both the objects may be better classified into the PE group or as extreme PE sources where the PDR is relatively small.

The F group is comprised of 11 objects that did not pass any spectral feature's PCA test. Seven may show signs of weak silicate absorption at 10 $\mu$m, 7 may have fine-structure lines, and all 11 may have weak PAH emission. It may be that these sources are better classified into other groups, but because PCA was used to quantify the strengths of spectral features, they have been dubbed `featureless.' Since they represent a small fraction of the total sample ($6\%$) and do not appear biased towards a particular feature, we do not consider their contribution to the above evolutionary analyses statistically important. The five sources without full spectral coverage are assigned to the U Group.

\section{Conclusion and Future Directions}

We have obtained high- and low-resolution IR spectra of 294 objects in the LMC. The sources were chosen from a catalog of IR-luminous objects developed by GC09 based on their massive YSO-like photometric and SED qualities. The \textit{Spitzer} IRS spectra cover the wavelength range of 5--37 $\mu$m and show a variety of spectral features including absorption from silicates and ices and emission lines from PAHs and ionized atoms. Of the 294 sources observed, 277 spectroscopically resemble embedded massive or intermediate-mass YSOs. The remaining sources are comprised of a mixture of Herbig Ae/Be stars (with O-rich emission disks) or highly evolved stars such as RSGs, AGBs, and post-AGBs.

We classified both the YSOs and non-embedded sources into distinct groups based upon their spectral features. Embedded objects were categorized automatically via the statistical process dubbed PCA. This allowed for an unbiased and quantitative classification scheme. Non-embedded sources -- primarily evolved stars, and not the primary focus of this paper -- were classified into groups by-eye. The embedded sources all rise into the red end of the spectrum indicating their continuum emission is dominated by dust. Most embedded sources are identified as massive YSOs still surrounded by the dense cloud from which they formed. Our catalog largely consists of sources with luminosities consistent with YSOs of at least $\sim$10$M_{\sun}$, and therefore represent the later stages of massive star formation, after they have accreted a significant fraction of their mass. Individual spectra within the large sample of sources act as snapshots of massive star formation and collectively display an evolution of massive YSO spectra. The youngest sources (S and SE group) are dominated by deep silicate absorptions and ice absorptions originating from the cold outer regions of the envelope (van Dishoeck 2004). Later stages of massive YSO evolution show a gradual heating of the circumstellar envelope evidenced by the lack of ice features. Older sources also display signs of envelope depletion in the form of little or no silicate absorption. As the envelope is depleted and accretion slows or ceases, UV radiation from the young star begins to dissociate and ionize the nearby molecular material forming compact HII regions and PDRs \citep[e.g.][]{hoar07}. Well-known indicators of compact HII regions and PDRs are present in the spectra of later-stage massive YSOs (P, PE, and E group) including fine-structure emission lines and PAH emission features.

The work described here is a first step in a series of investigations using this newly-formed YSO catalog into the evolution of massive YSOs and the interaction between them and their environment. Follow-up papers will address both the inter- and intra-group evolution of specific features. YSO evolution will also be tied to the state of the molecular cloud they were formed from and the nearby stellar population. Finally, work should be done to compare the spectra of these massive YSOs and their corresponding evolutionary stage with radiation transfer models to assess both the quality of the models and the accuracy of their spectral-based evolutionary placement.

\acknowledgments
The authors wish to thank the anonymous referee for useful comments, Charles Gammie for an early helpful discussion of PCA, and Karna Desai and Ryan Peterson for their work on some of the figures and tables presented here. This work is based in part on observations made with the \textit{Spitzer Space Telescope}, which is operated by the Jet Propulsion Laboratory, California Institute of Technology under a contract with NASA. This study made use of observations obtained at Cerro Tololo Inter-American Observatory, a division of the National Optical Astronomy Observatories, which is operated by the Association of Universities for Research in Astronomy, Inc. under cooperative agreement with the National Science Foundation. This publication has made use of the data products from the Two Micron All Sky Survey, which is a joint project of the University of Massachusetts and the Infrared Processing and Analysis Center/California Institute of Technology, funded by NASA and the NSF. The CO maps are the result of the NANTEN project which is based on the mutual agreement between Nagoya University and the Carnegie Institution of Washington. This research made use of the SIMBAD database, operated at CDS, Strasbourg, France and SAOImage DS9, developed by Smithsonian Astrophysical Observatory. Support for this work was provided by NASA through an award issued by JPL/Caltech. 

{\it Facilities:} \facility{Spitzer (IRS, IRAC, MIPS)}

\clearpage

\begin{figure}[t]
\figurenum{1a}
\begin{center}
\includegraphics[width=0.8\textwidth]{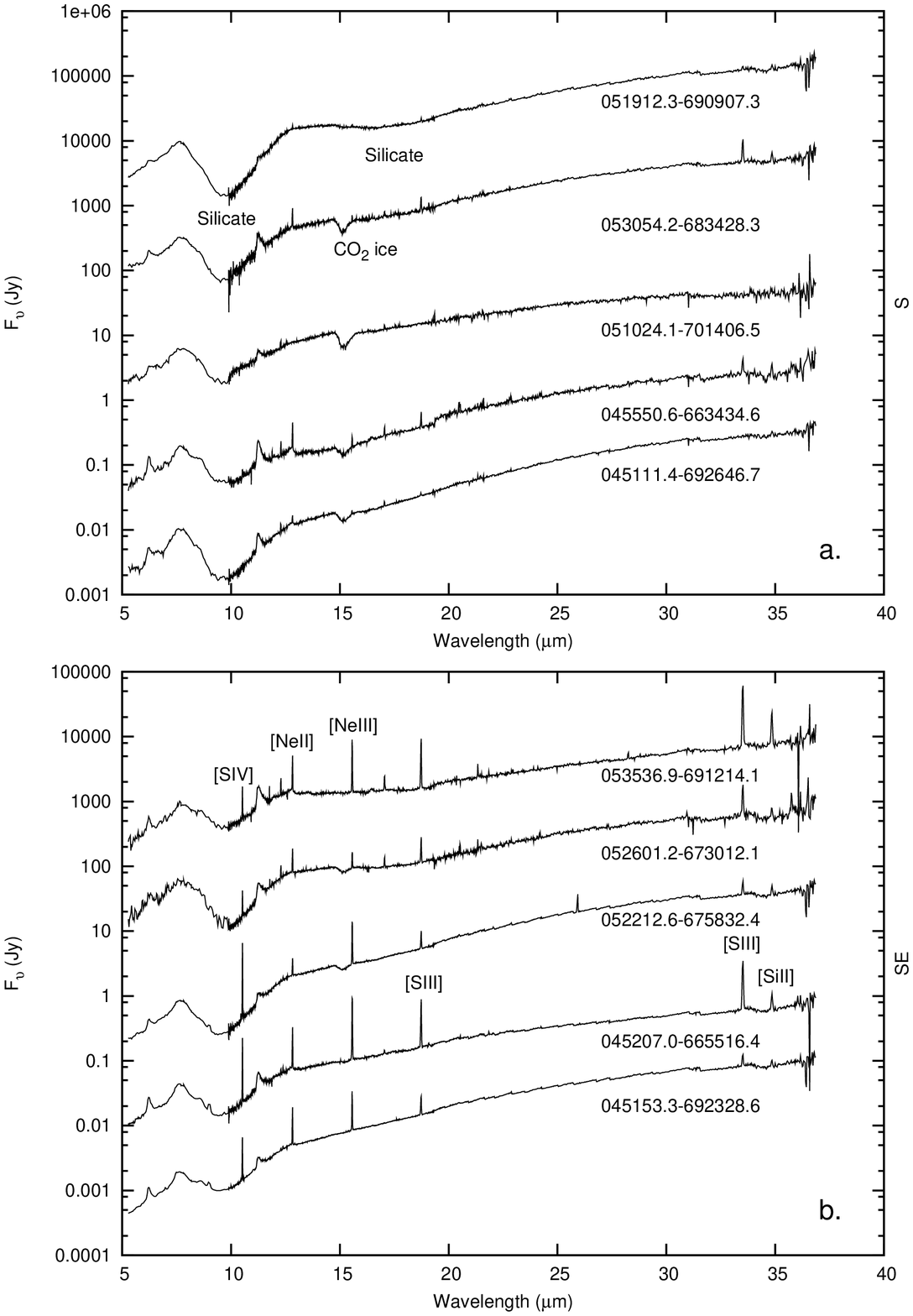}
\caption{Representative spectra of the S and SE groups of the embedded YSO sources. The source identification is given below each curve. Top: The broad silicate absorption features at 10 and 18 $\mu$m and the 15.2 $\mu$m CO$_{2}$ ice band are identified. The spectra from bottom to top have been scaled by the following multiplicative factors for clarity: 0.05, 1, 30, 800, and 10$^{4}$. Bottom: Prominent fine-structure lines are indicated including [SIII], [SIV], [NEII], [NEIII], and [SiII]. The spectra from bottom to top have been scaled by the following multiplicative factors for clarity: 0.001, 0.1, 1, 10$^{3}$, and 8$\times$10$^{3}$.}
\label{f1_1}
\end{center}
\end{figure}

\clearpage

\begin{figure}[t]
\figurenum{1b}
\begin{center}
\includegraphics[width=0.8\textwidth]{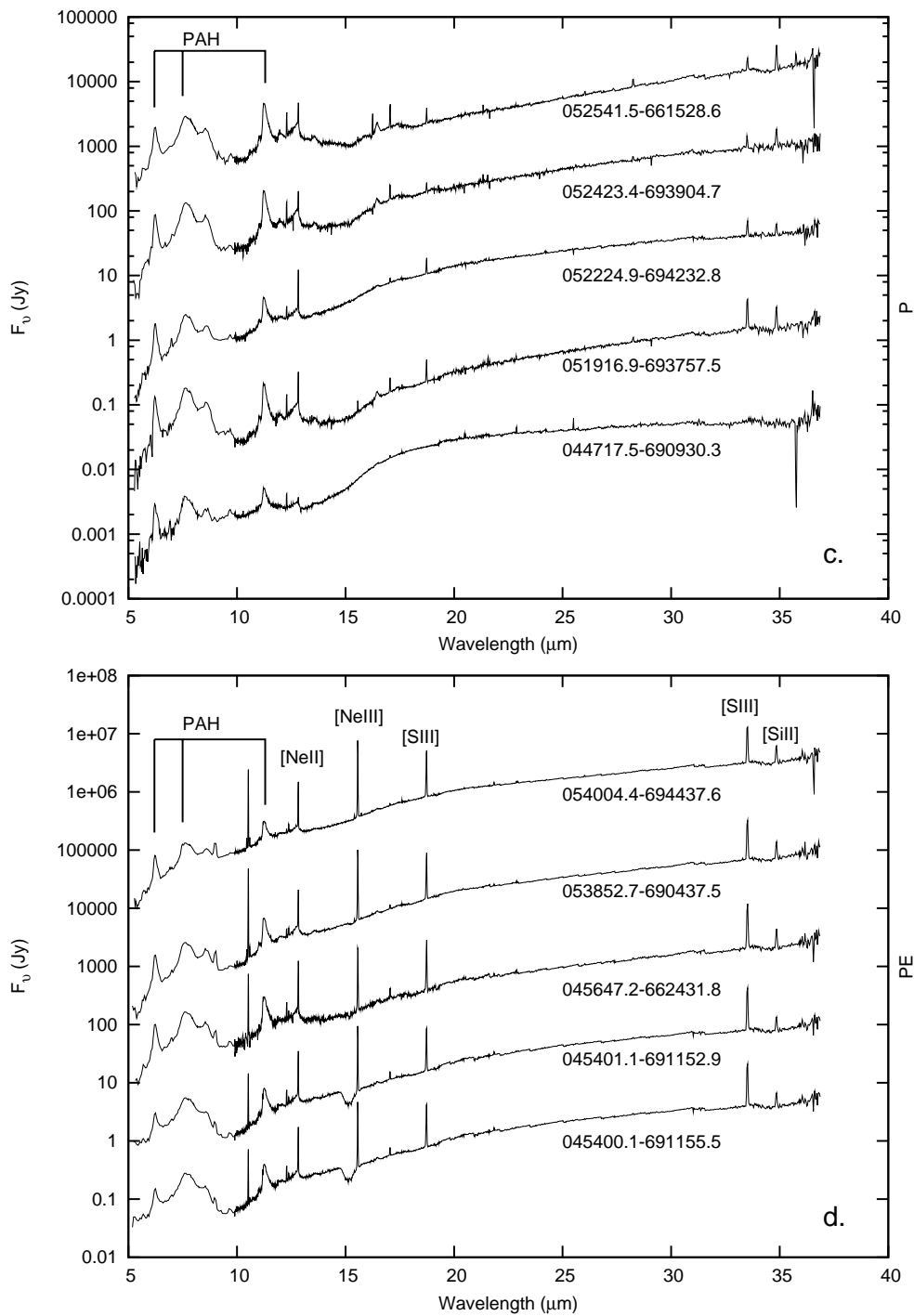}
\caption{Representative spectra of the P and PE groups of the embedded YSO sources. The source identification is given below each curve. Top: The prominent PAH emission features are identified. The spectra from bottom to top have been scaled by the following multiplicative factors for clarity: 0.05, 1, 10, 800, and 3$\times$10$^{4}$. Bottom: Prominent fine-structure lines are indicated along with three PAH emission features. The spectra from bottom to top have been scaled by the following multiplicative factors for clarity: 1, 20, 600, 6000, and 2$\times$10$^{5}$.}
\label{f1_2}
\end{center}
\end{figure}

\clearpage

\begin{figure}[t]
\figurenum{1c}
\begin{center}
\includegraphics[width=0.8\textwidth]{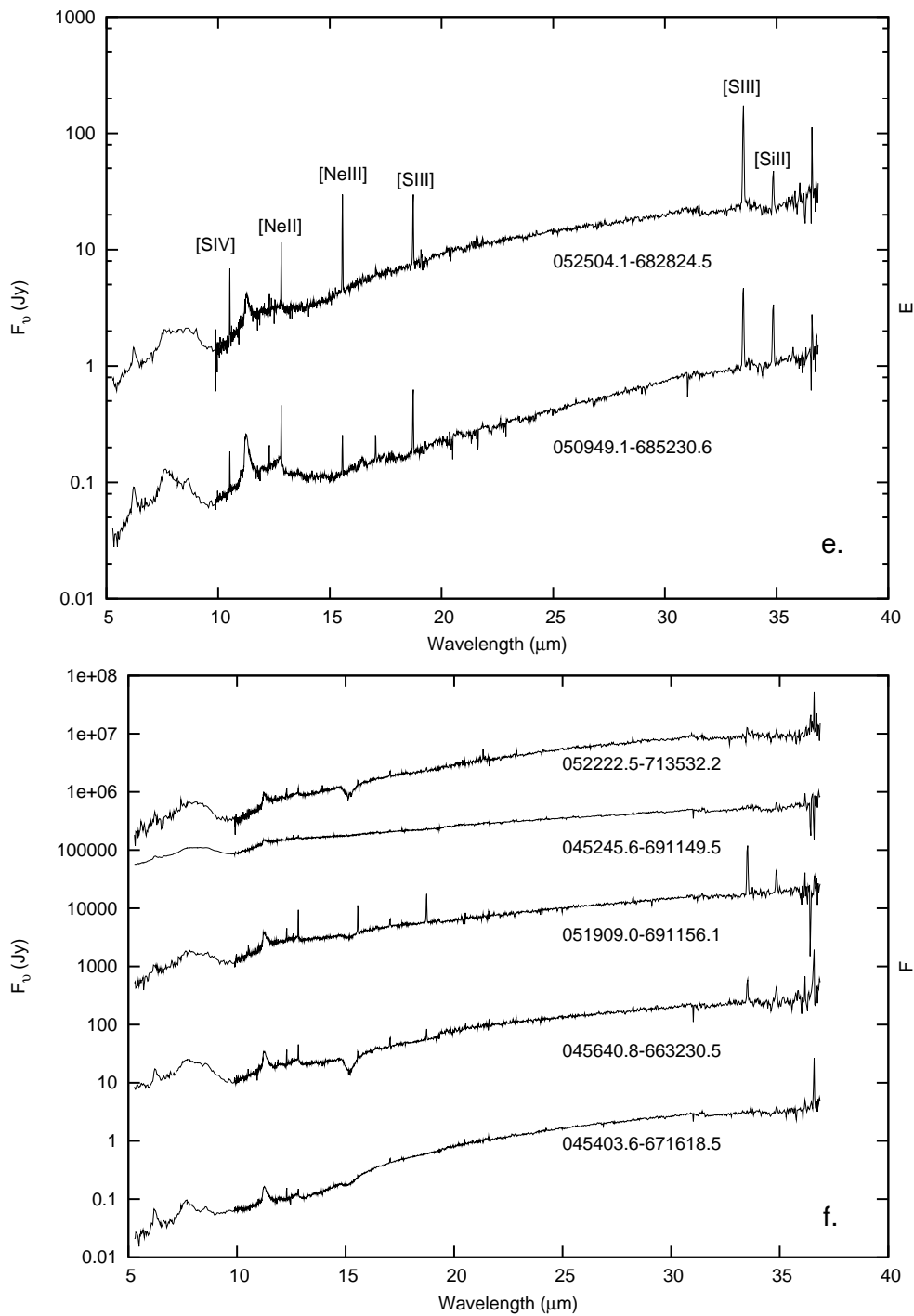}
\caption{The two spectra of the E group and the five spectra of the F group with complete spectral coverage. The source identification is given below each curve. Top: Prominent fine-structure lines are indicated including [SIII], [SIV], [NEII], [NEIII], and [SiII]. The spectra from bottom to top have been scaled by the following multiplicative factors for clarity: 1 and 10. Bottom: The spectra from bottom to top have been scaled by the following multiplicative factors for clarity: 1, 140, 24000, 10$^{5}$, and 10$^{7}$}
\label{f1_3}
\end{center}
\end{figure}

\clearpage

\begin{figure}[t]
\figurenum{1d}
\begin{center}
\includegraphics[width=0.8\textwidth]{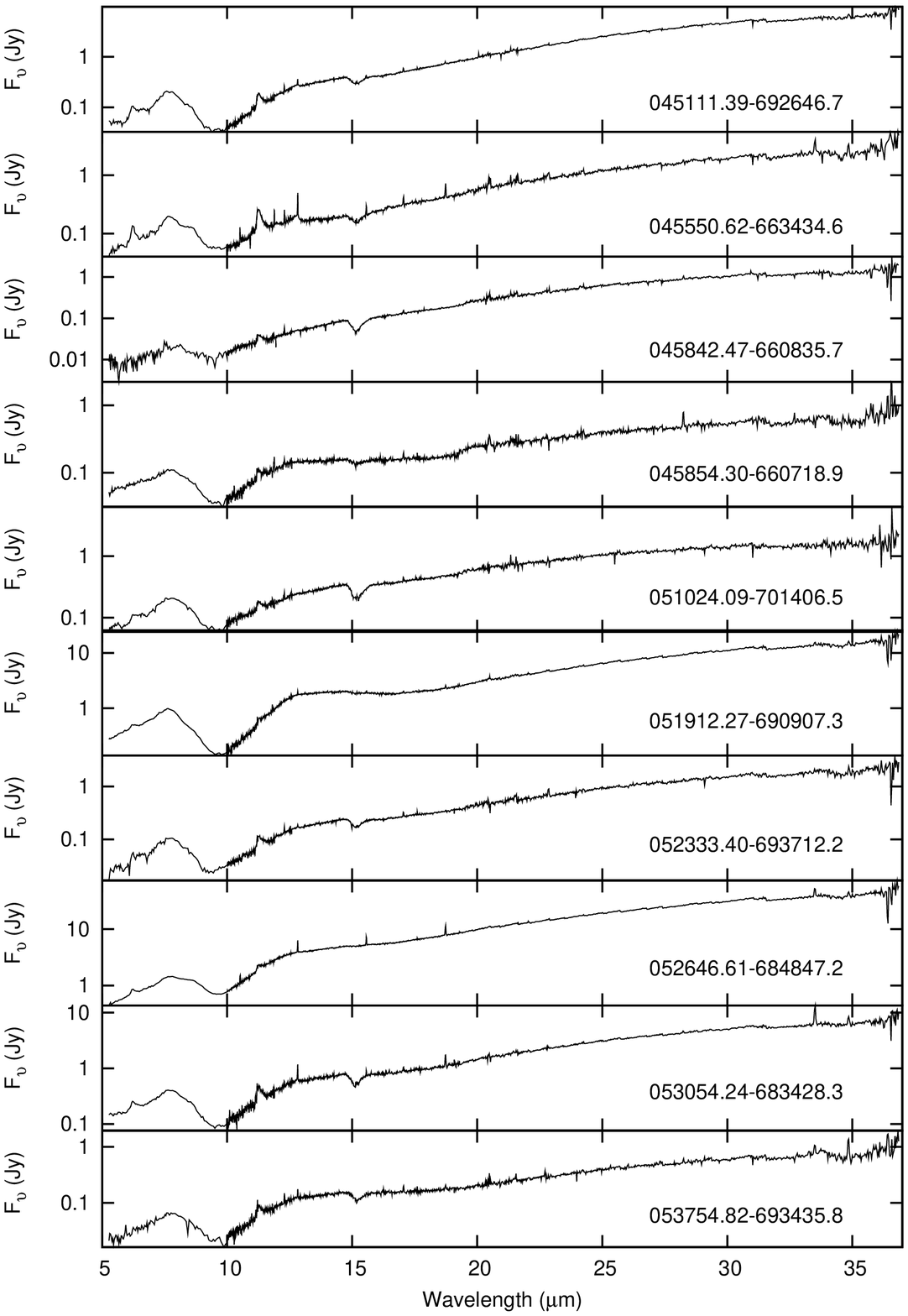}
\caption{Spectra of the complete source catalog from which selected sources are shown in Figures 1a--1c. Sources are ordered as in Table 1, and source identifications are given below each curve.}
\label{f1_4}
\notetoeditor{This figure is to appear in the electronic version of the paper only}
\end{center}
\end{figure}

\clearpage

\begin{figure}[t]
\figurenum{1e}
\begin{center}
\includegraphics[width=0.8\textwidth]{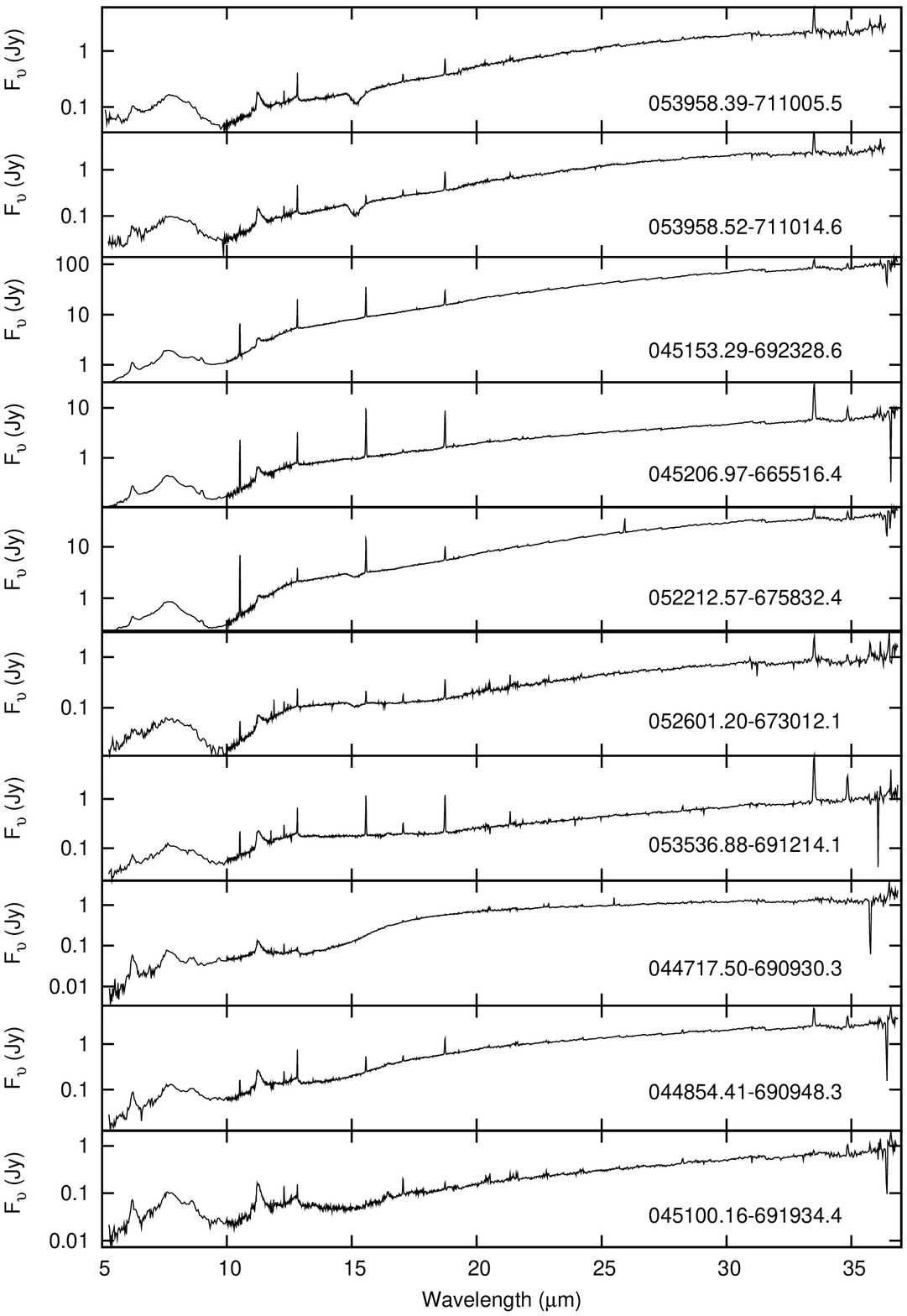}
\caption{Continuation of Figure 1d.}
\label{f1_5}
\notetoeditor{This figure is to appear in the electronic version of the paper only}
\end{center}
\end{figure}

\clearpage

\begin{figure}[t]
\figurenum{1f}
\begin{center}
\includegraphics[width=0.8\textwidth]{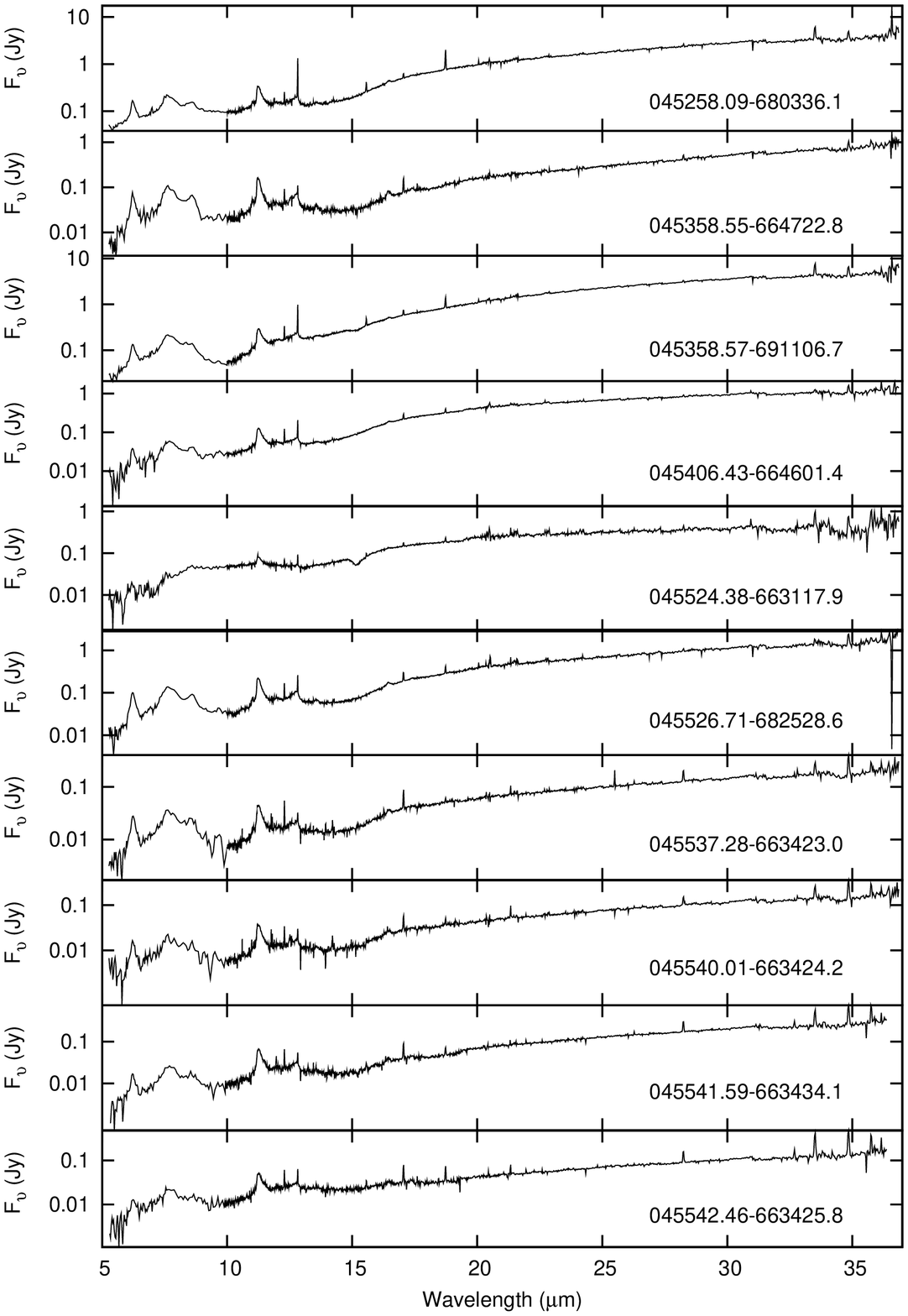}
\caption{Continuation of Figure 1d.}
\label{f1_6}
\notetoeditor{This figure is to appear in the electronic version of the paper only}
\end{center}
\end{figure}

\clearpage

\begin{figure}[t]
\figurenum{1g}
\begin{center}
\includegraphics[width=0.8\textwidth]{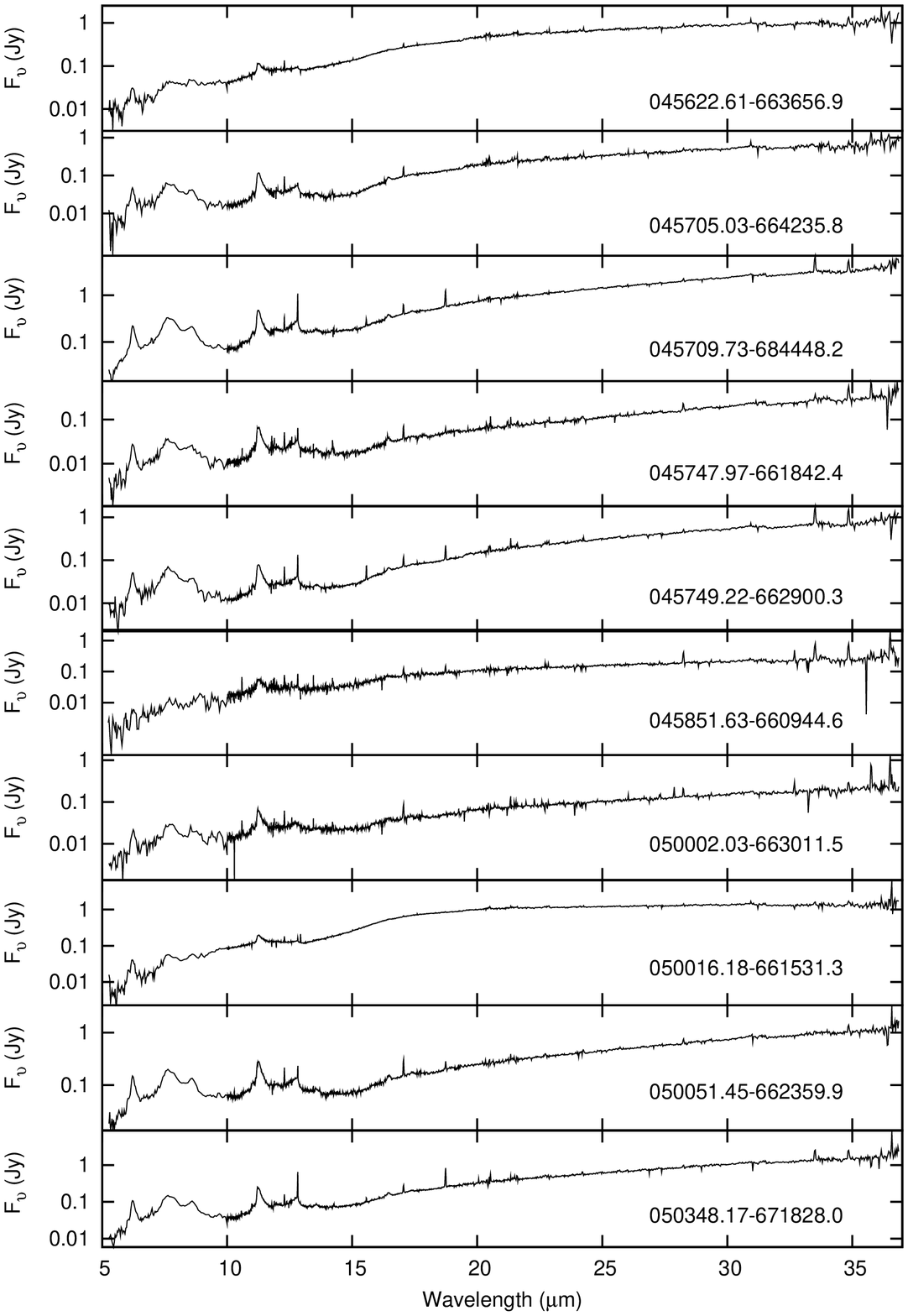}
\caption{Continuation of Figure 1d.}
\label{f1_7}
\notetoeditor{This figure is to appear in the electronic version of the paper only}
\end{center}
\end{figure}

\clearpage

\begin{figure}[t]
\figurenum{1h}
\begin{center}
\includegraphics[width=0.8\textwidth]{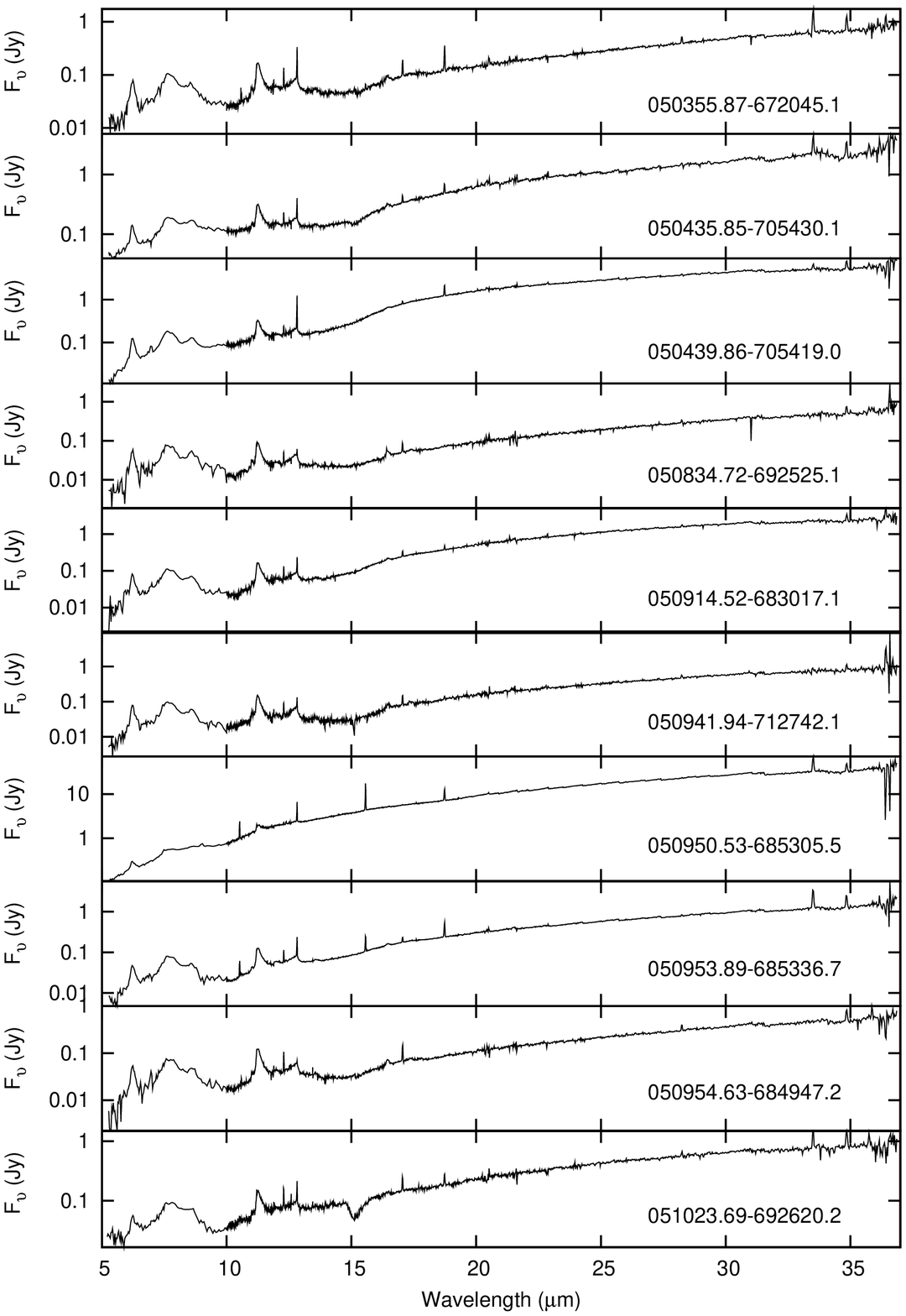}
\caption{Continuation of Figure 1d.}
\label{f1_8}
\notetoeditor{This figure is to appear in the electronic version of the paper only}
\end{center}
\end{figure}

\clearpage

\begin{figure}[t]
\figurenum{1i}
\begin{center}
\includegraphics[width=0.8\textwidth]{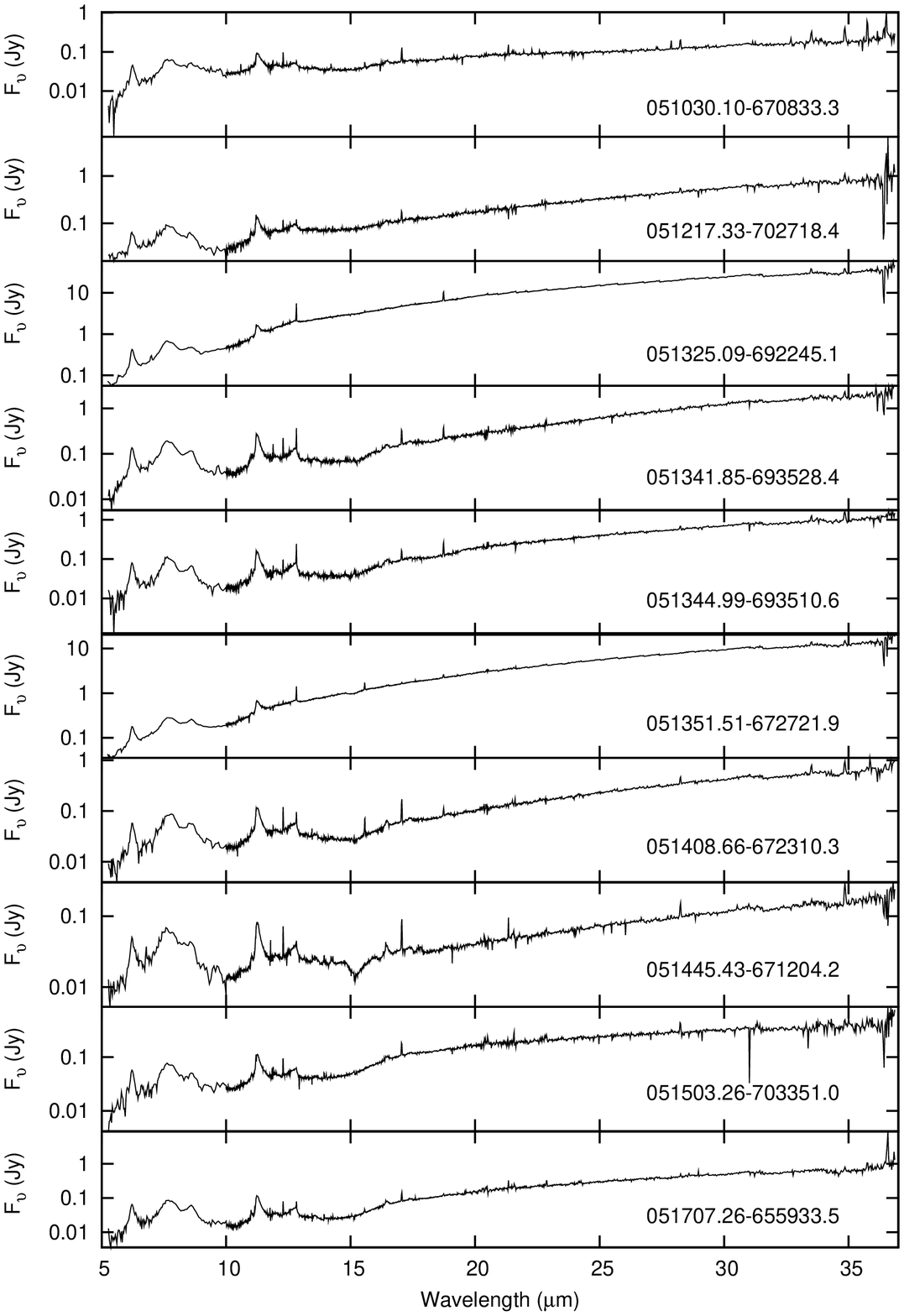}
\caption{Continuation of Figure 1d.}
\label{f1_9}
\notetoeditor{This figure is to appear in the electronic version of the paper only}
\end{center}
\end{figure}

\clearpage

\begin{figure}[t]
\figurenum{1j}
\begin{center}
\includegraphics[width=0.8\textwidth]{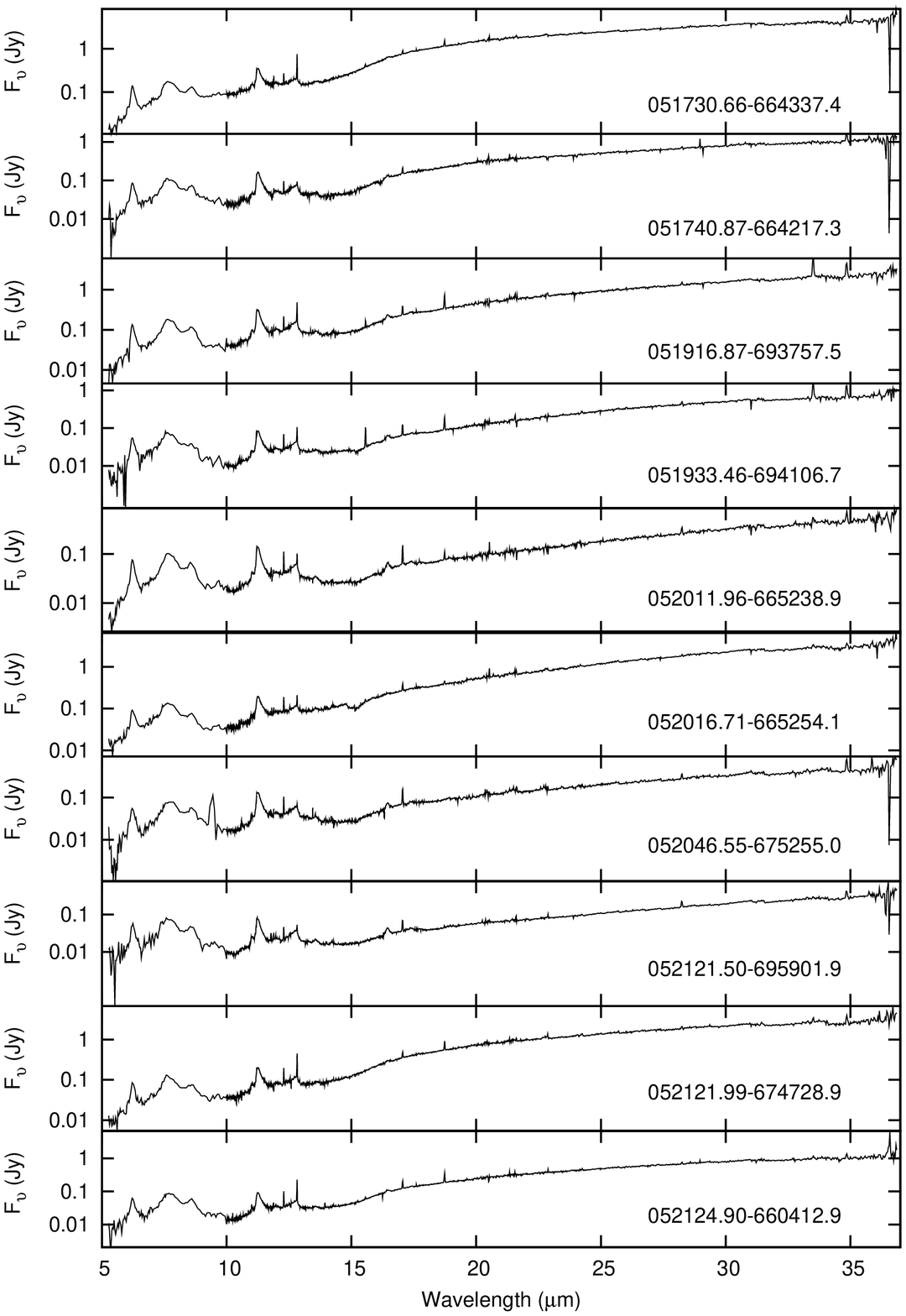}
\caption{Continuation of Figure 1d.}
\label{f1_10}
\notetoeditor{This figure is to appear in the electronic version of the paper only}
\end{center}
\end{figure}

\clearpage

\begin{figure}[t]
\figurenum{1k}
\begin{center}
\includegraphics[width=0.8\textwidth]{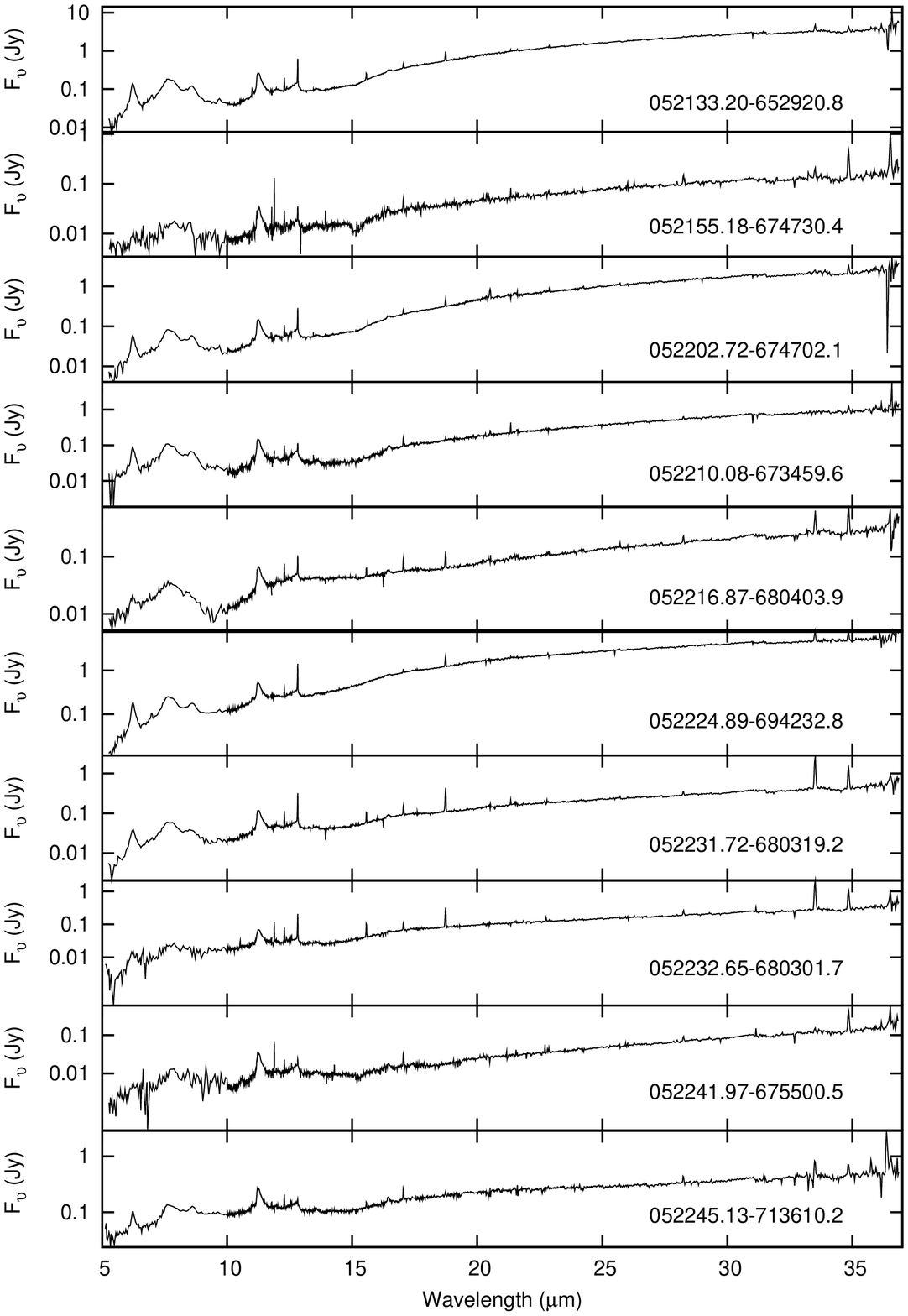}
\caption{Continuation of Figure 1d.}
\label{f1_11}
\notetoeditor{This figure is to appear in the electronic version of the paper only}
\end{center}
\end{figure}

\clearpage

\begin{figure}[t]
\figurenum{1l}
\begin{center}
\includegraphics[width=0.8\textwidth]{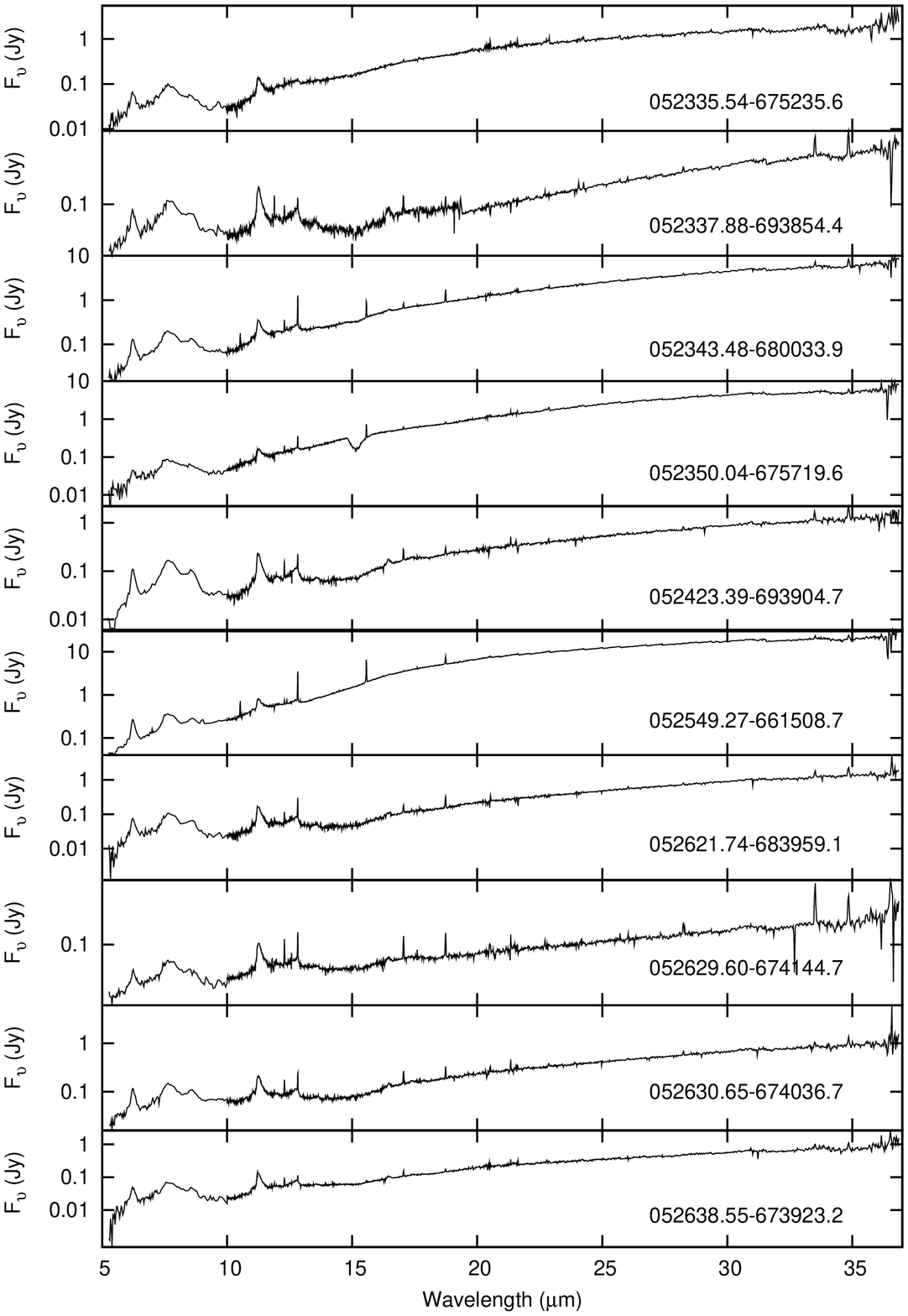}
\caption{Continuation of Figure 1d.}
\label{f1_12}
\notetoeditor{This figure is to appear in the electronic version of the paper only}
\end{center}
\end{figure}

\clearpage

\begin{figure}[t]
\figurenum{1m}
\begin{center}
\includegraphics[width=0.8\textwidth]{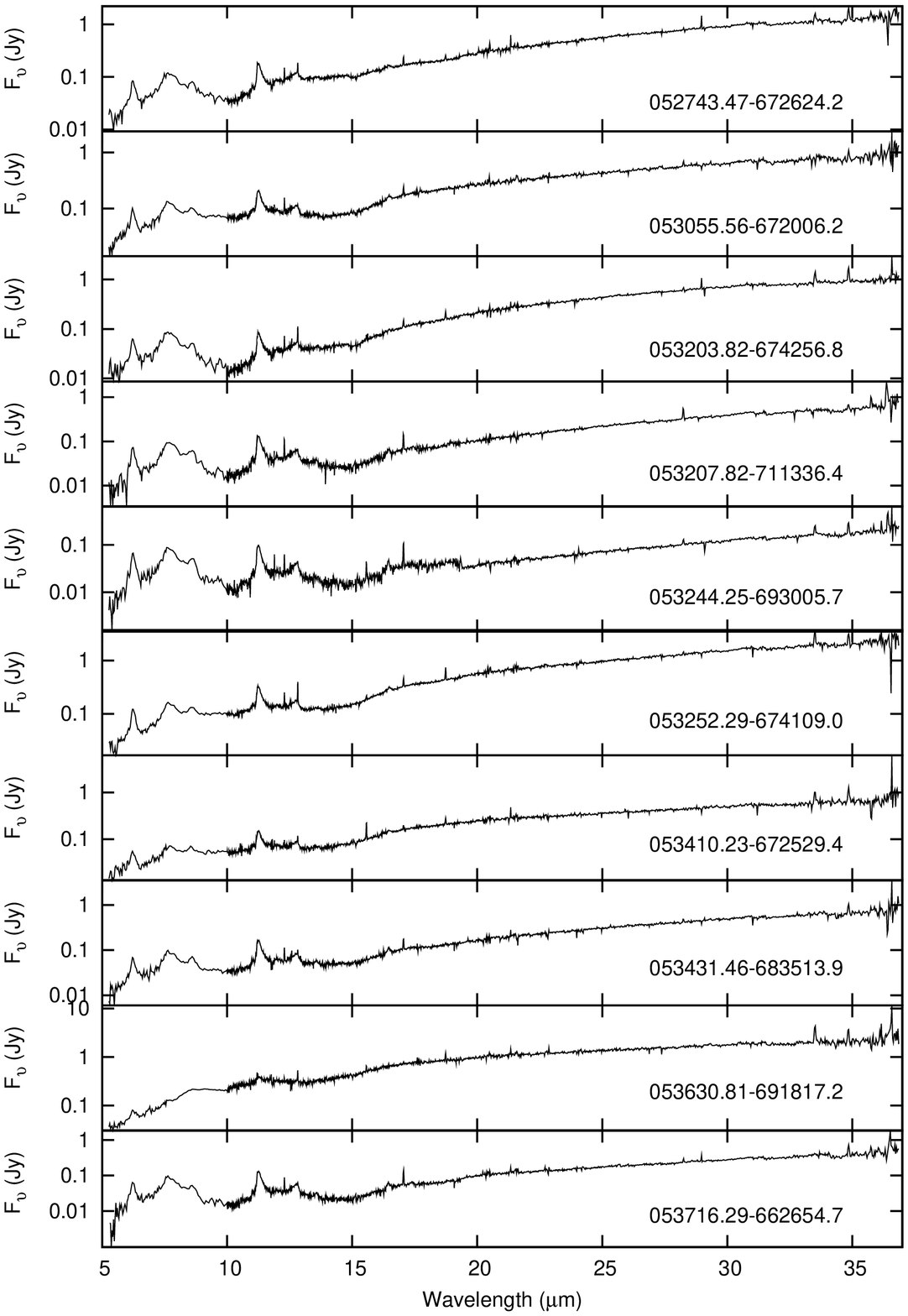}
\caption{Continuation of Figure 1d.}
\label{f1_13}
\notetoeditor{This figure is to appear in the electronic version of the paper only}
\end{center}
\end{figure}

\clearpage

\begin{figure}[t]
\figurenum{1n}
\begin{center}
\includegraphics[width=0.8\textwidth]{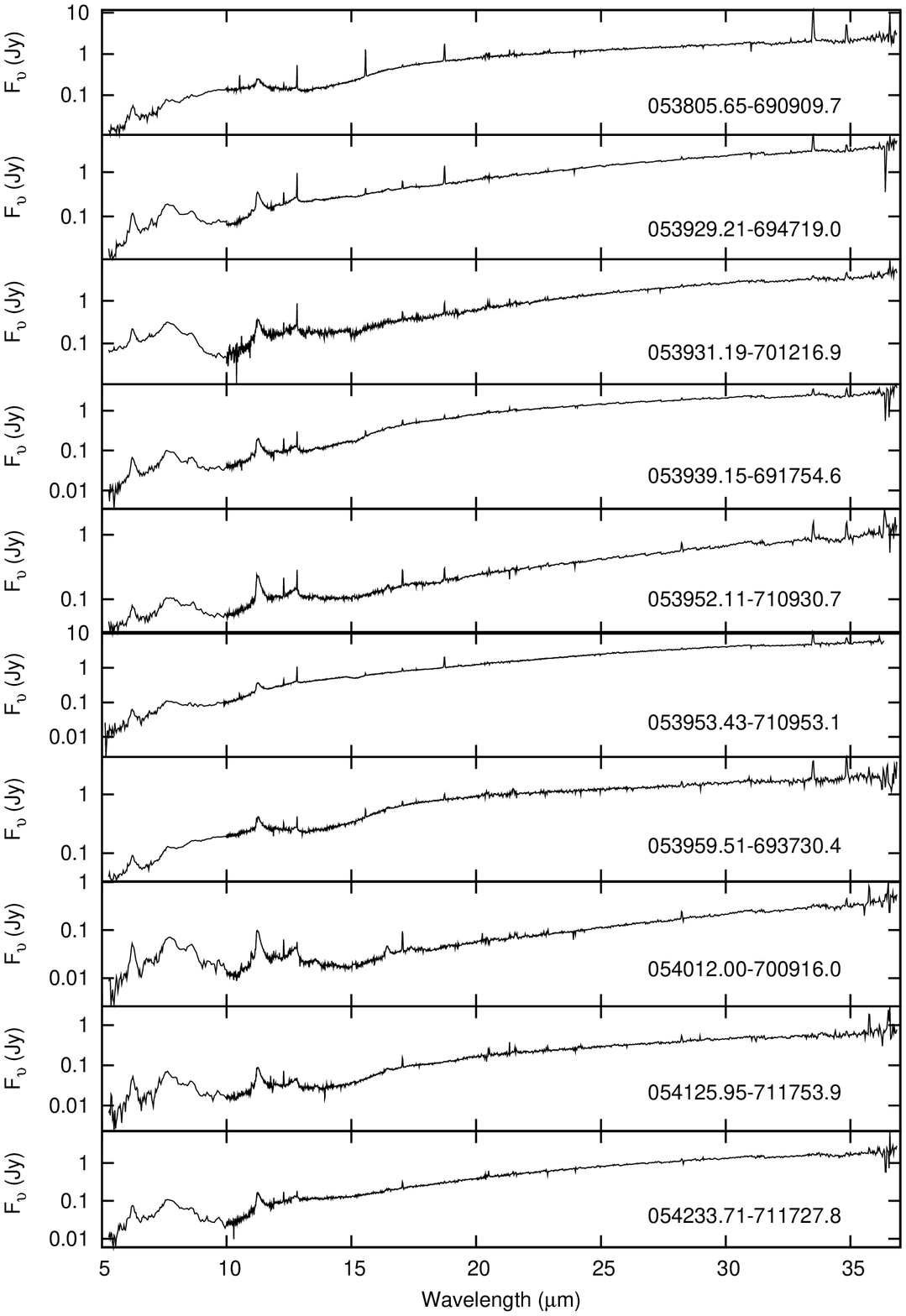}
\caption{Continuation of Figure 1d.}
\label{f1_14}
\notetoeditor{This figure is to appear in the electronic version of the paper only}
\end{center}
\end{figure}

\clearpage

\begin{figure}[t]
\figurenum{1o}
\begin{center}
\includegraphics[width=0.8\textwidth]{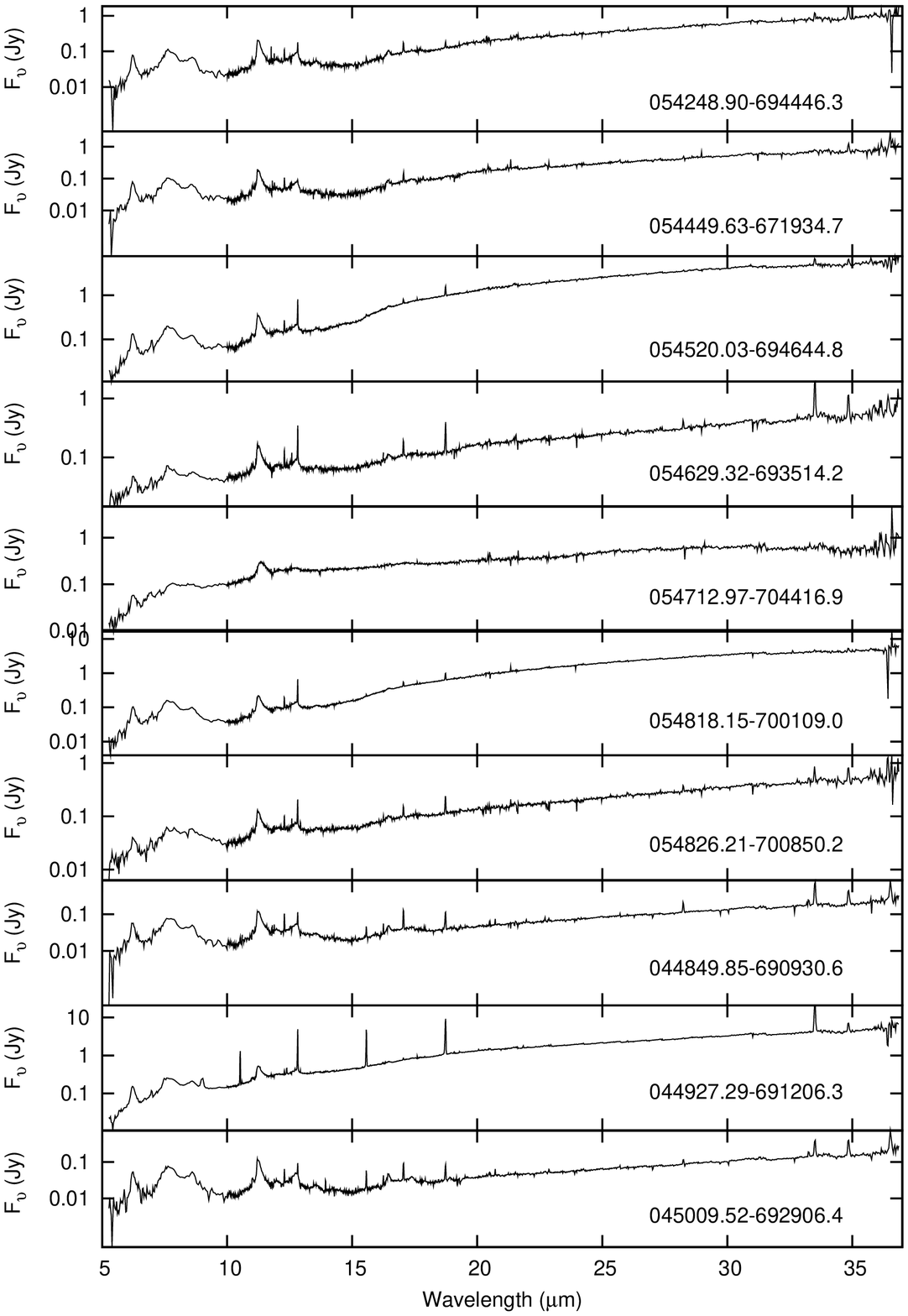}
\caption{Continuation of Figure 1d.}
\label{f1_15}
\notetoeditor{This figure is to appear in the electronic version of the paper only}
\end{center}
\end{figure}

\clearpage

\begin{figure}[t]
\figurenum{1p}
\begin{center}
\includegraphics[width=0.8\textwidth]{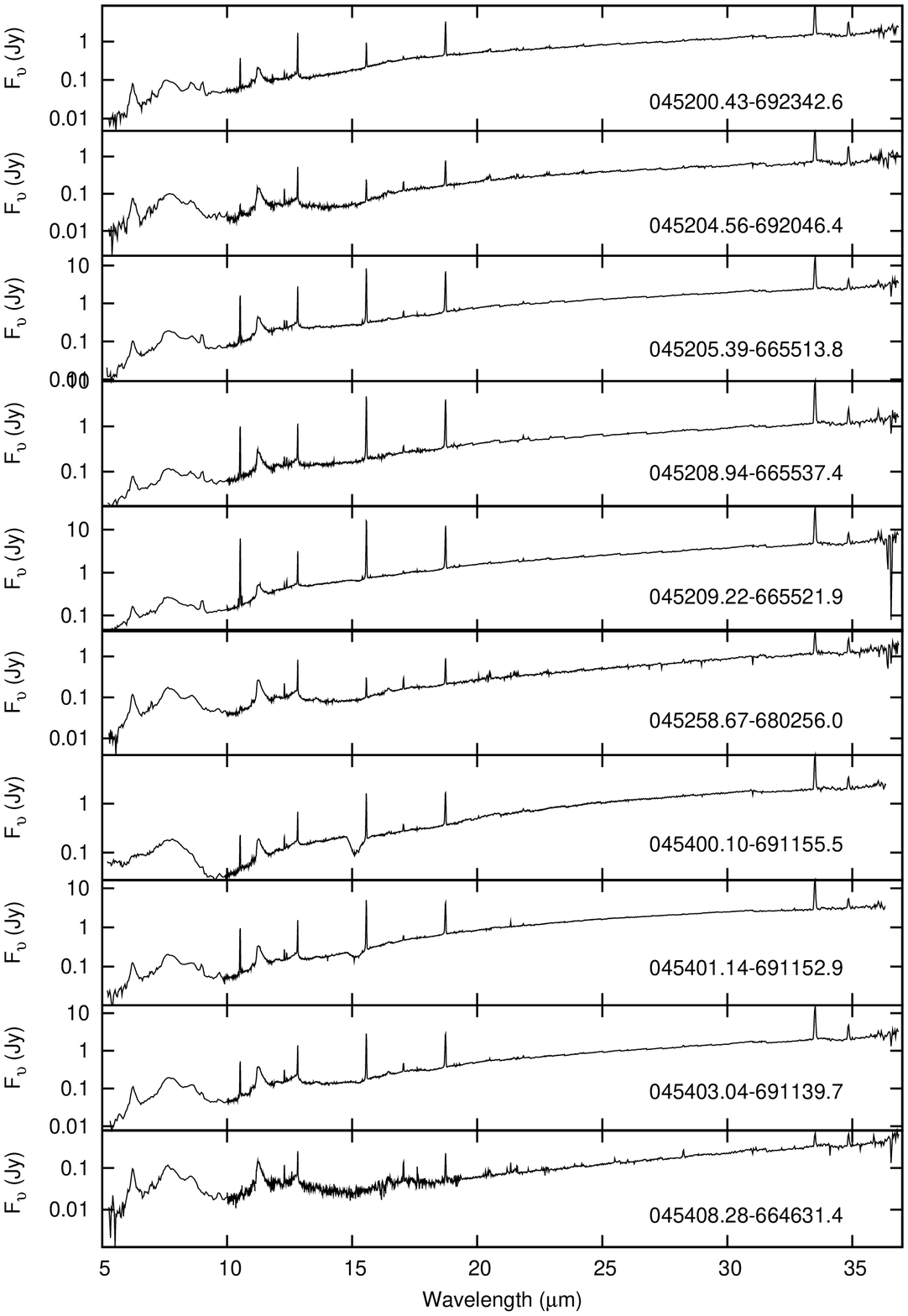}
\caption{Continuation of Figure 1d.}
\label{f1_16}
\notetoeditor{This figure is to appear in the electronic version of the paper only}
\end{center}
\end{figure}

\clearpage

\begin{figure}[t]
\figurenum{1q}
\begin{center}
\includegraphics[width=0.8\textwidth]{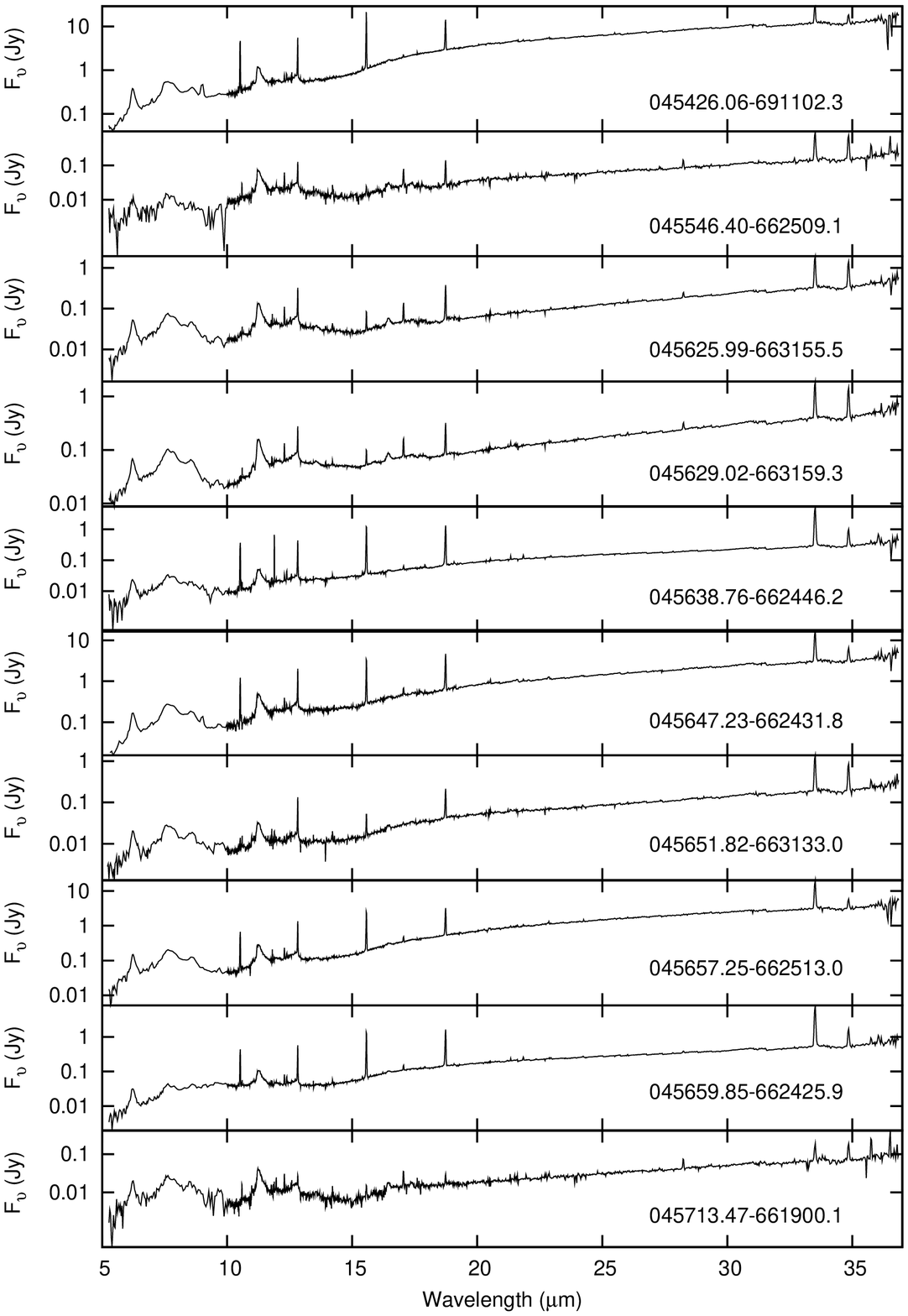}
\caption{Continuation of Figure 1d.}
\label{f1_17}
\notetoeditor{This figure is to appear in the electronic version of the paper only}
\end{center}
\end{figure}

\clearpage

\begin{figure}[t]
\figurenum{1r}
\begin{center}
\includegraphics[width=0.8\textwidth]{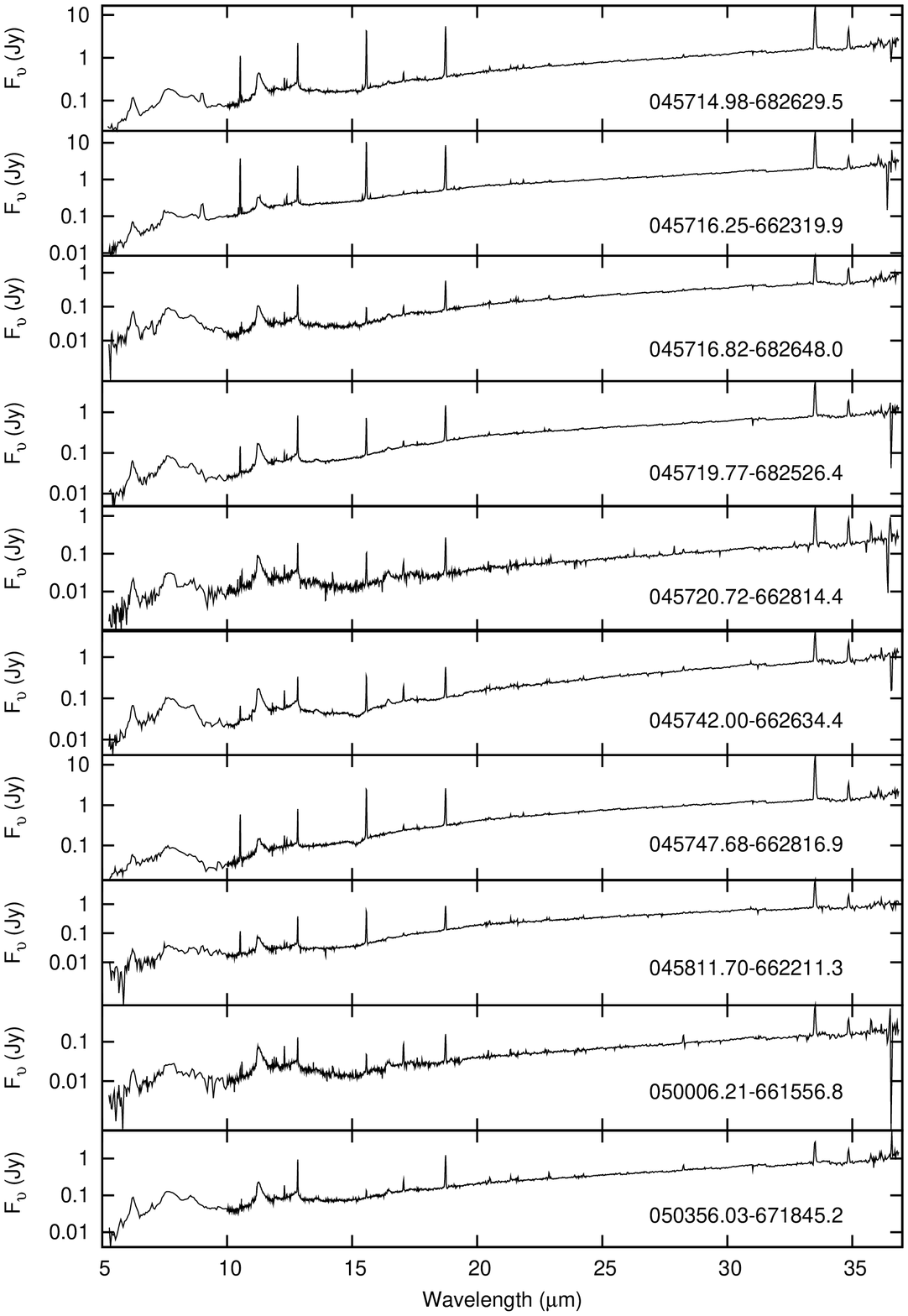}
\caption{Continuation of Figure 1d.}
\label{f1_18}
\notetoeditor{This figure is to appear in the electronic version of the paper only}
\end{center}
\end{figure}

\clearpage

\begin{figure}[t]
\figurenum{1s}
\begin{center}
\includegraphics[width=0.8\textwidth]{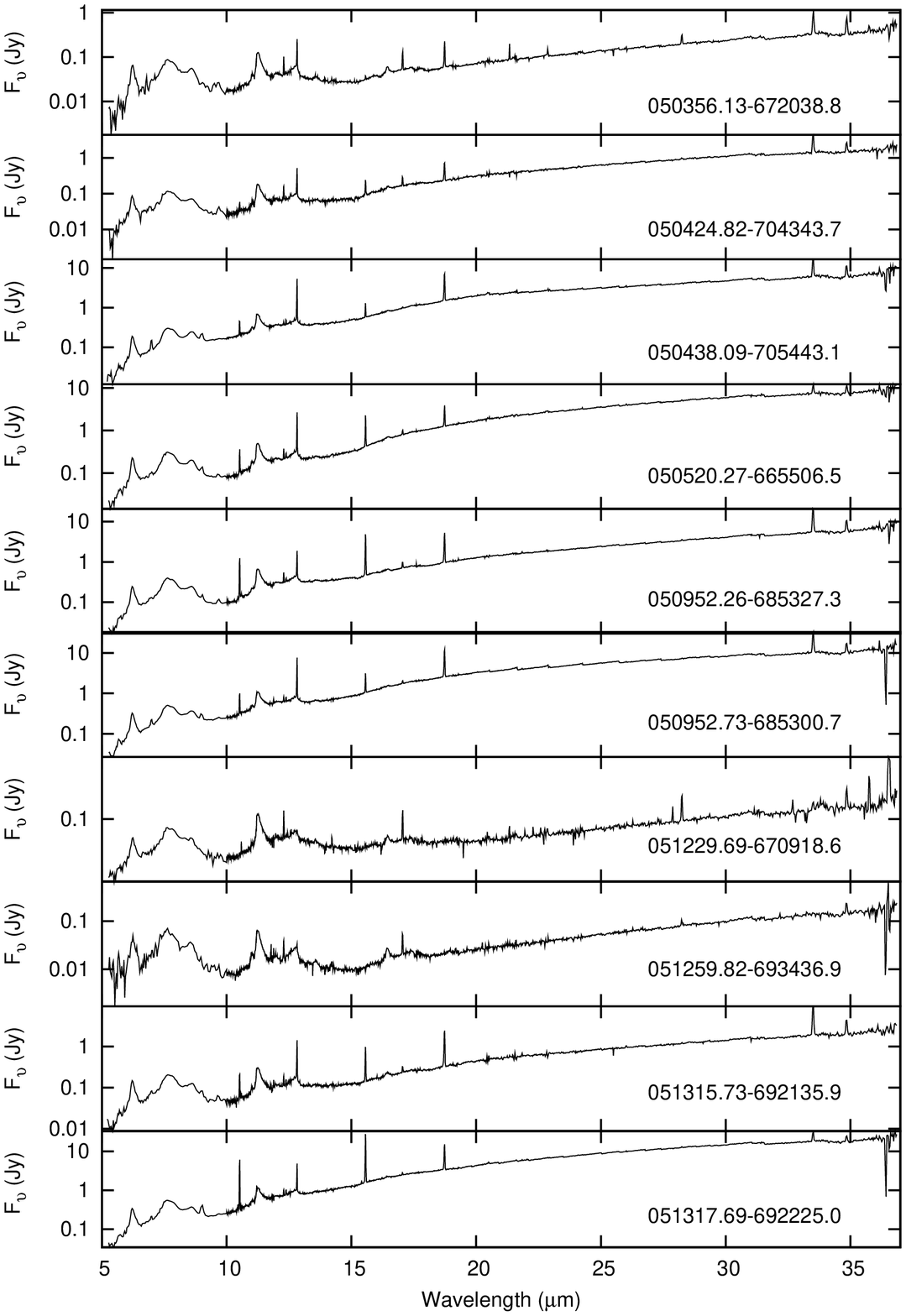}
\caption{Continuation of Figure 1d.}
\label{f1_19}
\notetoeditor{This figure is to appear in the electronic version of the paper only}
\end{center}
\end{figure}

\clearpage

\begin{figure}[t]
\figurenum{1t}
\begin{center}
\includegraphics[width=0.8\textwidth]{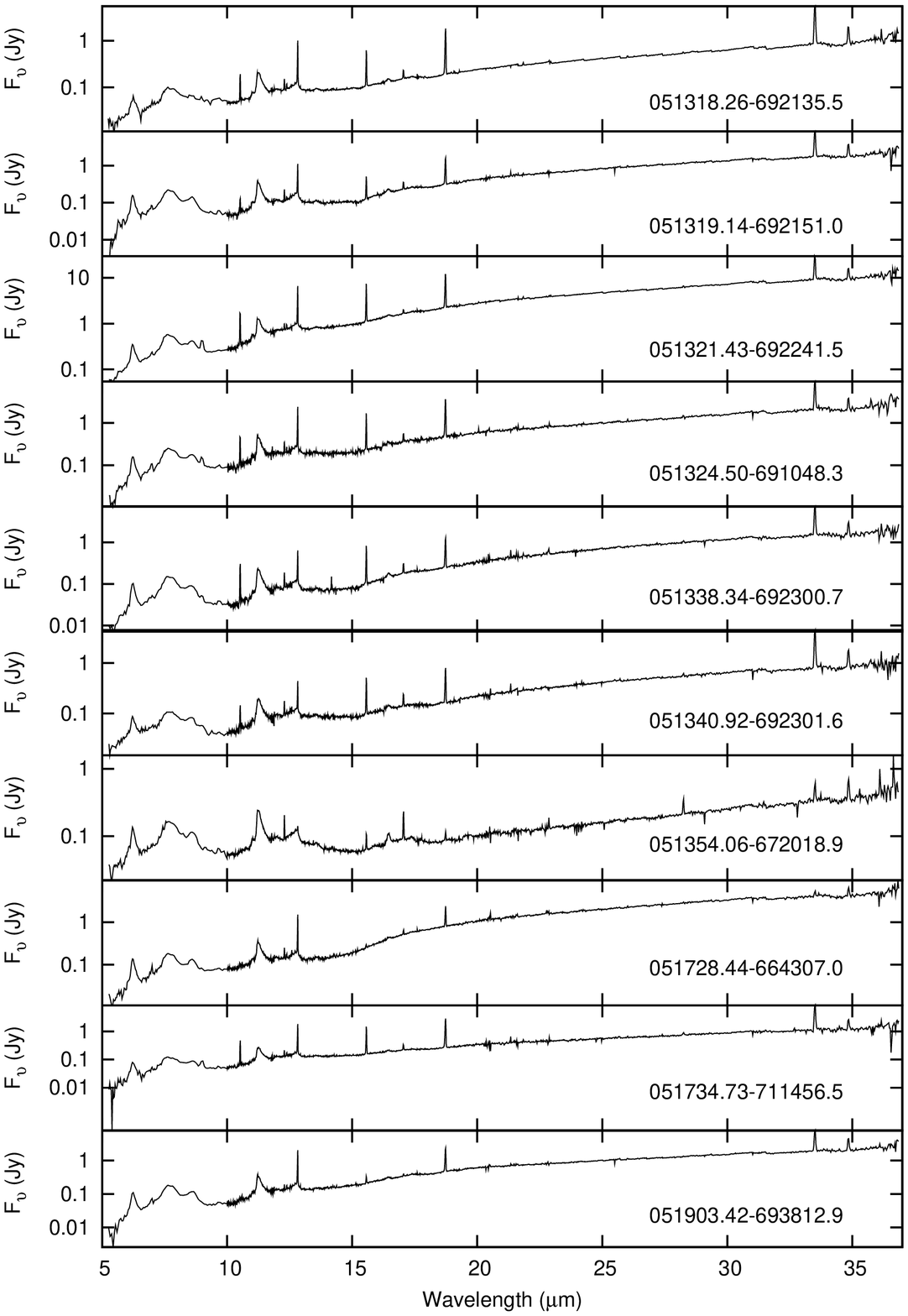}
\caption{Continuation of Figure 1d.}
\label{f1_20}
\notetoeditor{This figure is to appear in the electronic version of the paper only}
\end{center}
\end{figure}

\clearpage

\begin{figure}[t]
\figurenum{1u}
\begin{center}
\includegraphics[width=0.8\textwidth]{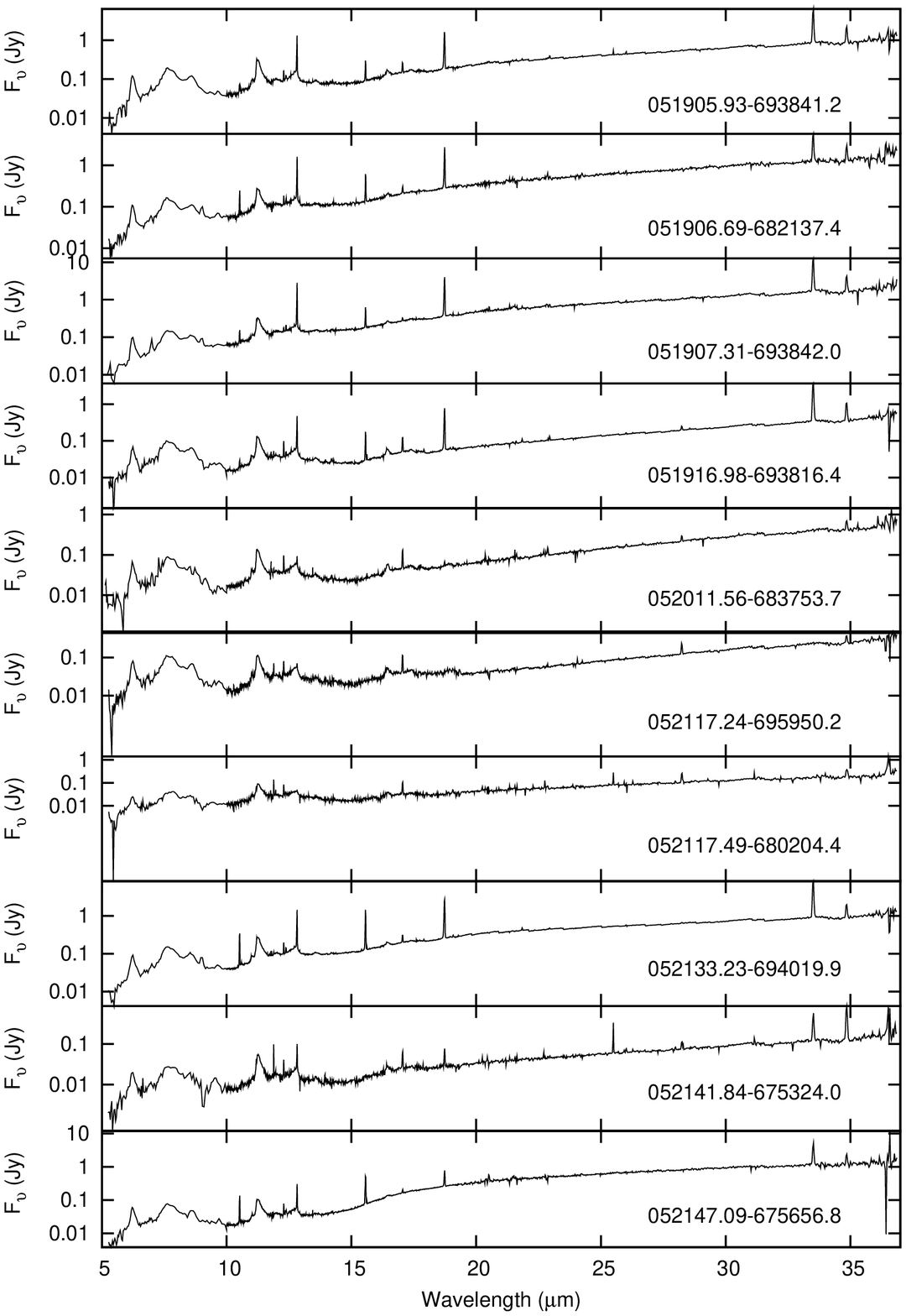}
\caption{Continuation of Figure 1d.}
\label{f1_21}
\notetoeditor{This figure is to appear in the electronic version of the paper only}
\end{center}
\end{figure}

\clearpage

\begin{figure}[t]
\figurenum{1v}
\begin{center}
\includegraphics[width=0.8\textwidth]{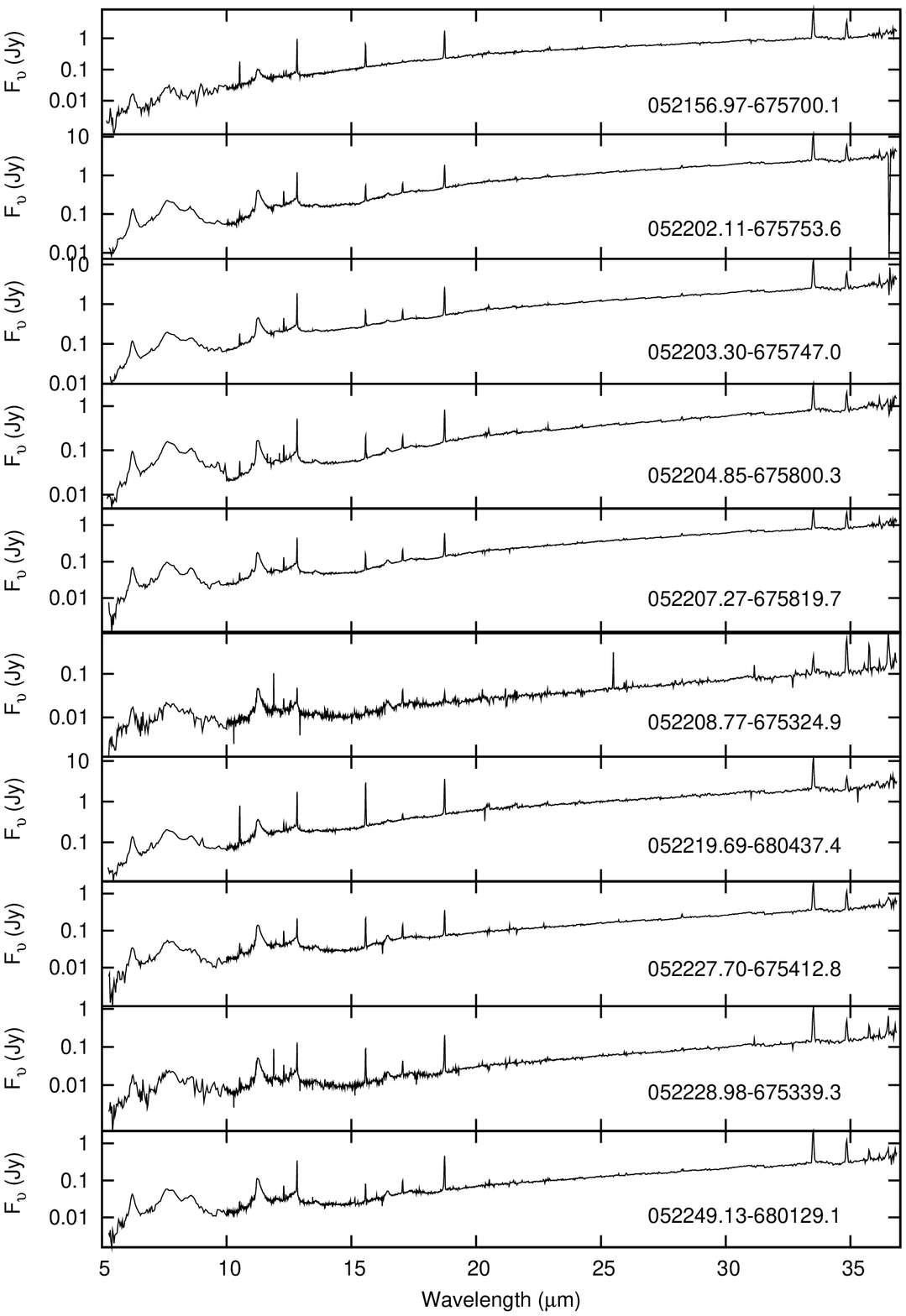}
\caption{Continuation of Figure 1d.}
\label{f1_22}
\notetoeditor{This figure is to appear in the electronic version of the paper only}
\end{center}
\end{figure}

\clearpage

\begin{figure}[t]
\figurenum{1w}
\begin{center}
\includegraphics[width=0.8\textwidth]{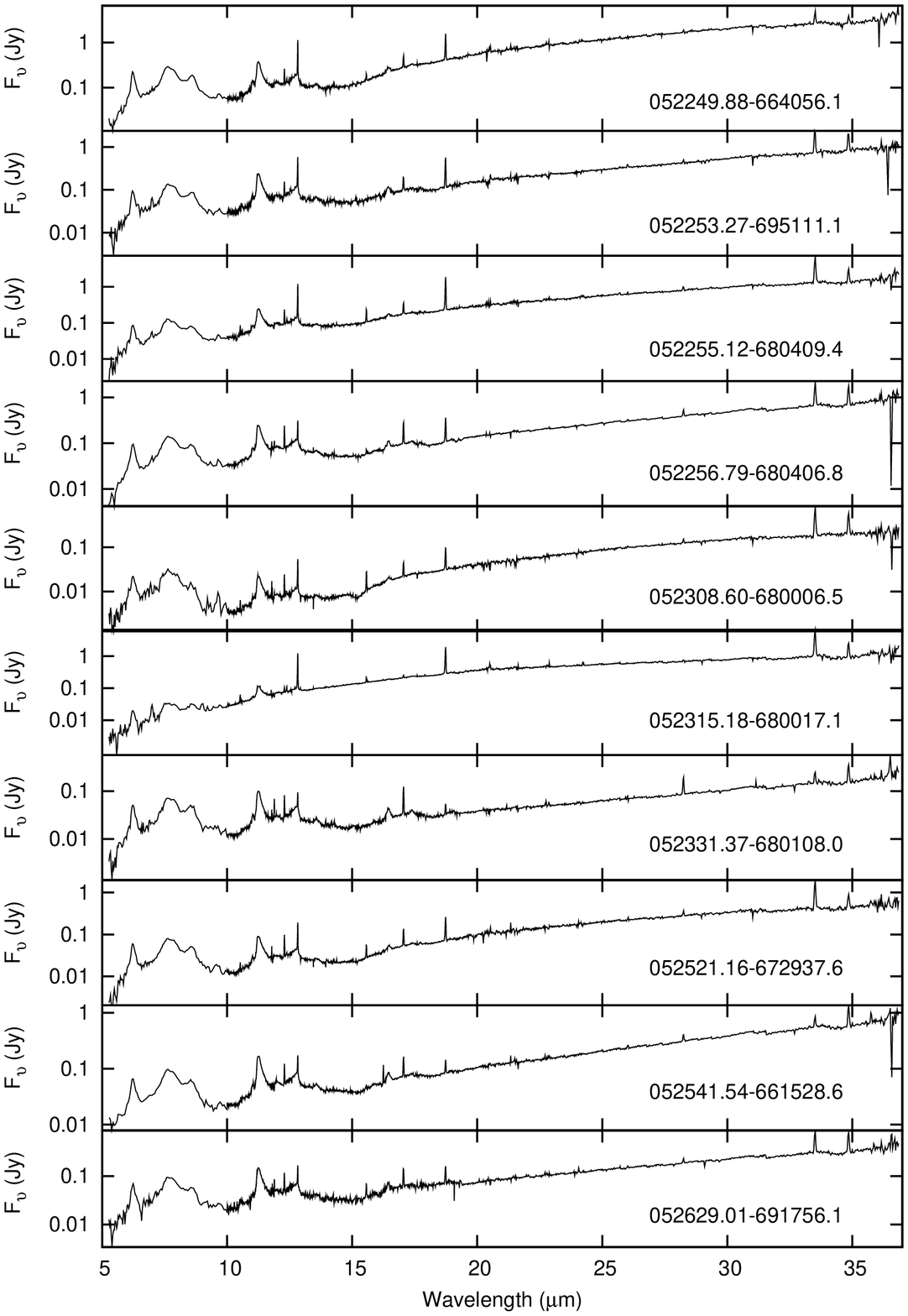}
\caption{Continuation of Figure 1d.}
\label{f1_23}
\notetoeditor{This figure is to appear in the electronic version of the paper only}
\end{center}
\end{figure}

\clearpage

\begin{figure}[t]
\figurenum{1x}
\begin{center}
\includegraphics[width=0.8\textwidth]{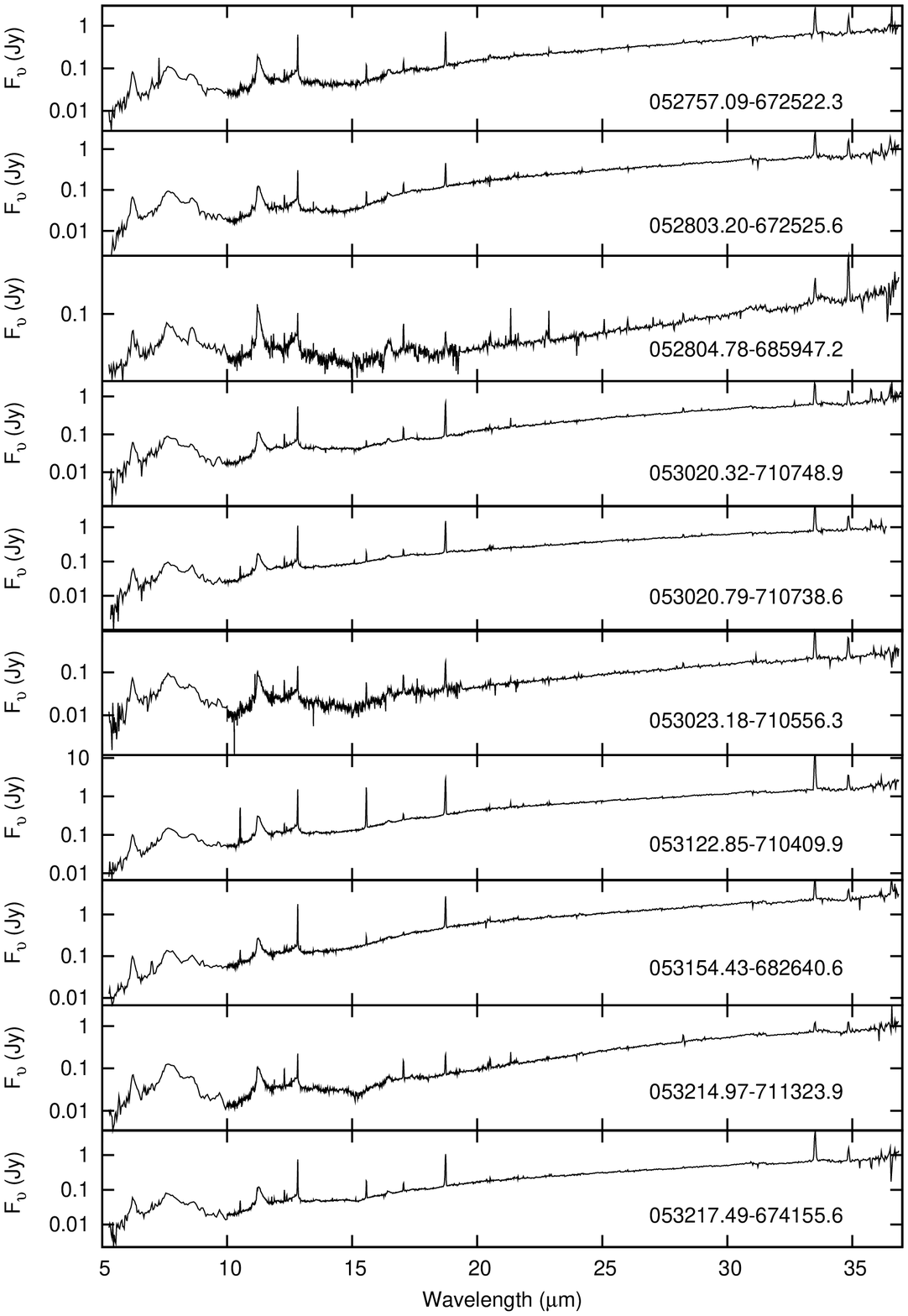}
\caption{Continuation of Figure 1d.}
\label{f1_24}
\notetoeditor{This figure is to appear in the electronic version of the paper only}
\end{center}
\end{figure}

\clearpage

\begin{figure}[t]
\figurenum{1y}
\begin{center}
\includegraphics[width=0.8\textwidth]{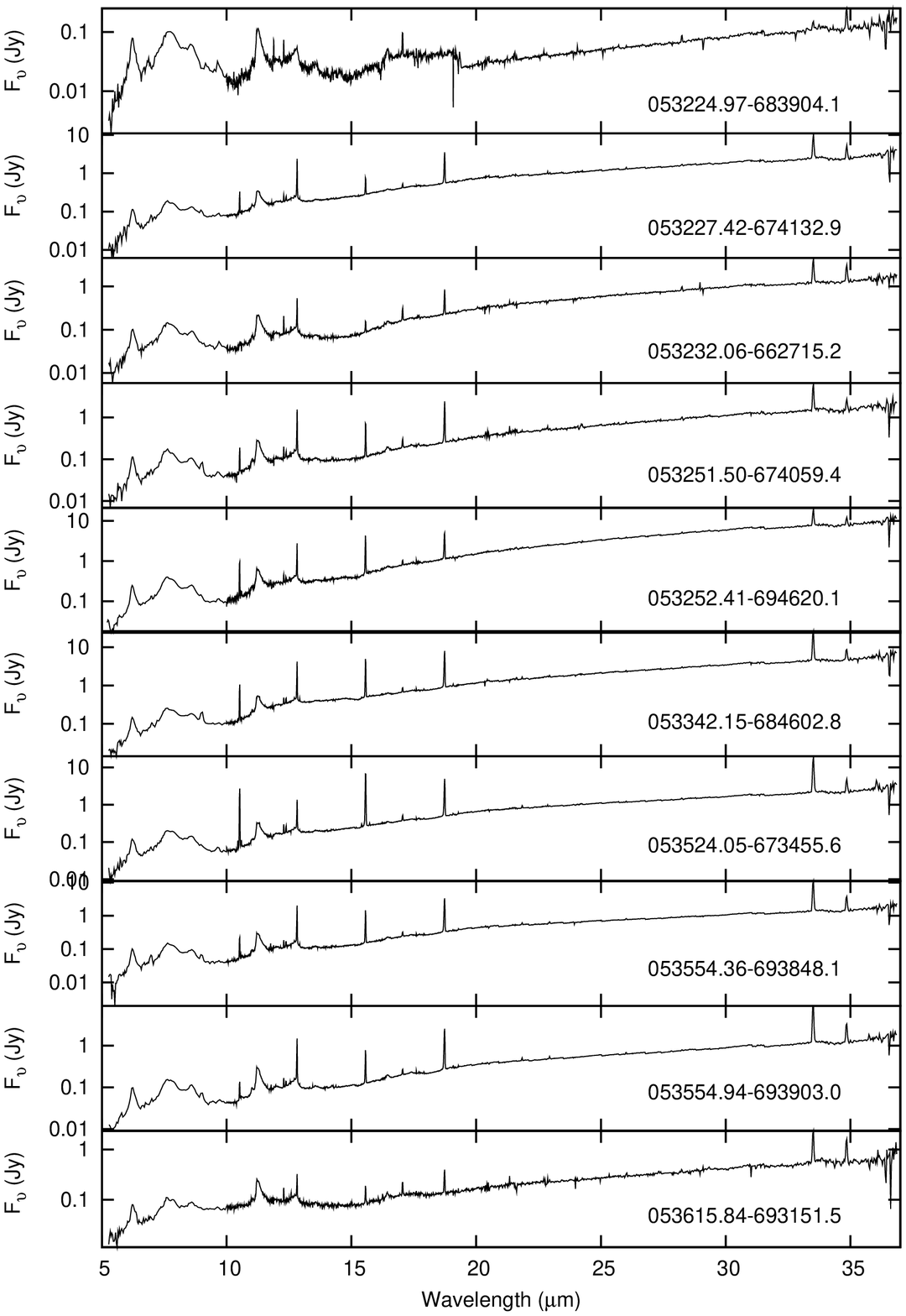}
\caption{Continuation of Figure 1d.}
\label{f1_25}
\notetoeditor{This figure is to appear in the electronic version of the paper only}
\end{center}
\end{figure}

\clearpage

\begin{figure}[t]
\figurenum{1z}
\begin{center}
\includegraphics[width=0.8\textwidth]{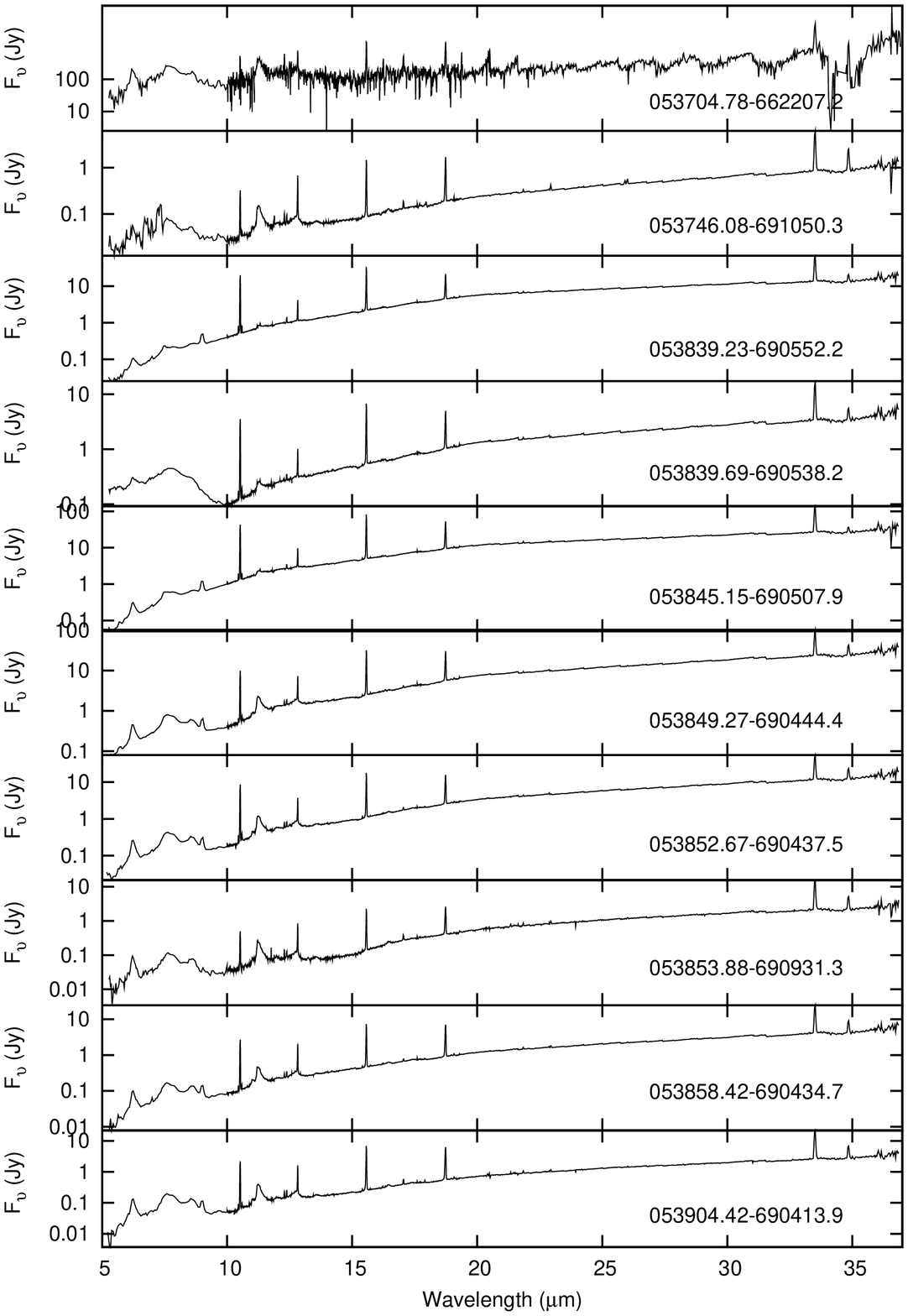}
\caption{Continuation of Figure 1d.}
\label{f1_26}
\notetoeditor{This figure is to appear in the electronic version of the paper only}
\end{center}
\end{figure}

\clearpage

\begin{figure}[t]
\figurenum{1aa}
\begin{center}
\includegraphics[width=0.8\textwidth]{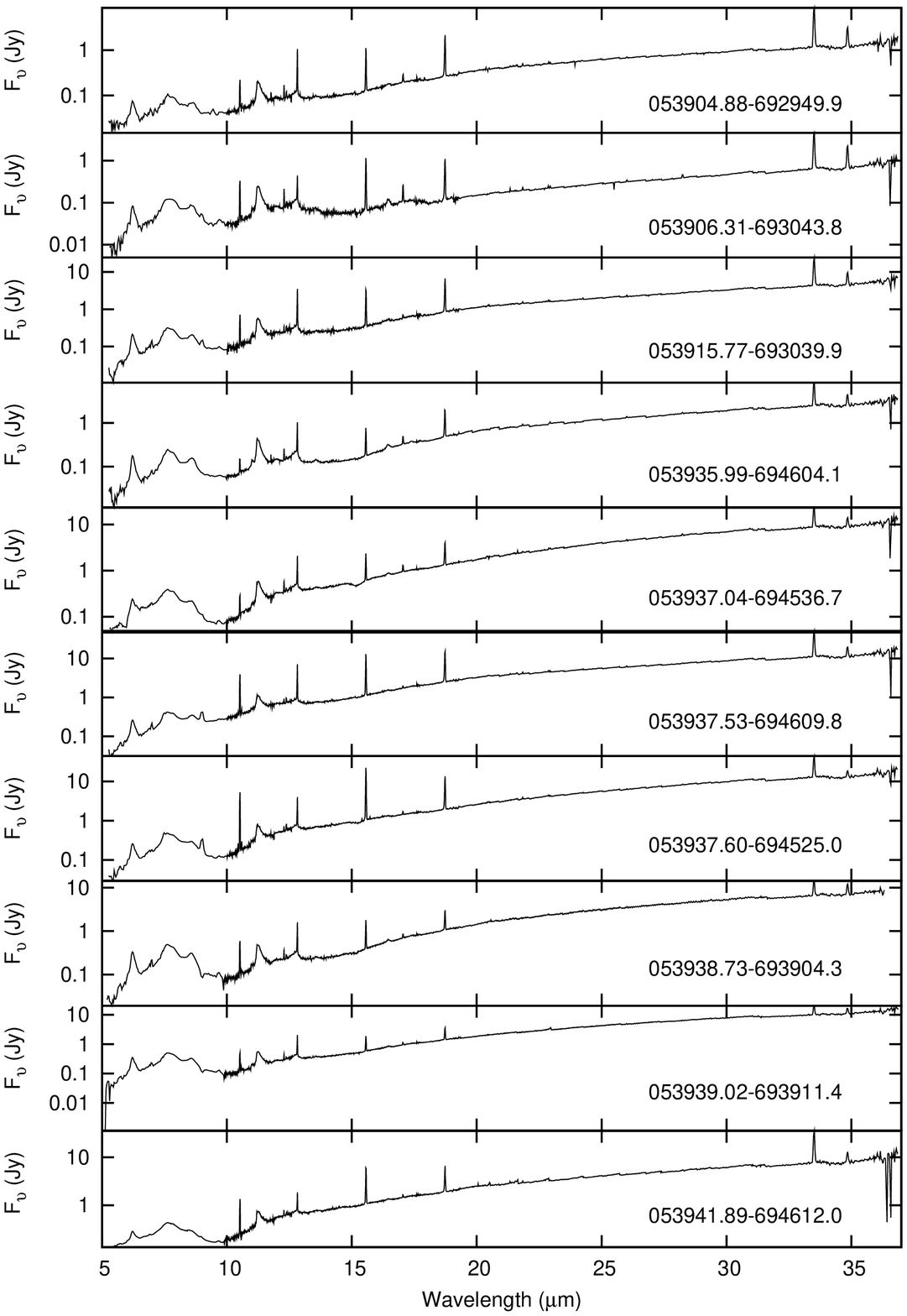}
\caption{Continuation of Figure 1d.}
\label{f1_27}
\notetoeditor{This figure is to appear in the electronic version of the paper only}
\end{center}
\end{figure}

\clearpage

\begin{figure}[t]
\figurenum{1bb}
\begin{center}
\includegraphics[width=0.8\textwidth]{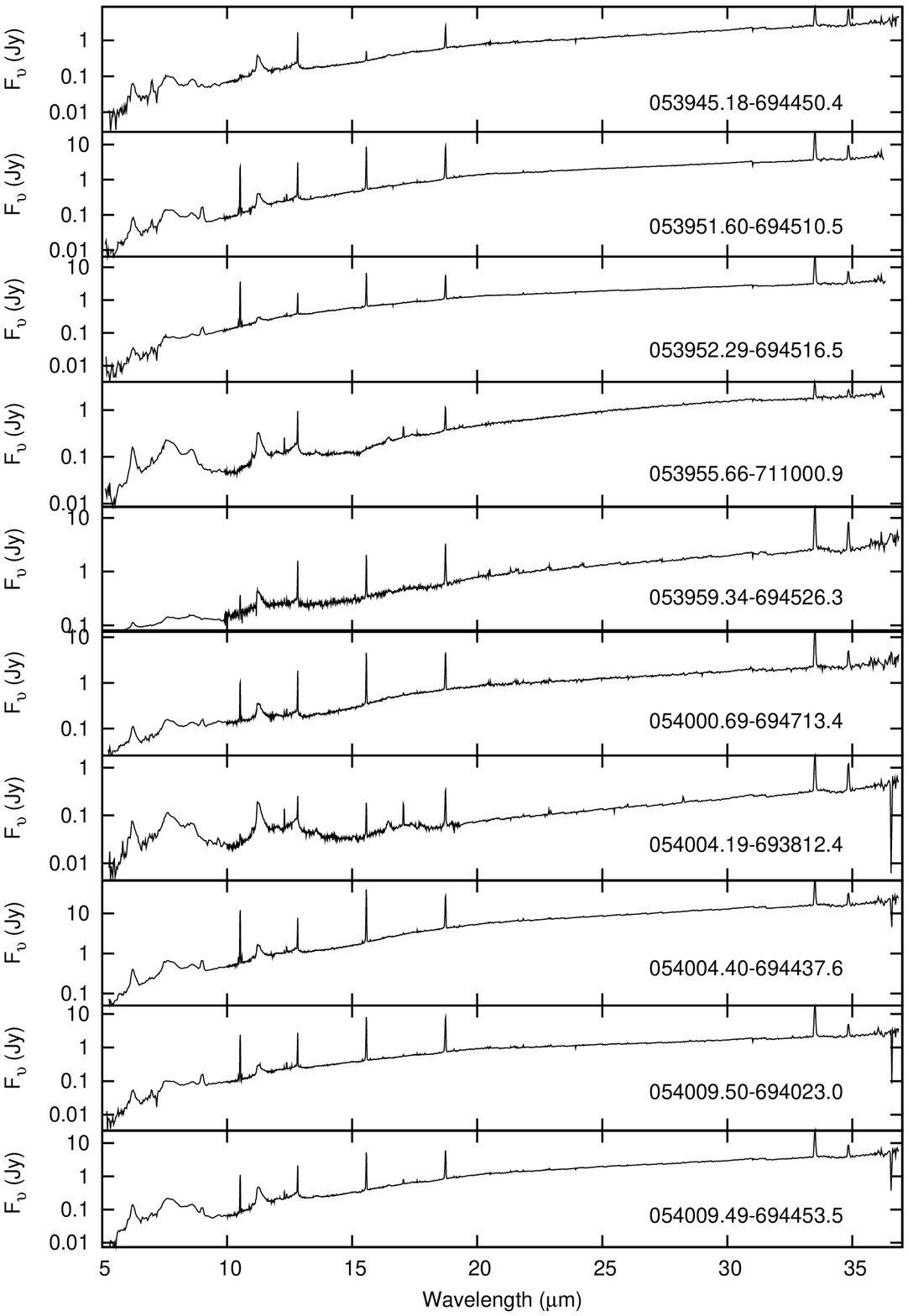}
\caption{Continuation of Figure 1d.}
\label{f1_28}
\notetoeditor{This figure is to appear in the electronic version of the paper only}
\end{center}
\end{figure}

\clearpage

\begin{figure}[t]
\figurenum{1cc}
\begin{center}
\includegraphics[width=0.8\textwidth]{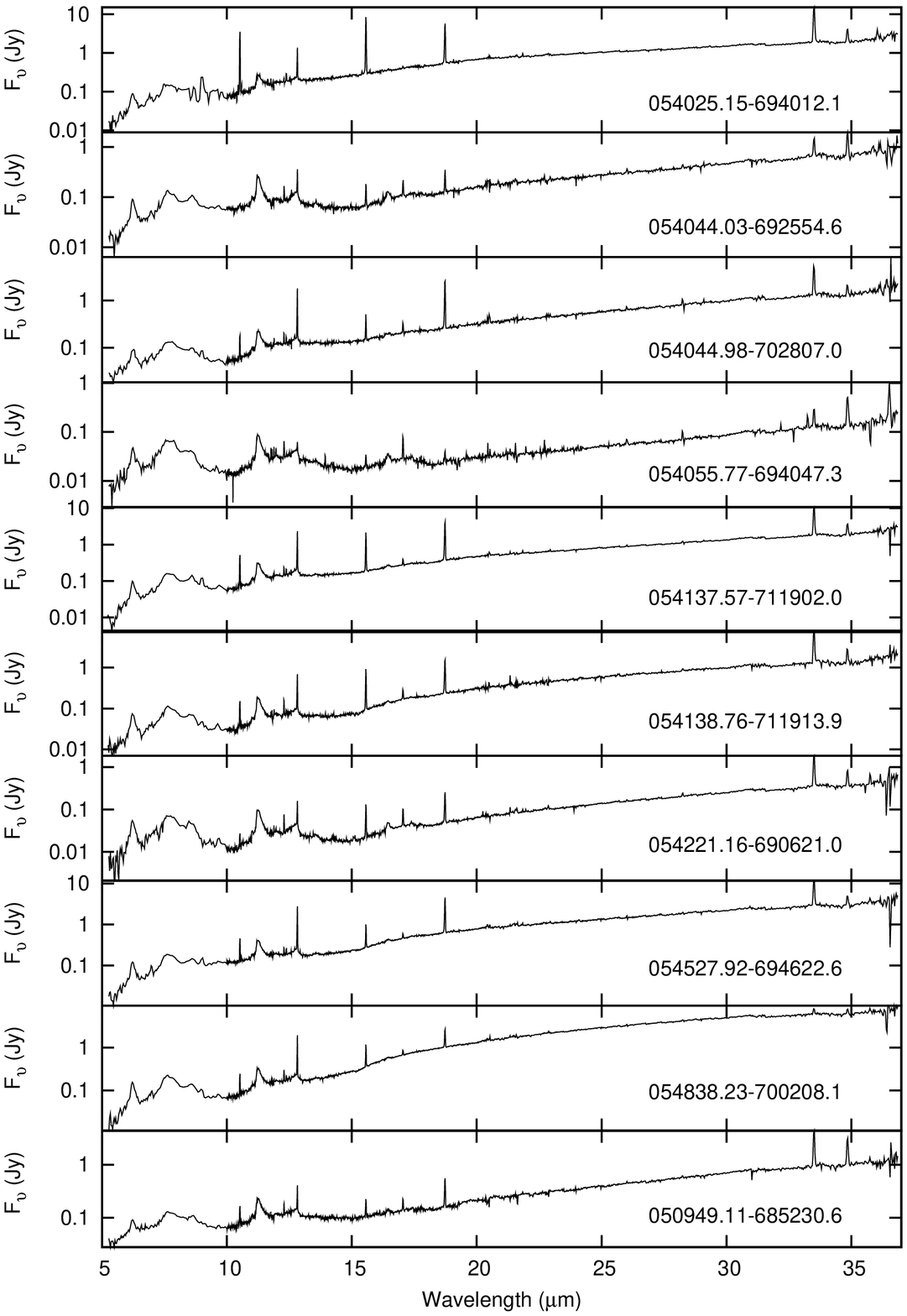}
\caption{Continuation of Figure 1d.}
\label{f1_29}
\notetoeditor{This figure is to appear in the electronic version of the paper only}
\end{center}
\end{figure}

\clearpage

\begin{figure}[t]
\figurenum{1dd}
\begin{center}
\includegraphics[width=0.8\textwidth]{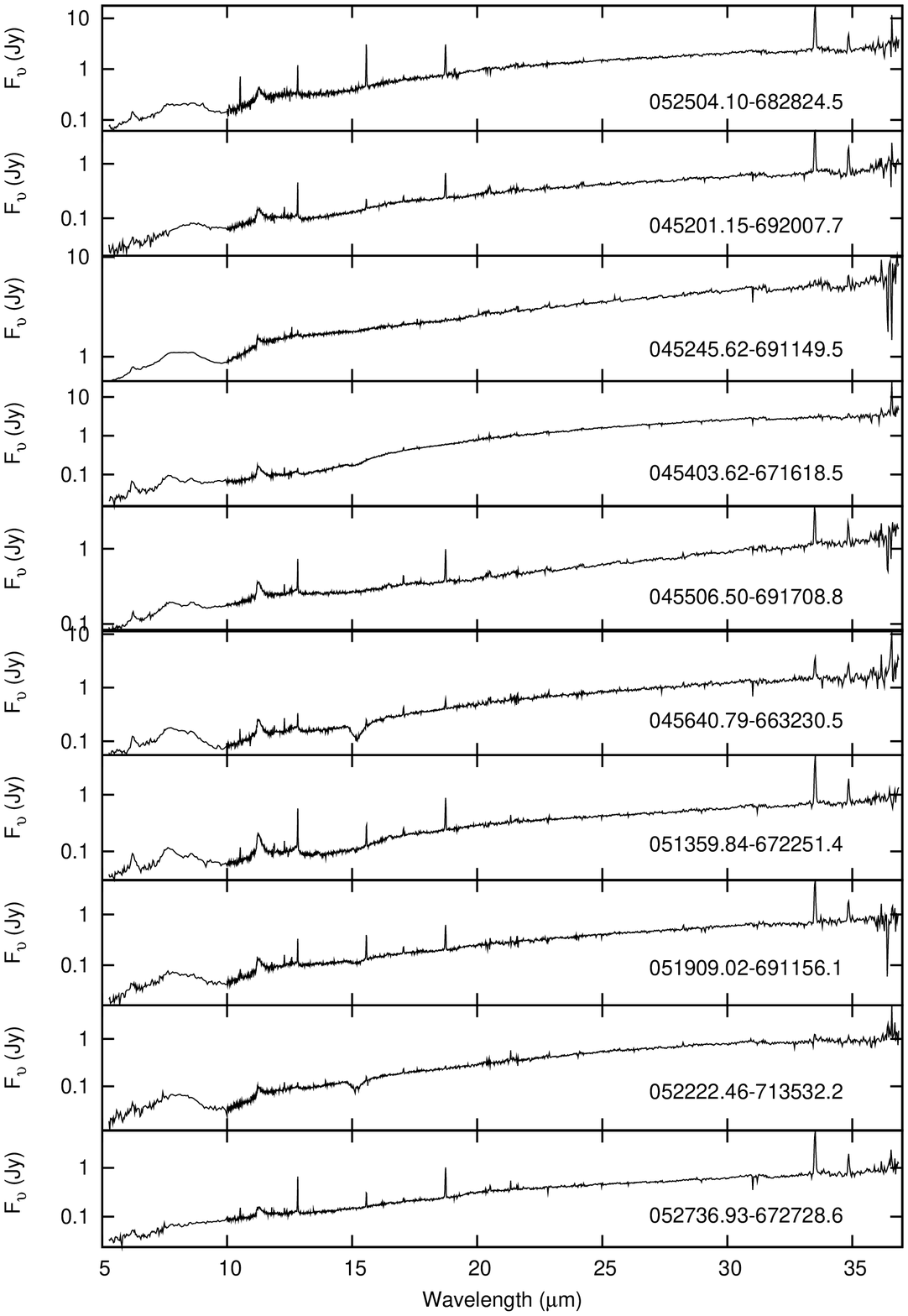}
\caption{Continuation of Figure 1d.}
\label{f1_30}
\notetoeditor{This figure is to appear in the electronic version of the paper only}
\end{center}
\end{figure}

\clearpage

\begin{figure}[t]
\figurenum{1ee}
\begin{center}
\includegraphics[width=0.8\textwidth]{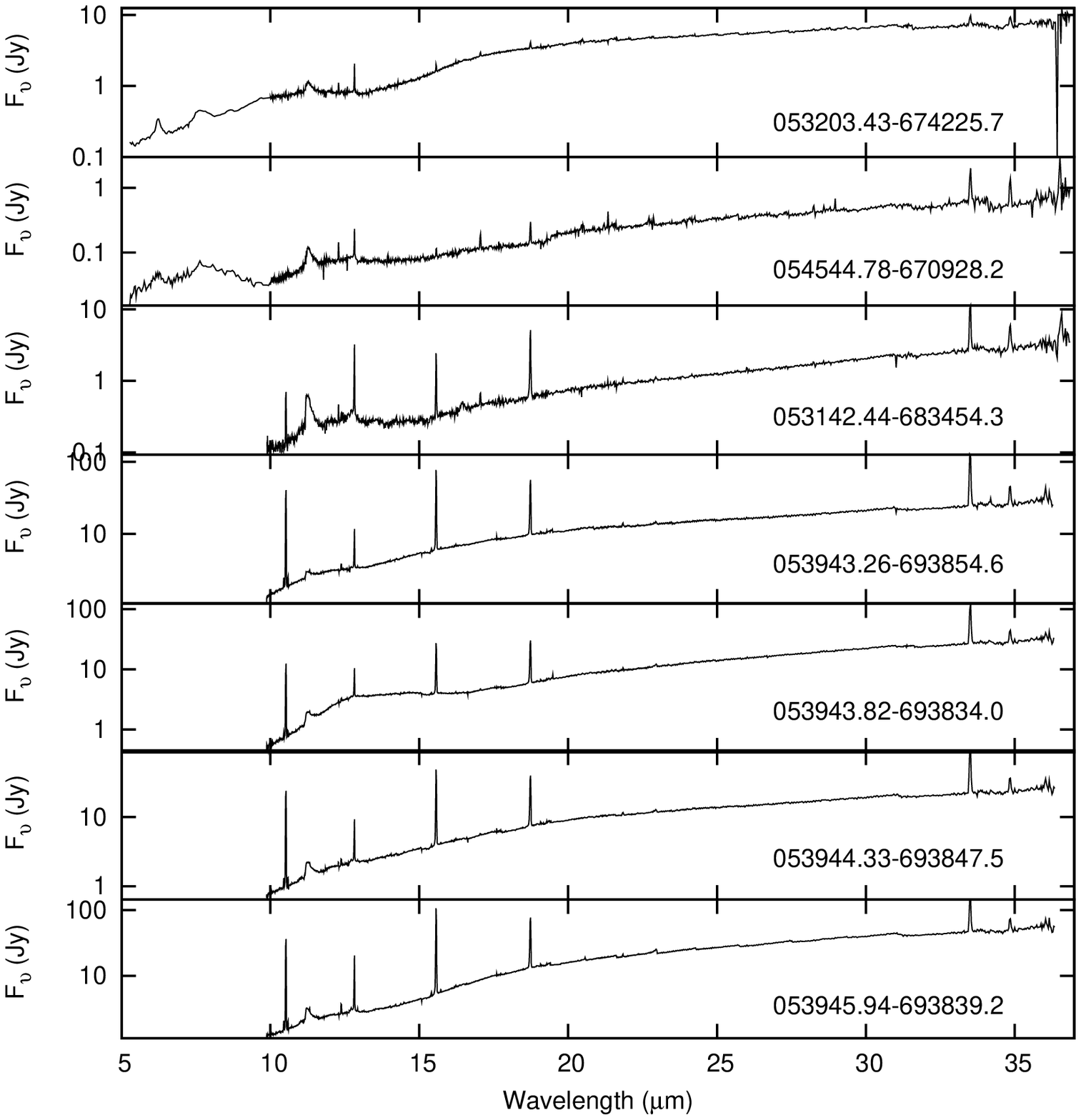}
\caption{Continuation of Figure 1d.}
\label{f1_31}
\notetoeditor{This figure is to appear in the electronic version of the paper only}
\end{center}
\end{figure}

\clearpage

\begin{figure}[t]
\figurenum{2a}
\begin{center}
\includegraphics[width=0.7\textwidth]{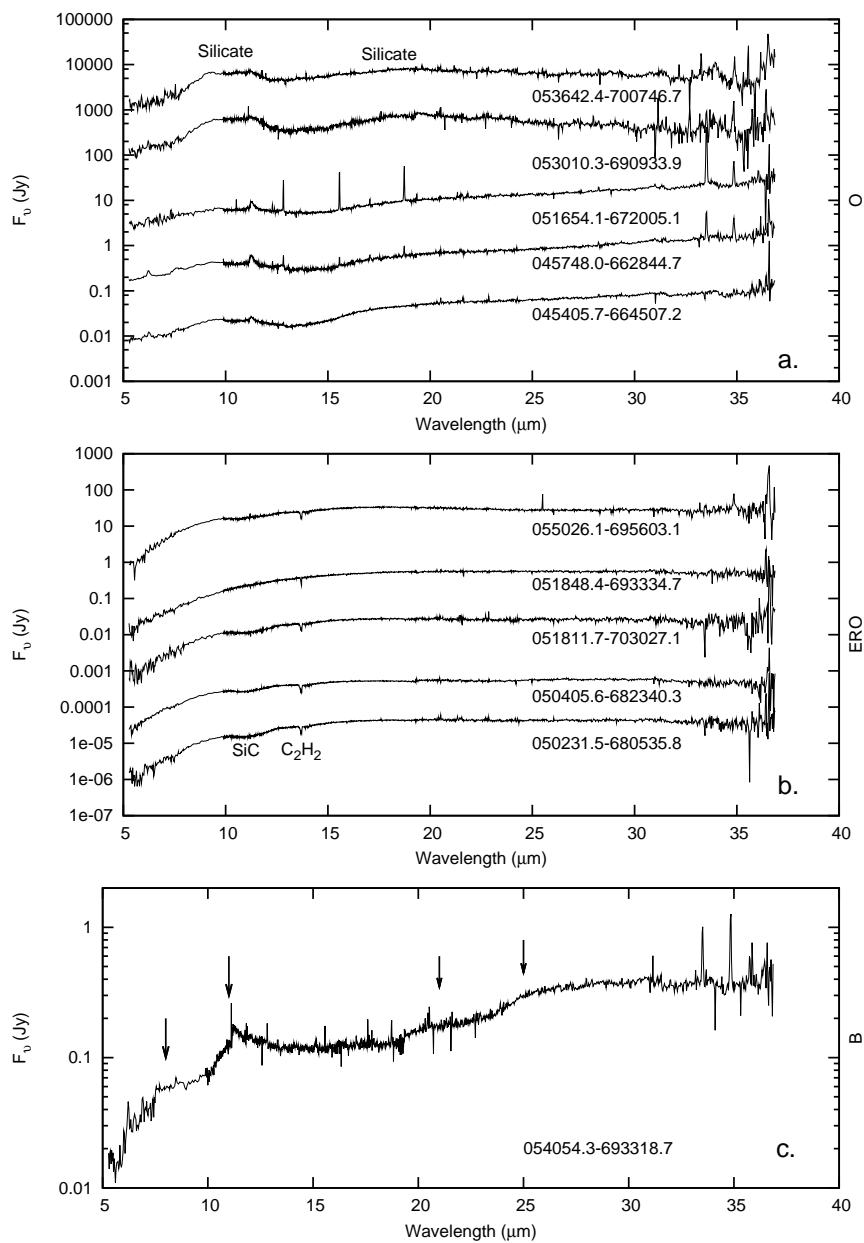}
\caption{Representative spectra of the O, ERO, and B groups in panels a., b., and c., respectively. The source identification is given below each curve. Top: Silicate emission features at 10 and 18 $\mu$m are indicated. The spectra from bottom to top have been scaled by the following multiplicative factors for clarity: 0.1, 1, 100, 4000, and 8$\times$10$^{4}$. Middle: Absorption features coincident with known SiC and C$_{2}$H$_{2}$ bands are indicated. The spectra from bottom to top have been scaled by the following multiplicative factors for clarity: 10$^{-4}$, 0.1, 1, 100, and 10$^{-3}$. Bottom: Arrows indicate the locations of unidentified broad emission features that create the step-like spectral shape.}
\label{f2_1}
\end{center}
\end{figure}

\clearpage

\begin{figure}[t]
\figurenum{2b}
\begin{center}
\includegraphics[width=0.8\textwidth]{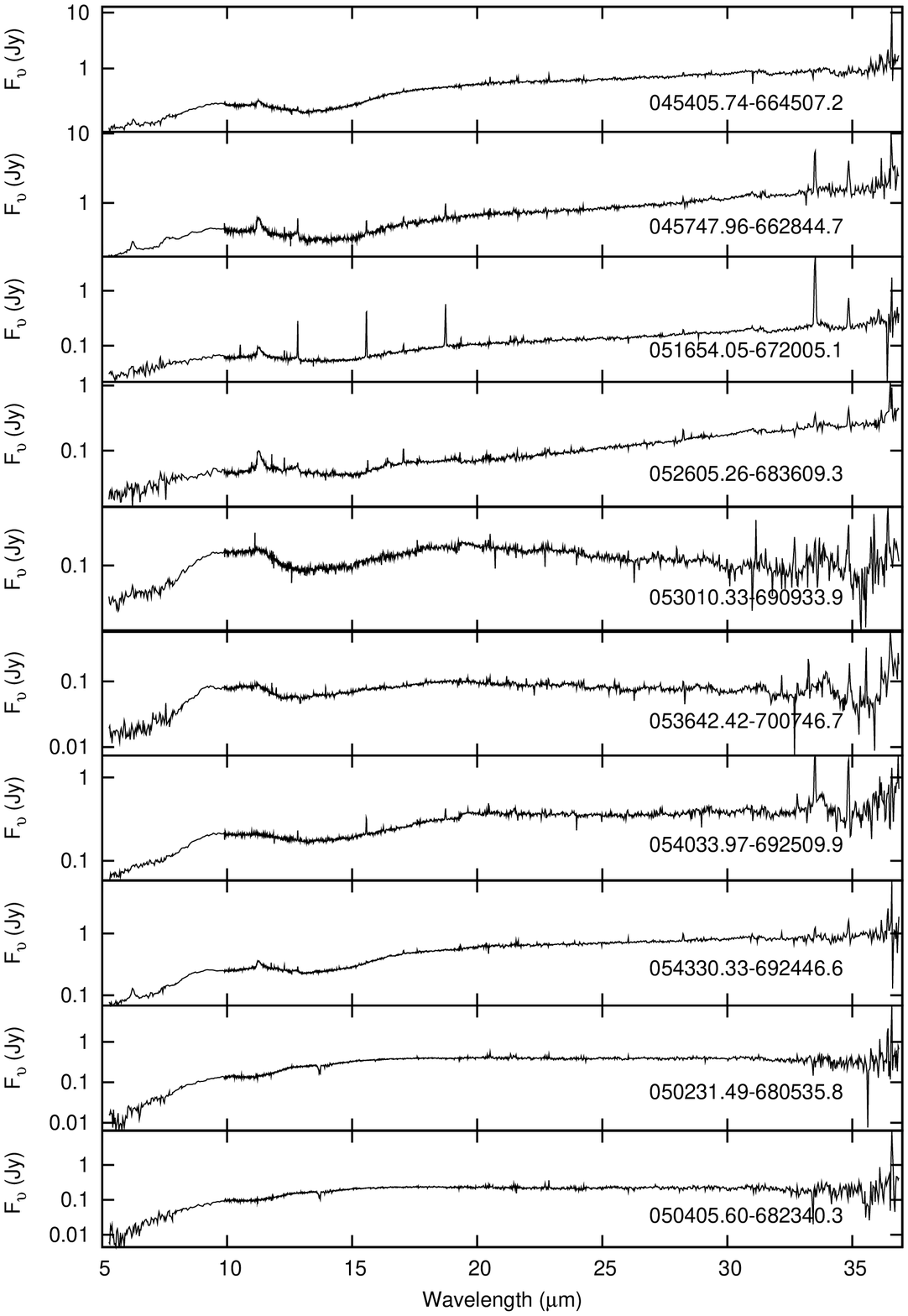}
\caption{Spectra of the complete source catalog from which selected sources are shown in Figure 2. Sources are ordered as in Table 1, and source identifications are given below each curve.}
\label{f2_2}
\notetoeditor{This figure is to appear in the electronic version of the paper only}
\end{center}
\end{figure}

\clearpage

\begin{figure}[t]
\figurenum{2c}
\begin{center}
\includegraphics[width=0.8\textwidth]{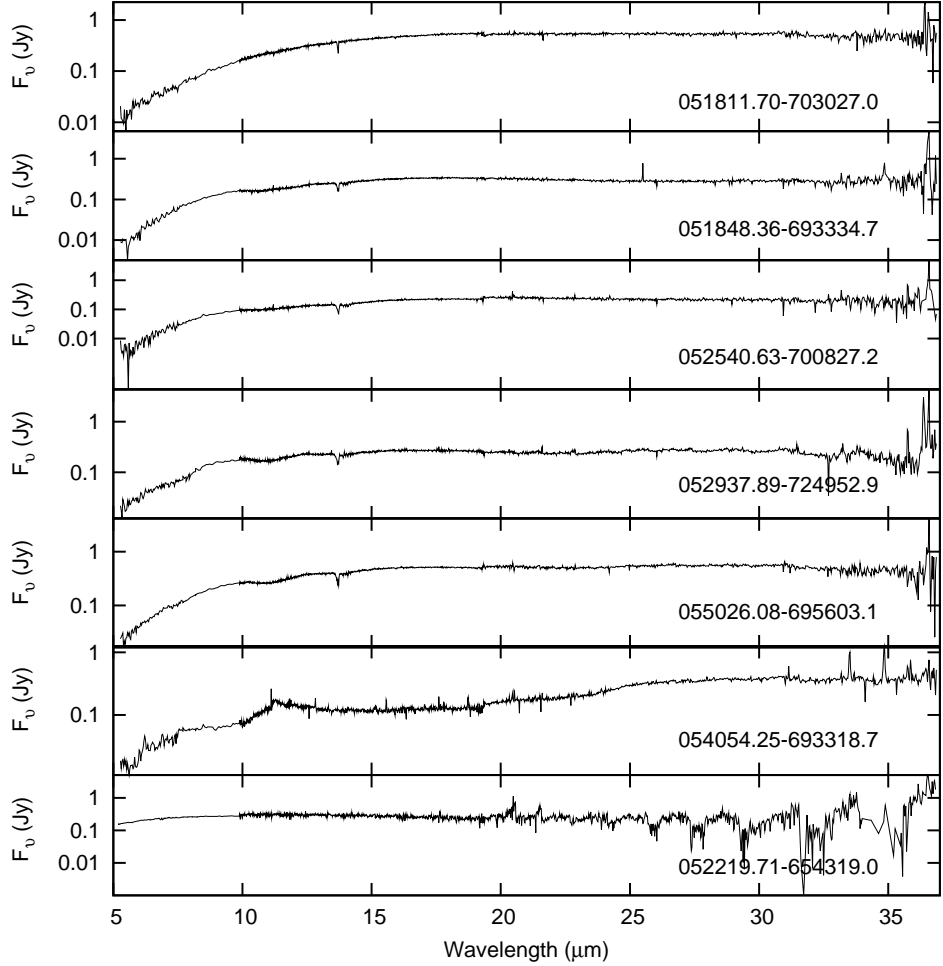}
\caption{Continuation of Figure 2b.}
\label{f2_3}
\notetoeditor{This figure is to appear in the electronic version of the paper only}
\end{center}
\end{figure}

\clearpage

\begin{figure}[t]
\figurenum{3}
\begin{center}
\includegraphics[width=0.8\textwidth]{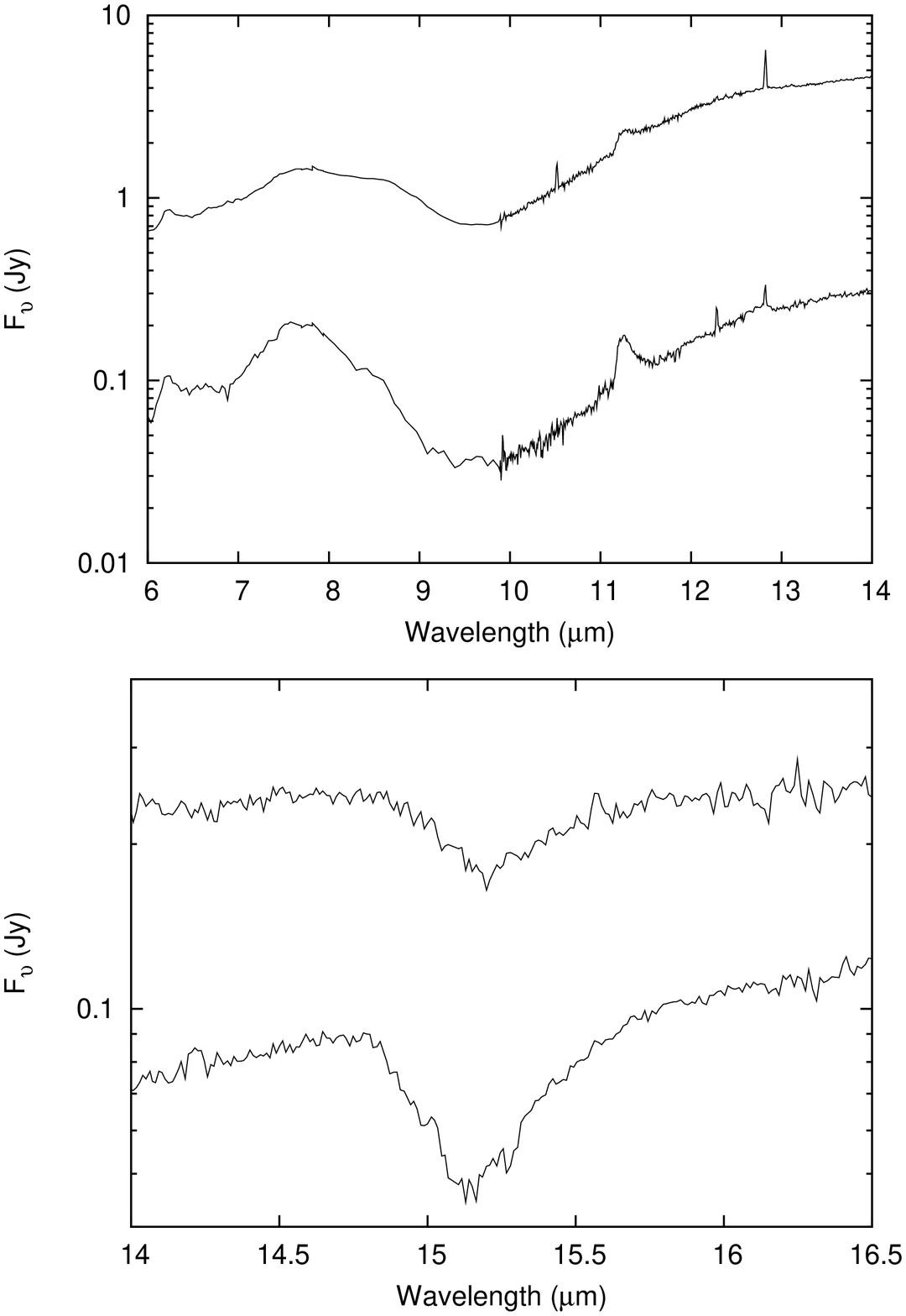}
\caption{Top panel: Spectra of 052646.61--684847.2 (top) and 045111.39--692646.7 (bottom) showing the variation in silicate absoption band profiles. Bottom panel: Spectra of 053754.82--693435.8 (top) and 045842.47--660835.7 (bottom) showing the variation in CO$_{2}$ ice absorption band profiles.  The flux values of 053754.82--693435.8 have been scaled by a factor of 2 for clarity.}
\label{f3}
\end{center}
\end{figure}

\clearpage

\begin{figure}[t]
\figurenum{4}
\begin{center}
\includegraphics[width=0.8\textwidth]{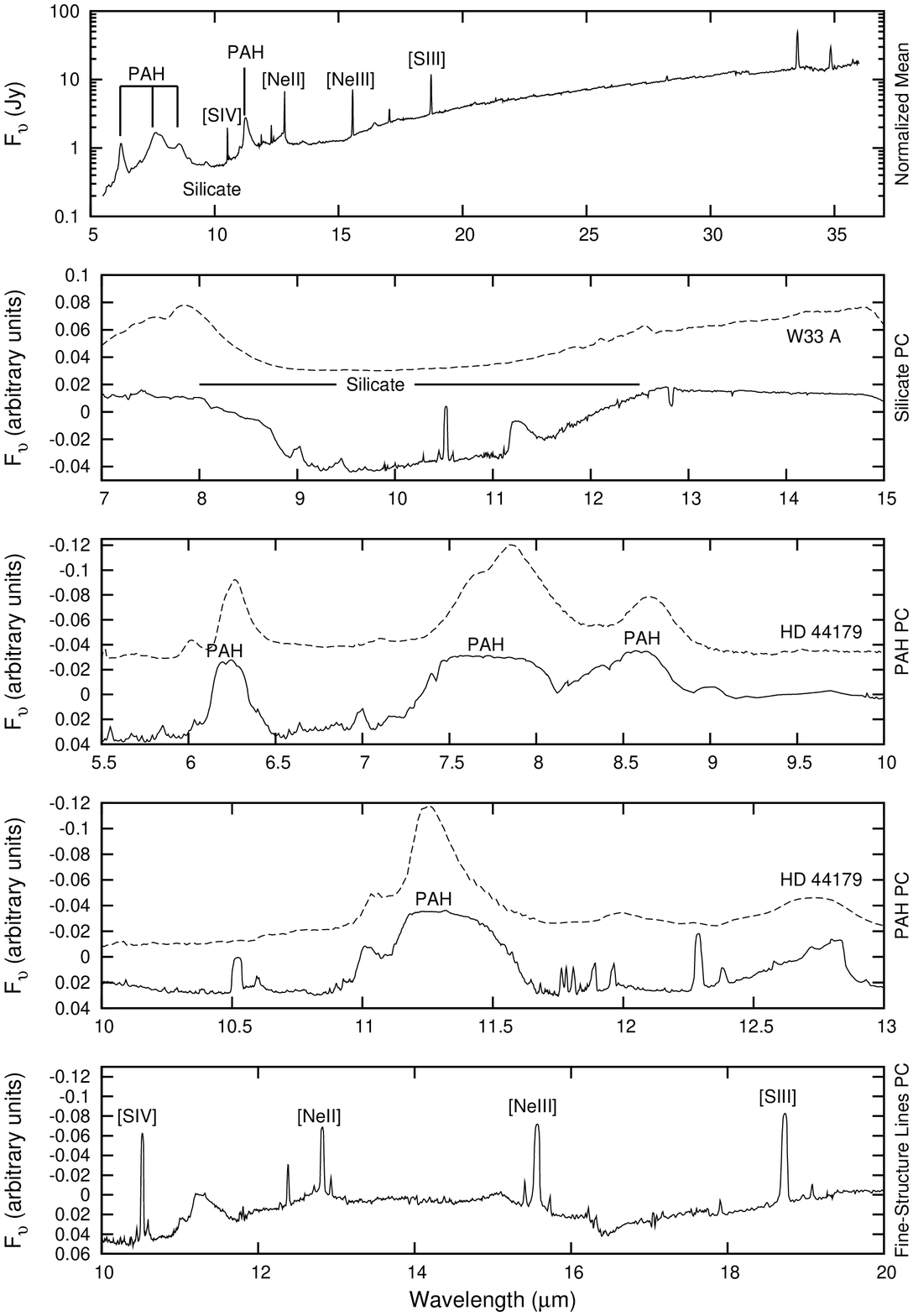}
\tiny
\caption{Top panel: the full normalized mean spectra of all the embedded YSOs. Bottom panels: the solid curves show the four principal components used in our analysis. The dashed curves are smoothed ISO data from objects with IR spectra dominated by the species represented within each particular principal component. The silicate panel shows an ISO spectrum from the highly embedded protostar W33A, while the PAH panels have the baseline-subtracted ISO spectrum of HD 44179, the post-AGB star at the center of the Red Rectangle proto-planetary neblua. Prominent features are identified in each panel.}
\label{f6}
\end{center}
\end{figure}

\clearpage

\begin{figure}[t]
\figurenum{5}
\begin{center}
\includegraphics[width=0.65\textwidth]{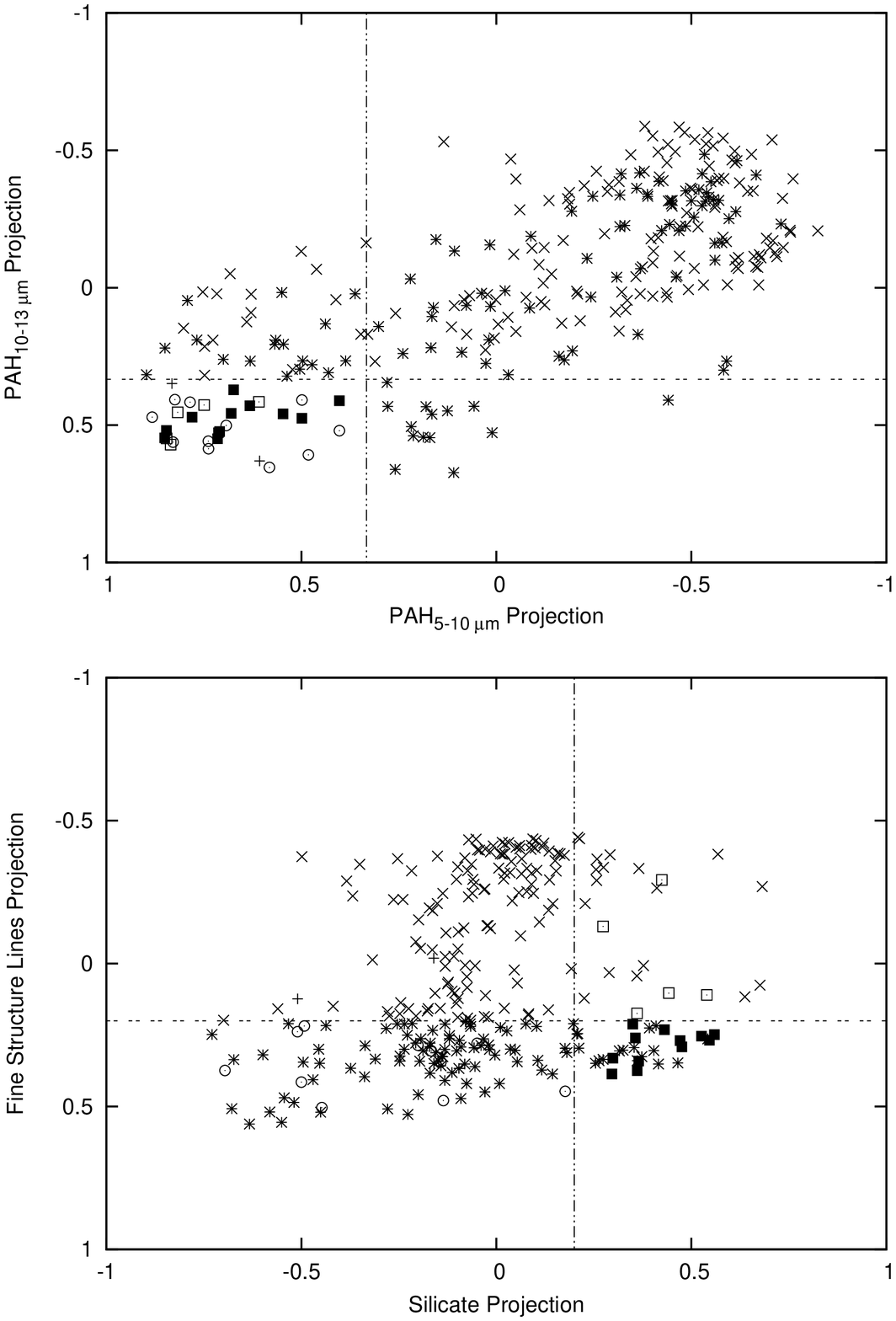}
\caption[\small]{Top: Projections of each embedded YSO's spectrum onto both the PAH$_{5-10\mu m}$ and PAH$_{10-13\mu m}$ principal components. S group sources are indicated by solid squares, SE objects by open squares, P group by stars, PE group by X's, E group are crosses, and F objects are open circles. The vertical and horizontal dotted lines indicate the cut-offs between sources which pass and fail the PAH$_{5-10\mu m}$ and PAH$_{10-13\mu m}$ tests, respectively. Objects with projections to the right of the vertical line or above the horizontal line are considered to have PAH emission. Bottom: Projections of each embedded YSO's spectrum onto both the silicate and fine-structure line principal components. The vertical and horizontal dotted lines indicate the cut-offs between sources which pass and fail the silicate and fine-structure line tests, respectively. Objects with projections to the right of the vertical line have silicate absorption, while those above the horizontal line are considered to have fine-structure line emission. Symbols are identical to the top panel.}
\label{f5}
\end{center}
\end{figure}

\clearpage

\begin{deluxetable}{lccccccccccc}
\rotate
\tablecolumns{12}
\tabletypesize{\tiny}
\tablewidth{588.40025pt}
\tablecaption{Source Classification}
\tablehead{ 
\multicolumn{3}{c}{} & \multicolumn{4}{c}{Presence of Feature} & \colhead{} & \multicolumn{4}{c}{Environment} \\
\cline{4-7} 
\cline{9-12}
\colhead{Object ID} & \colhead{KSPW} & \colhead{GC09} & \colhead{Silicate} & \colhead{PAH$_{5-10\mu m}$} & \colhead{PAH$_{10-13\mu m}$} & \colhead{Fine-Structure} & \colhead{Possible} & \colhead{Molecular} & \colhead{Compact } & \colhead{Diffuse H$\alpha$} & \colhead{Diffuse} \\
\colhead{} & \colhead{Class\tablenotemark{a}} & \colhead{Class} & \colhead{Absorption} & \colhead{Emission} & \colhead{Emission} & \colhead{Line Emission} & \colhead{Multiple} & \colhead{Cloud} & \colhead{H$\alpha$} & \colhead{([SIII] concen-} & \colhead{Dust} \\
\colhead{} & \colhead{} & \colhead{} & \colhead{(projection\tablenotemark{b})} & \colhead{(projection\tablenotemark{b})} & \colhead{(projection\tablenotemark{b})} & \colhead{(projection\tablenotemark{b})} & \colhead{} & \colhead{} & \colhead{} & \colhead{tration ratio\tablenotemark{c})} & \colhead{}
}

\startdata
\cutinhead{S Group}
045111.39-692646.7 &    5.SA        &  C  & yes (0.546) & no (0.402) & no (0.411) & no (0.268) &    \checkmark     &    \checkmark    &    \checkmark    &    \checkmark    &    \checkmark   \\ 
045550.62-663434.6 &    5.U         &  C  & yes (0.361) & no (0.674) & no (0.372) & no (0.375) &    \checkmark     &    \checkmark    &    \checkmark    &    \checkmark    &    \checkmark   \\ 
045842.47-660835.7 &    5.SA        &  C  & yes (0.299) & no (0.71) & no (0.524) & no (0.331) &                   &    \checkmark    &                  &                  &    \checkmark   \\ 
045854.30-660718.9 &    5.SA        &  C  & yes (0.35) & no (0.846) & no (0.519) & no (0.212) &                   &    \checkmark    &                  &                  &    \checkmark   \\ 
051024.09-701406.5 &    5.SA        &  C  & yes (0.365) & no (0.715) & no (0.55) & no (0.341) &                   &                  &                  &    \checkmark    &    \checkmark   \\ 
051912.27-690907.3 &    5.SA        &  C  & yes (0.559) & no (0.632) & no (0.429) & no (0.249) &    \checkmark     &    \checkmark    &                  &    \checkmark    &    \checkmark   \\ 
052333.40-693712.2 &    5.SA        &  C  & yes (0.476) & no (0.547) & no (0.459) & no (0.292) &    \checkmark     &                  &                  &                  &    \checkmark   \\ 
052646.61-684847.2 &    5.SAe       &  C  & yes (0.356) & no (0.85) & no (0.547) & no (0.26) &    \checkmark     &    \checkmark    &    \checkmark    &    \checkmark    &    \checkmark   \\
\enddata
\tablenotetext{a}{Primary GC09 classifications for the entire source sample. When applicable, the primary classification is followed by the secondary classification. GC09 classifications are as follows: C -- candidate YSO; A -- AGB or post-AGB star (non-ERO); D -- diffuse source; G -- background galaxy; S -- normal star; E -- extremely red object. Sources without a GC09 classification are assigned a U -- unknown -- classification.}
\tablenotetext{b}{Dot-product of the source's spectrum with the eigenvector containing the pertaining spectral feature.}
\tablenotetext{c}{Ratio of the equivalent width per pixel of the [SIII] 18.7 $\mu$m fine-structure line of the full slit extraction to the equivalent width per pixel of the same line taken at the 1-pixel wide `off-source' slit position. Sources without values do not have a measureable [SIII] line.}
\end{deluxetable}

\clearpage

\begin{deluxetable}{lc|cccccccc}
\tabletypesize{\small}
\tablecolumns{10}
\tablewidth{0pt}
\tablecaption{Comparison between GC09 and Our Classification}
\tablehead{ 
\multicolumn{2}{c}{Our Classification}  & \colhead{} & \multicolumn{7}{c}{GC09 Primary \& Secondary Classification} \\
\cline{1-2} \cline{4-10} 
\colhead{Group} & \colhead{Total} & \colhead{} & \colhead{YSO} & \colhead{AGB} & \colhead{Galaxy} & \colhead{Diffuse} & \colhead{Stellar} & \colhead{ERO} & \colhead{Unknown}\\
\colhead{} & \colhead{} & \colhead{} & \colhead{} & \colhead{(non-ERO)} & \colhead{} & \colhead{} & \colhead{} & \colhead{} & \colhead{}}

\startdata
Total& \nodata  && \bf{269} & \bf{3 (1)} & \bf{0 (1)} & \bf{2 (2)} & 0\bf{ (5)} & \bf{7}     & \bf{13}\\
\cline{1-10}
&&&&&&&&&\\
YSO  & \bf{277}      && 262 & 1 (1) & 0 (1) & 2 (2) & 0 (4) & 0     & 12\\
S    & \bf{12  }     && 12  & 0     & 0     & 0     & 0     & 0     & 0\\
SE   & \bf{5   }     && 5   & 0     & 0     & 0     & 0     & 0     & 0\\
P    & \bf{100   }    && 96  & 1     & 0     & 0     & 0     & 0     & 3 \\
PE   & \bf{142}      && 131 & 0     & 0 (1) & 2 (2) & 0 (3) & 0     & 9 \\
E    & \bf{2   }     && 2   & 0     & 0     & 0     & 0     & 0     & 0\\
\enddata
\tablecomments{GC09 primary classification counts are given for every GC09 class. When applicable, secondary classification counts are provided in parentheses. The total number of sources in the spectral groups assigned in this paper are provided in the \textit{column} labeled `Total,' and the total number of sources in each GC09 class are provided in the \textit{row} labeled `Total.' Unbolded numbers show the overlap betweeb the two classification schemes.}
\tablenotetext{1}{Objects in the U group are not able to be classified due to incomplete spectral coverage.}
\end{deluxetable}

\end{document}